\documentclass[preprint,12pt]{elsarticle}

\usepackage{fullpage}
\usepackage{amsmath}
\usepackage{epsfig}    
\usepackage{amsfonts}
\usepackage{amssymb}
\usepackage{amsmath}
\usepackage{subfigure}
\usepackage{comment}
\usepackage{soul} 
\usepackage{hyperref}
\usepackage{color}
\usepackage{mathtools}
\usepackage{algorithmic}

\usepackage{textcomp}
\usepackage{siunitx}
\usepackage{epstopdf}
\usepackage{url}
\usepackage{caption}\usepackage{tikz}
\usepackage{tkz-tab}
\usetikzlibrary{automata,arrows,positioning,calc}
\usetikzlibrary{shapes,snakes}
\usetikzlibrary{arrows}
\usepackage{multirow}
\usepackage{booktabs}
\usepackage{hyperref}
\usepackage{scalerel}
\usepackage[toc,page]{appendix}

\usepackage[super]{nth}
\usepackage{graphicx,adjustbox}
\usepackage{caption}
\usepackage{setspace}
\usepackage{xspace}
\usepackage{siunitx}
\usepackage{soul}
\usepackage{url}
\usepackage{tablefootnote}
\usepackage[linesnumbered,ruled]{algorithm2e}
\usepackage{booktabs}
\usepackage{hyperref}
\usepackage[normalem]{ulem}
\usepackage{footnote}
\usepackage[misc,geometry]{ifsym} 
\makesavenoteenv{tabular}

\let\OldTexttrademark\texttrademark
\renewcommand{\texttrademark}{\OldTexttrademark\xspace}%

\usepackage{amsthm}
\theoremstyle{plain}
\newtheorem{assumption}{Assumption}

\usepackage{tabularx,colortbl}
\definecolor{mygray}{gray}{0.9}

\journal{Computer Communications}

\begin{document}
	
\begin{frontmatter}
	
\title{Spatial Reuse in IEEE 802.11ax WLANs}

\author[label1]{Francesc Wilhelmi \corref{cor1}} % \corref{cor1}}%\ead{francisco.wilhelmi@upf.edu}
\author[label1]{Sergio~Barrachina-Mu\~noz}
\author[label2]{Cristina~Cano}
\author[label3]{Ioannis~Selinis}
\author[label1]{Boris~Bellalta}
\address[label1]{Wireless Networking Research Group (WN-UPF), 08002 Barcelona, Spain}
\address[label2]{Wireless Networks Research Group (WINE-UOC), 08860 Barcelona, Spain}
\address[label3]{Institute for Communications Systems (ICS-UoS), GU2 7XH Surrey, United Kingdom}
\cortext[cor1]{Corresponding Author: Francesc Wilhelmi (\href{francisco.wilhelmi@upf.edu}{francisco.wilhelmi@upf.edu})}

\begin{abstract}
Dealing with massively crowded scenarios is one of the most ambitious goals of next-generation wireless networks. With this goal in mind, the IEEE 802.11ax amendment includes, among other techniques, the Spatial Reuse (SR) operation. The SR operation encompasses a set of unprecedented techniques {that are expected to significantly boost Wireless Local Area Networks (WLANs) performance in dense environments}. In particular, the main objective of the SR operation is to maximize the utilization of the medium by increasing the number of parallel transmissions. Nevertheless, due to the novelty of the operation, its performance gains remain largely unknown. In this paper, we first provide a gentle tutorial of the SR operation included in the IEEE 802.11ax. Then, we analytically model SR and delve into the new kinds of MAC-level interactions among network devices. Finally, we provide a simulation-driven analysis to showcase the potential of SR in various deployments, comprising different network densities and traffic loads. Our results show that the SR operation can significantly improve the medium utilization, especially in scenarios under high interference conditions. Moreover, our results demonstrate the non-intrusive design characteristic of SR, which allows enhancing the number of simultaneous transmissions with a low impact on the environment. We conclude the paper by giving some thoughts on the main challenges and limitations of the IEEE 802.11ax SR operation, including research gaps and future directions.
\end{abstract}
	

\end{frontmatter}
		
\newpage	
		
\section{Introduction}
\label{section:intro}

Due to the popularity and ease of deployment of IEEE Wireless Local Area Networks (WLANs), it is becoming increasingly common to find multiple Basic Service Sets (BSSs) within the same overlapping areas. Unfortunately, the most typical channel access mechanism based on Carrier Sense Multiple Access (CSMA) was not designed to support a massive number of contending devices, thus resulting in low performance.

Several amendments have been conceived over the past few years to improve the performance of WLANs. Earlier IEEE 802.11 standards, e.g., 11n (2009) and 11ac (2013), defined the concepts of High Throughput (HT) and Very High Throughput (VHT) devices, respectively. These standards defined new functionalities to be included at that time, such as Channel Bonding (CB). More recently, the Task Group ax (TGax) was created to develop the IEEE 802.11ax-2021 (11ax) standard \cite{tgax2019draft}, which belongs to the group of standards for next-generation WLANs (e.g., IEEE 802.11aq, IEEE 802.11ad, IEEE 802.11ay). Through the definition of High Efficiency (HE) WLANs, the 11ax aims to improve network efficiency in dense deployments. To that purpose, it includes several novel techniques, such as Orthogonal Frequency Division Multiple Access (OFDMA), Downlink/Uplink Multi-User Multiple-Input-Multiple-Output (DL/UL MU-MIMO), and the Spatial Reuse (SR) operation. We refer the reader to the works in \cite{bellalta2016ieee, afaqui2016ieee, qu2018survey, khorov2018tutorial} for an overview of the major novelties proposed in the IEEE 802.11ax standard.

\begin{figure}[ht!]
	\centering
	\includegraphics[width=.6\columnwidth]{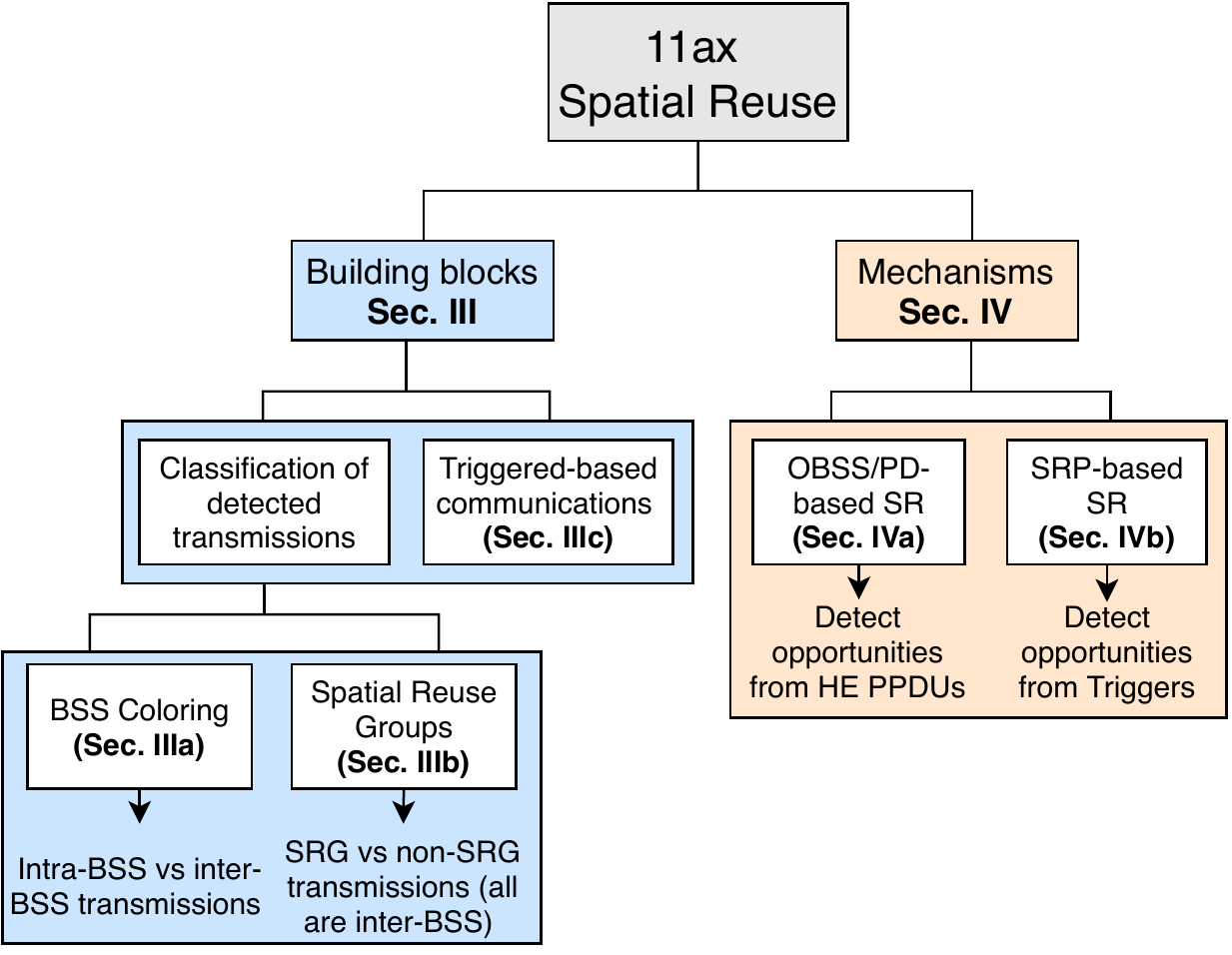}
	\caption{Summary of the 11ax SR operation.}
	\label{fig:sr_summary}
\end{figure}

In this paper, we focus on the 11ax SR operation \cite{merlin2009methods}, which seeks to increase the number of parallel transmissions and therefore improve spectral efficiency. In order to do so, the amendment introduces Carrier Sense Threshold (CST) adjustment for the detected inter-BSS transmissions,\footnote{In the following, we will use intra-BSS or inter-BSS to refer to the transmissions detected from the same or a different BSS, respectively.} which is achieved through two different mechanisms: \emph{i)} \textcolor{black}{Overlapping Basic Service Set (OBSS)} Packet Detect (PD)-based SR, and \emph{ii)} Parametrized Spatial Reuse (PSR). The main difference between the two mechanisms lies in the degree of collaboration among BSSs for identifying SR-based opportunities (further details are provided in Sections \ref{section:enablers_sr_11ax} and \ref{section:operation_sr_11ax}). Both mechanisms include Transmission Power Control (TPC) to limit the additional interference produced by simultaneous transmissions. Figure~\ref{fig:sr_summary} summarizes the components that constitute the 11ax SR operation, which \textcolor{black}{is} described in detail throughout this paper.

\textcolor{black}{The SR operation has been introduced as a mechanism to increase the number of transmissions held in an OBSS, and thus spectral efficiency. Nevertheless, little is known about the actual potential of SR.} The fact is that \textcolor{black}{the performance of a device implementing SR} depends on multiple factors, such as the network topology, the type of propagation effects, or the type of radio used by devices \cite{guo2003spatial, zhu2004adapting}. Moreover, sensitivity adjustment and power control may result in asymmetric links that can potentially lead to \textcolor{black}{unfair} situations \cite{mhatre2007interference}. In some cases, dynamic sensitivity and transmission power adjustment have been shown to significantly increase the network performance and contribute to reducing the effects of the well-known hidden and exposed terminal problems \cite{zhou2005balancing}. However, in some other cases, these problems may be exacerbated \cite{wilhelmi2019potential}. Modifying either the CST or the transmit power can worsen the hidden/exposed terminal problems by generating flow starvation and asymmetries.

\begin{figure}[ht!]
	\centering
	\includegraphics[width=0.4\columnwidth]{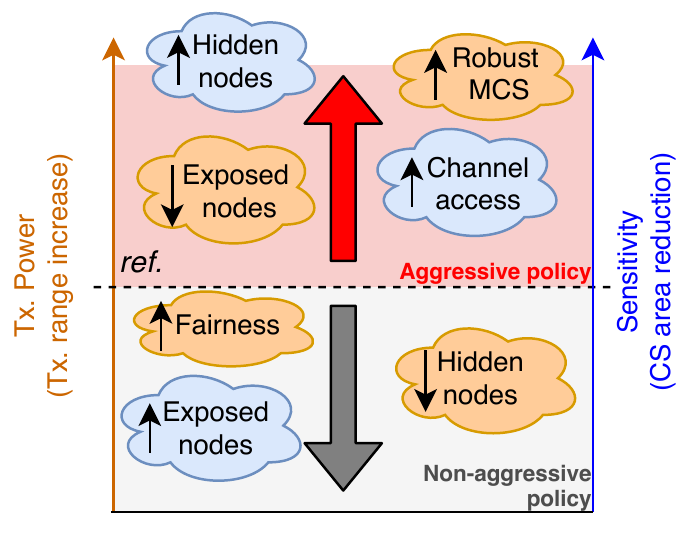}
	\caption{Effects of different policies concerning sensitivity adjustment and transmission power control.}
	\label{fig:policies_sr}
\end{figure}

Figure~\ref{fig:policies_sr} \textcolor{black}{summarizes} the effect of increasing and decreasing both the transmission power and the sensitivity in WLANs. For instance, increasing the sensitivity may contribute to accessing the channel more often since the \textcolor{black}{Carrier Sense (CS) area} is reduced. However, this can lead to observing a higher number of collisions by hidden nodes. Moreover, using a more aggressive channel access policy may expose the receivers to a higher level of interference, thus requiring a more robust Modulation and Coding Scheme (MCS).

%As discussed, dealing with the spatial dimension has different implications and leads to complex inter-BSS interactions that are hard to predict beforehand. 
The 11ax SR operation is one of the least studied features for next-generation WLANs, and only a few works have devised its potential. Firstly, the authors of \cite{mori2014performance} evaluated the benefits of using dynamic sensitivity thresholds for inter-BSS transmissions, given a fixed transmit power. Secondly, the work in \cite{qu2018survey} exhaustively surveyed the 11ax amendment, thus providing an overview of the first drafted SR operation. Moreover, it provided some results of applying SR in both indoor and outdoor scenarios, showing a higher potential for indoor deployments. Similarly, the authors of \cite{shen2018research} introduced the SR operation as described in the 11ax amendment. Besides, they provided a performance evaluation based on the adjustment of the inter-BSS sensitivity threshold. Their results showed significant gains when applying SR, especially for dense scenarios. 

% Contributions
Unlike in \cite{mori2014performance, qu2018survey, shen2018research}, in this paper, we delve into the SR operation in more detail since we consider the two mechanisms included in the 11ax amendment. In addition, our analysis of the 11ax SR is not limited to the technical information included in the amendment. Instead, we accompany our descriptions with illustrative use cases, bringing a new perspective that allows devising the real utility behind the operation. Thus, we go beyond the definition of the specification, shedding light on its purpose, benefits, and challenges. Our aim in this paper is to provide a comprehensive tool for researchers interested in the topic and to analyze the potential of the SR operation in future WLANs. Besides, we focus on the potential gaps in the standard to be filled by the research community. This paper's main contributions lie in the description, analysis, and evaluation of the 11ax SR operation. In particular:
\begin{enumerate}
	\item We provide a gentle, exhaustive, and comprehensive overview of the SR operation included in the 11ax amendment.
	\item We analytically model the 11ax SR operation for the sake of capturing and understanding the new kinds of inter-BSS interactions that result from it. The results of this model are verified with the Komondor 11ax-based simulator \cite{barrachina2019komondor}.
	\item We study the potential performance gains of the 11ax SR operation through simulations. 
	\item We delve into the gaps and gray areas existing in the current 11ax SR operation and devise future research directions in the field.
\end{enumerate}

% Document structure
The remainder of this document is structured as follows. Section \ref{section:previous_work_sr} surveys the related work on SR in WLANs. Section \ref{section:enablers_sr_11ax} describes the specifications and procedures that enable 11ax SR, whereas Section \ref{section:operation_sr_11ax} details the operation itself. Section~\ref{section:analytical_model} presents an analysis-based study of 11ax SR in simple scenarios. The analysis is extended in Section~\ref{section:performance_evaluation} through the simulation of dense deployments. Section \ref{section:ways_forwad} identifies the gaps and research opportunities found within the 11ax SR operation and explores potential ways forward. Finally, Section~\ref{section:conclusions} provides some concluding remarks.

% ----------------------------------
% -
% 	-- Previous Work on SR --
% -
% ----------------------------------
\section{Spatial Reuse techniques in IEEE 802.11 WLANs}%\section{Related Work on Spatial Reuse in IEEE 802.11 WLANs}
\label{section:previous_work_sr}

\textcolor{black}{The SR operation relies on dynamic Clear Channel Assessment CS (CCA/CS) adjustment to increase the number of transmission opportunities (TXOPs) in an OBSS. The CCA/CS mechanism is triggered at a Wi-Fi device upon detecting the preamble of another device's transmission.\footnote{\textcolor{black}{Similarly, the Energy Detect (ED) threshold is applied for non-802.11 signals.}} Notice that detected transmissions (above the physical sensitivity threshold) may not be properly decoded if the received signal is poor. In contrast, for decoded transmissions above the CCA/CS threshold, either physical or virtual carrier sense operation sets the medium to busy. Moreover, the capture effect is used when detecting multiple signals, thus allowing to lock onto the strongest signal instead of experiencing packet collisions.}

\begin{figure}[ht!]
	\centering
	\includegraphics[width=.9\textwidth]{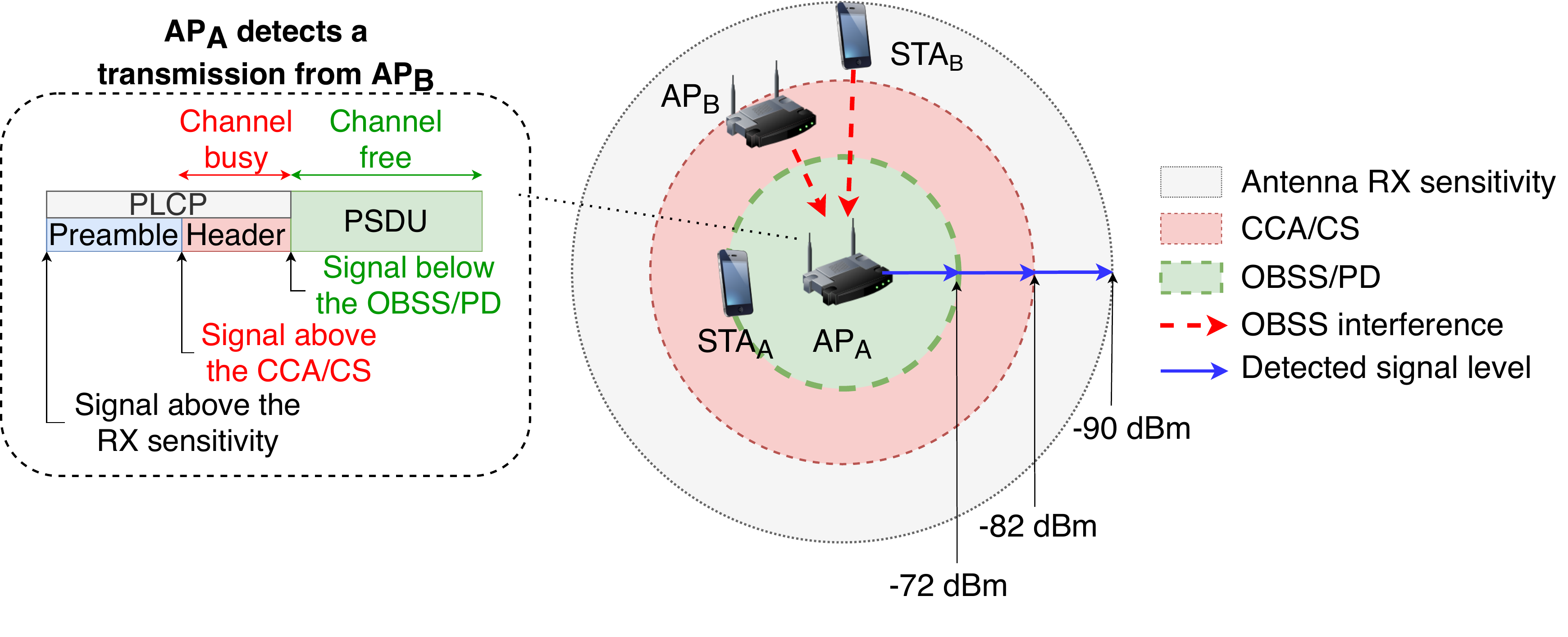}
	\caption{\textcolor{black}{Transmission detection and decoding in IEEE 802.11 devices. Unlike typical coverage representations in wireless networks, we illustrate the CS area of a given device (instead of the interference it generates). The transmission power is fixed and all the devices use the same frequency channel.}}
	\label{fig:spatial_reuse_11ax}
\end{figure}

\textcolor{black}{The aforementioned concepts are illustrated in Figure~\ref{fig:spatial_reuse_11ax}, where the Access Point (AP) in the center (namely, AP$_\text{A}$), may detect incoming signals above the antenna's receiver sensitivity (in gray), but only decode those above the CCA/CS threshold (in red). In addition, due to the 11ax SR operation, an OBSS/PD threshold can be used to ignore AP$_\text{B}$'s transmissions (in green), thus enhancing channel utilization. Moreover, for TXOPs detected by using the OBSS/PD threshold, a transmit power limitation is enforced.}
%If the new incoming frame arrives while the receiver is receiving the preamble of another frame and the SIR of the new incoming frame is above a fixed margin, then the current frame is dropped and the receiver locks onto the new incoming frame. If more than one PPDU arrives at a receiver within a PPDU capture window of A nsec, then the strongest of the arriving PPDUs shall be the one captured (A is TBD but with one option of 800ns).  The PPDU capture window of A nsec starts at the first arrival PPDU with rx power higher than rx sensitivity. At the same time just after the end of the preamble, a preemption window (the window size is TBD in [0, PPDU duration]) begins during which time if a new PPDU arrives with rx power at least PdB above the current reception, then the current reception is terminated and the PPDU capture phase is re-entered with this new PPDU (P is expected to be have a value of about 9 dB)

The problem of dynamic sensitivity and transmission power adjustment has been previously addressed in multiple ways. \textcolor{black}{Firstly}, we find centralized solutions such as the ones proposed in \cite{li2011achieving, jamil2016novel, nakahira2014centralized}, where the SR operation is controlled and mandated from the APs. Among these, we highlight \cite{jamil2016novel}, which uses a method based on Neural Networks (NN) to compute the best combination of sensitivity and transmit power to be used by all the BSSs in a given scenario. Nonetheless, centralized approaches require coordination and extra overhead, which is usually impractical.

\textcolor{black}{Secondly}, SR has been addressed through a decentralized perspective in \cite{chevillat2005dynamic, tang2011improving, chau2017effective, wilhelmi2019collaborative, wilhelmi2019potential}. Most of the decentralized strategies rely on collecting feedback on several performance metrics (e.g., sensed interference, packets lost). While works such as \cite{chevillat2005dynamic, tang2011improving, chau2017effective} propose adaptive mechanisms to adjust the CST and the transmission power, some others like \cite{wilhelmi2019collaborative, wilhelmi2019potential} provide probabilistic approaches based on Reinforcement Learning (RL) for finding the best possible configuration. 

Concerning IEEE 802.11ax WLANs, the Dynamic Sensitivity Control (DSC) scheme was proposed to be included in the standard, but it was never incorporated. The performance of DSC was evaluated in \cite{afaqui2015evaluation, afaqui2016dynamic, kulkarni2015taming}. Furthermore, the authors of \cite{selinis2016evaluation, selinis2017exploiting} combined DSC with BSS color schemes to devise further improvements in WLANs.

The current 11ax SR operation has nonetheless been studied to a lower extent. Based on the OBSS/PD-based SR operation, the work in \cite{selinis2018control} proposed a new mechanism to adjust the OBSS/PD threshold.\footnote{The OBSS/PD threshold refers to the sensitivity to be used for detected inter-BSS transmissions.} This mechanism, so-called Control OBSS/PD Sensitivity Threshold (COST), differs from DSC in terms of the information available in 11ax nodes. In this case, nodes need to be aware of changes in the neighboring BSSs. \textcolor{black}{Besides, \cite{ropitault2018evaluation} proposed a method for adjusting the OBSS/PD threshold in 11ax WLANs, which is based on the Received Signal Strength Indicator (RSSI).}

Unlike previous works, we focus on the IEEE 802.11ax SR operation defined in Draft v4.0 and delve into its potential through analytical modeling and a simulation tool. Moreover, we identify potential gaps and research opportunities \textcolor{black}{concerning} the amendment.

% ----------------------------------
% -
% 	-- IEEE 802.11ax Preliminaries --
% -
% ----------------------------------
\section{IEEE 802.11ax Spatial Reuse Operation: Building Blocks}
\label{section:enablers_sr_11ax}
Before delving into the 11ax SR mechanisms, we first describe the enabling concepts and features. In particular, the 11ax SR operation can be understood through \textbf{BSS coloring} and \textbf{Spatial Reuse Groups (SRG)}. Besides, we introduce the \textbf{Triggered-based (TB) transmissions} upon which the PSR operation is based.

%% BSS COLORING
\subsection{BSS coloring}	
\label{section:bss_coloring}	
BSS coloring is a key enabler of the 11ax SR operation, whereby HE nodes can rapidly identify the source of a given \textcolor{black}{detected} transmission. \textcolor{black}{When implementing BSS coloring,} a given device can effectively determine whether the channel is occupied by another device belonging to the same BSS (intra-BSS transmission, same color) or from a different one (inter-BSS transmission, different color). \textcolor{black}{The BSS color is determined by the AP (an integer between 1 and 63) and is included in the preambles of Wi-Fi frames.}\footnote{The \texttt{BSS color} field is included in the Physical Layer Convergence Procedure (PLCP) header. See Appendix \ref{section:frames} for further details.} It remains static until the AP considers to change it\textcolor{black}{, which can be provoked when a BSS color overlap is noticed (i.e., two different BSSs use the same color).} The method for selecting a new color is out of the scope of the 11ax amendment, but the advertising procedure is defined. An HE AP may announce a new BSS color via the \texttt{BSS Color Change Announcement} element, carried in Beacon, Probe Response, and (Re)Association Response frames. 

% Intra-BSS and Inter-BSS frames
\subsubsection{BSS color-based channel access rules}
\label{section:bss_color_channel_access}
When detecting a transmission, an HE node can distinguish between intra and inter-BSS frames by rapidly inspecting the \texttt{BSS color} field that is carried in every HE PLCP Protocol Data Unit (PPDU).\footnote{If the \texttt{BSS color} is not announced, frames can be classified according to the \texttt{GROUP\_ID} and \texttt{PARTIAL\_AID} in VHT PPDUs, or the MAC address in the MAC header of legacy frames (i.e., predecessor amendments of the IEEE 802.11ac).} 

In practice, once a frame is detected (the received power is above the receiver's sensitivity), a device locks onto it and starts decoding the fields of the different headers. \textcolor{black}{If the preamble and PHY headers can be successfully decoded (the received signal of interest is above the CCA/CS), the frame is forwarded to the MAC layer for further processing (e.g., determining the transmission destination, or setting virtual carrier sensing). Otherwise, the packet is considered as interference.} In particular, HE nodes applying SR use the default CCA/CS threshold (i.e., -82 dBm) upon detecting intra-BSS frames. On the contrary, when inter-BSS frames are detected, more aggressive PD thresholds can be applied to increase the number of parallel transmissions. Those PD thresholds are termed \textbf{non-SRG OBSS/PD} and \textbf{SRG OBSS/PD}. The SRG OBSS/PD is used when spatial reuse groups are allowed, which is discussed in detail in Section \ref{section:srg}.

\begin{figure}[ht!]
	\centering
	\subfigure[Scenario]{\label{fig:11ax_bss_coloring_a}\includegraphics[width=0.35\columnwidth]{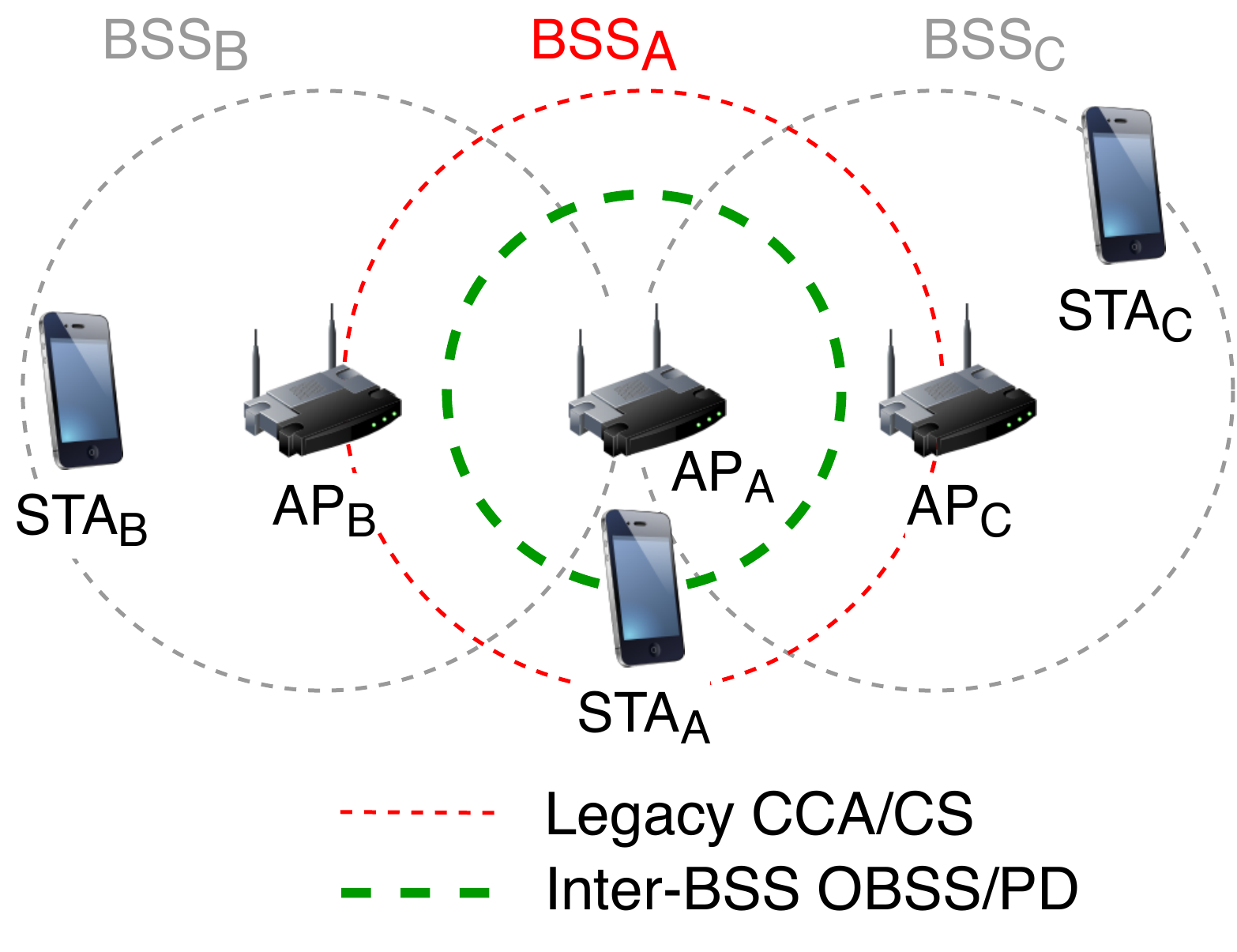}}
	\hspace{1cm}
	\subfigure[Packets exchange]{\label{fig:11ax_bss_coloring_b}\includegraphics[width=0.45\columnwidth]{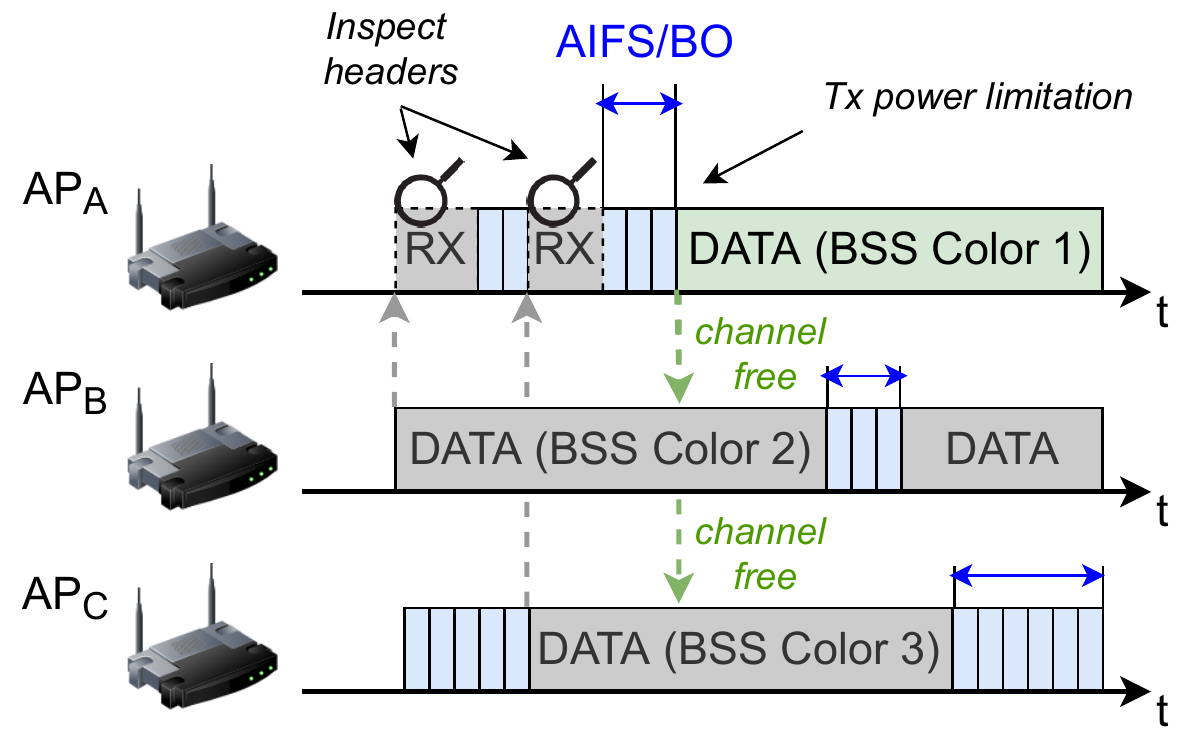}}
	\caption{Channel access rules based on BSS coloring. In (b), the propagation delay is considered to be negligible.}
\end{figure}

To illustrate how BSS coloring can help at enhancing SR, let us consider the scenario shown in Figure~\ref{fig:11ax_bss_coloring_a}. We consider that $\text{AP}_\text{A}$ is prone to suffer from flow starvation if it uses the default CCA/CS value, since it is in the range of both $\text{AP}_\text{B}$ and $\text{AP}_\text{C}$. Note that simultaneous downlink transmissions can be held in both $\text{BSS}_\text{B}$ and $\text{BSS}_\text{C}$ because the transmitters are not in the range of each other. The flow-in-the-middle starvation generated to $\text{BSS}_\text{A}$ can be overcome by applying an OBSS/PD threshold higher than the CCA/CS for inter-BSS frames. The way $\text{BSS}_\text{A}$ can ignore inter-BSS transmissions is illustrated in Figure~\ref{fig:11ax_bss_coloring_b}, where $\text{AP}_\text{A}$ first identifies the source of a detected transmission by inspecting its headers (notice that the frame reception procedure has been simplified for the sake of illustration). Then, after detecting the source of the ongoing transmissions (indicated by the color), the non-SRG OBSS/PD threshold is applied. If the OBSS/PD threshold is high enough to ignore inter-BSS transmissions, $\text{AP}_\text{A}$ can reset the PHY and continue the backoff procedure. Besides, it is important to remark the importance of the capture effect for preventing losses from frame collisions \cite{lee2007experimental}. A widely adopted assumption to characterize IEEE 802.11 WLANs is that the Signal-to-Noise-plus-Interference Ratio (SINR) of a receiver device is sufficiently high to overcome interfering devices \cite{durvy2007modeling}.

\subsubsection{Two NAVs}
\label{section:two_navs}
The SR operation provides significant changes in the virtual carrier sensing procedure since two different Network Allocation Vectors (NAVs), i.e., \emph{intra-BSS NAV} and \emph{inter-BSS NAV}, have to be maintained for intra and inter-BSS frames, respectively. Accordingly, a given transmitter can decrease its backoff counter only if both NAV timers are set to zero. Otherwise, it must remain idle for at least the duration of the ongoing transmission(s),\footnote{The duration used for setting the NAV is indicated in the Duration field of RTS/CTS frames or Physical layer Service Data Units (PSDU).} which had previously activated the virtual carrier sensing. 

\textcolor{black}{The reason for holding two NAVs is twofold: \emph{1)} increase the number of parallel transmissions by mitigating inter-BSS interference, and \emph{2)} avoid potential misbehavior caused by Contention Free End (CF-End) control frames, whereby the NAV can be canceled. For the latter, notice that either intra or inter-BSS frames can announce the end of a CFP (Contention-Free Period) by sending a CF-End frame. In dense deployments at which hidden and exposed terminal problems are exacerbated, this may lead to experiencing a high number of collisions.}

\begin{figure*}[ht!]
	\centering
	\subfigure[Scenario]{\label{fig:fig_5_a}\includegraphics[width=0.35\columnwidth]{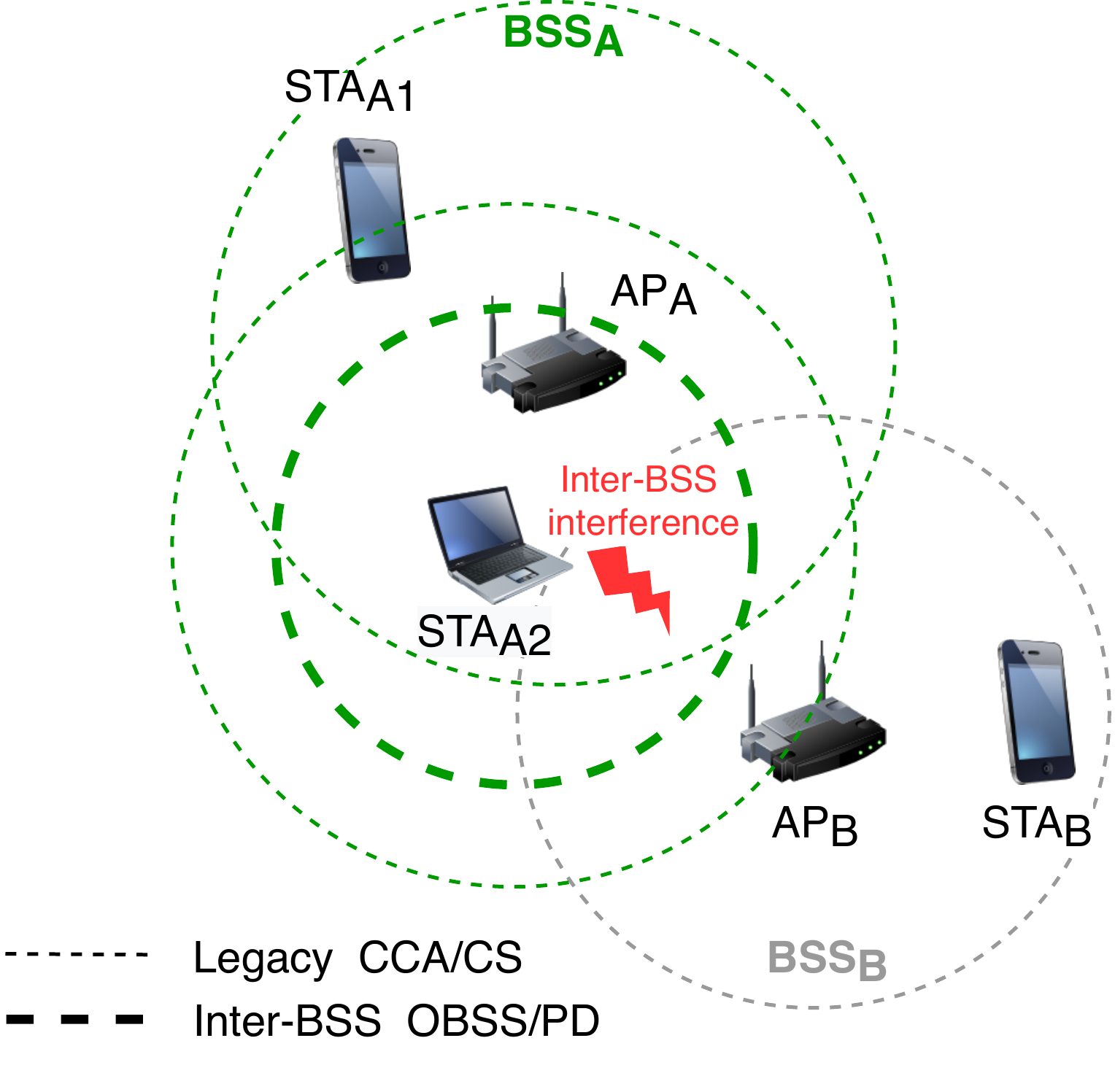}}
	\hspace{1cm}
	\subfigure[Packets exchange]{\label{fig:fig_5_b}\includegraphics[width=.5\columnwidth]{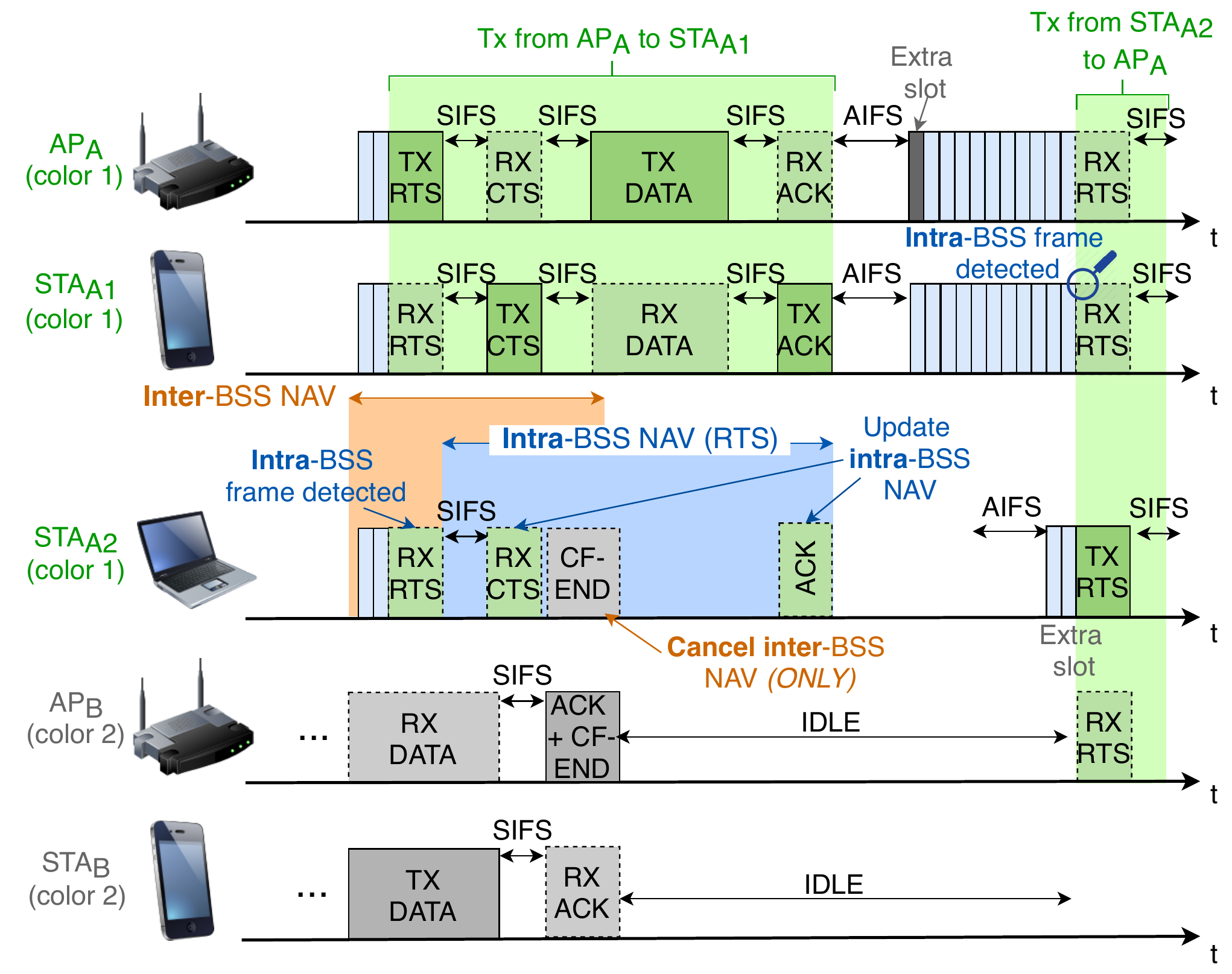}}
	\caption{\textcolor{black}{An example of the \textit{Two NAVs} operation in an OBSS.}}
	\label{fig:two_navs}
\end{figure*}

\textcolor{black}{To illustrate the \emph{two NAVs} concept, let us consider the deployment shown in Figure~\ref{fig:fig_5_a}, where packets are exchanged as illustrated in Figure~\ref{fig:fig_5_b}. In this scenario, $\text{BSS}_\text{A}$ and $\text{BSS}_\text{B}$ can transmit simultaneously in case of using the OBSS/PD threshold. Nevertheless, provided that a single NAV was used, STA$_{\text{A2}}$ is prone to misbehave when AP$_\text{B}$ sends a CF-End frame for indicating the end of the virtual carrier sense period. On the contrary, due to the two NAVs operation, STA$_{\text{A2}}$ can identify the source of the CF-End, thus canceling the inter-BSS NAV only.}

% Two NAVs

%%% SR groups
\subsection{Spatial Reuse Groups}
\label{section:srg}
To further enhance network efficiency, the 11ax amendment provides a mechanism for differentiating between two types of inter-BSS frames; that is to say, belonging or not to the same SRG. These groups can be formed by BSSs to achieve a more sophisticated SR operation. For instance, more aggressive channel access policies can be used for transmissions within the same SRG, \textcolor{black}{in case that the nodes of the same SRG could support higher levels of interference}. \textcolor{black}{Alternatively}, it could be the other way around. A conservative policy can be employed for the sake of minimizing collisions by hidden nodes. \textcolor{black}{Although the formation of SRGs is out of the amendment's scope,} differentiating between two OBSS/PD thresholds can be useful for capturing more subtle inter-BSS interactions. Note, as well, that SRGs could be formed online to address some issues detected by an entity controlling a set of APs (e.g., belonging to the same operator).

\textcolor{black}{For supporting SRGs,} the involved HE nodes must have indicated support for it. \textcolor{black}{HE STAs enable the SRG feature} upon the reception of an activating \texttt{Spatial Reuse Parameter Set} element (further described in Appendix \ref{section:srps}) from their AP. Then, for the following detected PPDUs, both HE APs and STAs may differentiate between SRG and non-SRG PPDUs. Note that 11ax devices can also identify the source of non-HE transmissions. Therefore, not only HE devices are supported but also legacy devices. The way of classifying SRG frames is backward compatible with previous IEEE 802.11 amendments. Technically speaking, SRG identification is made as follows:
\begin{itemize}
	\item For HE PPDUs, an HE STA inspects the \texttt{BSS color} and checks if it belongs to the same SRG. This information is kept on the \texttt{SRG BSS Color Bitmap} of the \texttt{Spatial Reuse Parameter Set}, which stores the different BSS colors that belong to the same SRG. The AP of a given BSS is responsible for maintaining the SRG BSS Color Bitmap up to date, and to inform STAs \textcolor{black}{upon noticing any changes}.
	\item For VHT PPDUs, same SRG is indicated if the \texttt{GROUP\_ID} parameter (included in the \texttt{RXVECTOR}\footnote{The RXVECTOR constitutes a set of parameters that the PHY layer delivers to the MAC on receiving a PPDU.}) has a value of 0, and the bit in the \texttt{SRG Partial BSSID Bitmap} field corresponding to the numerical value of \texttt{PARTIAL\_AID}\footnote{The \texttt{PARTIAL\_AID} is an identifier which, similarly to the BSS color, is used by IEEE 802.11ac WLANs to identify the source of a given transmission quickly.} (also included in the \texttt{RXVECTOR}) is set to 1. 
	\item Finally, regarding other types of PPDUs, they are classified as SRG PPDUs if the BSSID information from a MAC Protocol Data Unit (MPDU) of the PPDU is correctly received and the bit in the \texttt{SRG Partial BSSID Bitmap} field corresponding to the numerical value of BSSID is 1.
\end{itemize}

\begin{figure}[ht!]
	\centering
	\includegraphics[width=0.4\columnwidth]{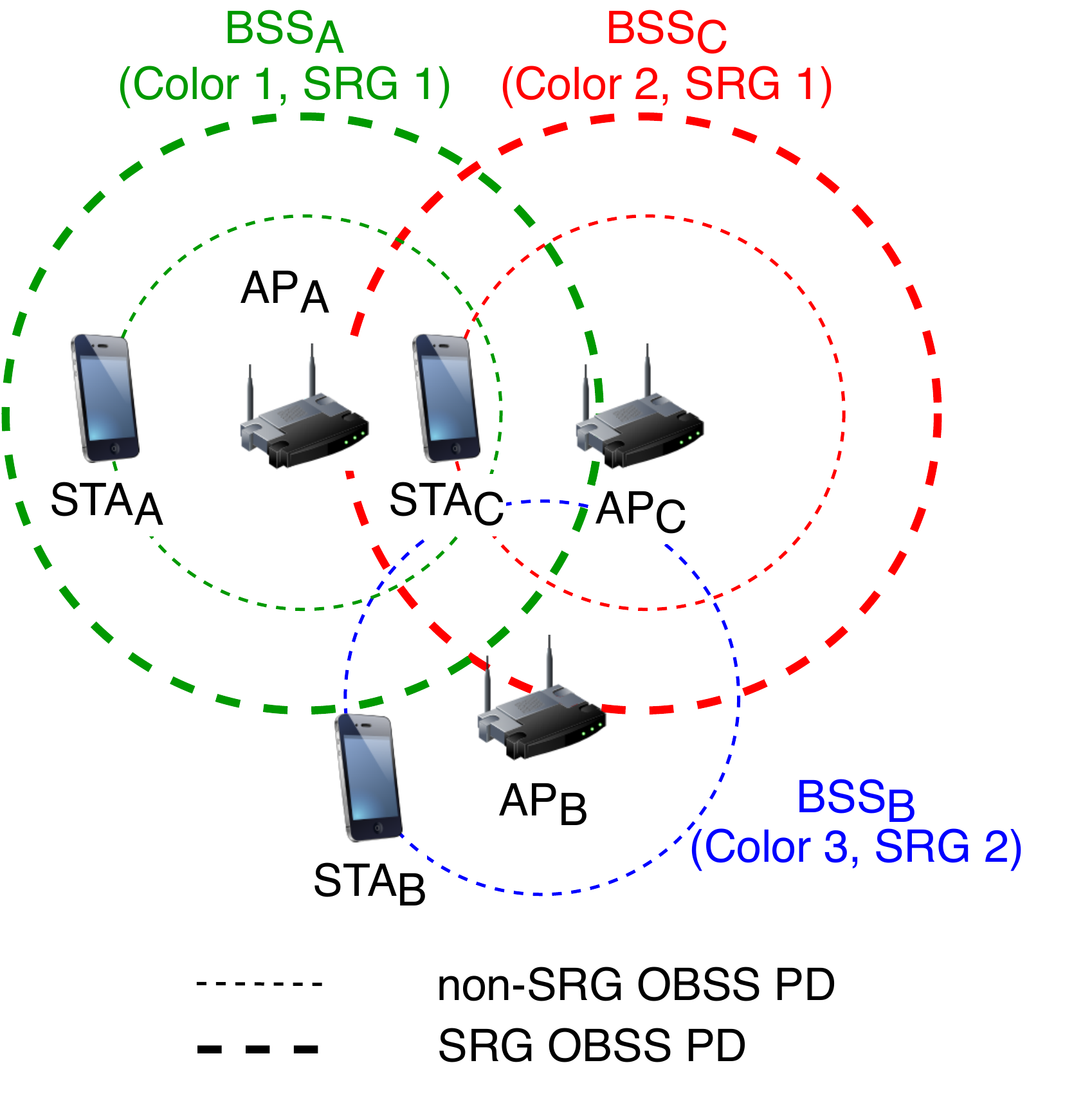}
	\caption{Spatial Reuse Groups in an OBSS.}
	\label{fig:fig_6}
\end{figure}

Figure~\ref{fig:fig_6} showcases the utility of SRG in a deployment at which using the OBSS/PD threshold homogeneously allows increasing parallel transmissions but leads to packet losses. In particular, $\text{STA}_\text{C}$ cannot properly decode the information sent by $\text{AP}_\text{C}$ when $\text{AP}_\text{A}$ is transmitting. To solve this, $\text{BSS}_\text{A}$ and $\text{BSS}_\text{C}$ can form a group and employ a more conservative SRG OBSS/PD threshold to avoid simultaneous transmissions between the two BSSs. Notice that the non-SRG OBSS/PD threshold (which is more aggressive) can be still employed for transmissions held by any pair of BSSs involving $\text{BSS}_\text{B}$, thus increasing network efficiency. %5As shown, not only the formation of SRGs is a complex task, but also the definition of both non-SRG and SRG OBSS/PD thresholds.

% SRG and non-SRG frames
\subsubsection{SRG-based Channel Access Rules}
\label{section:srg_channel_access}
Differentiating between SRGs may provide further SR enhancements than considering only one type of inter-BSS frame. \textcolor{black}{Although the specific utilization of SRGs is also out of the 11ax amendment scope}, we devise several situations where its application can be useful. As previously pointed out, one possibility is to establish groups for BSSs whose transmissions need to be protected. In other words, an HE STA detecting an SRG frame can implement a more conservative channel access policy. Conversely, a more aggressive policy can be applied for non-SRG PPDUs, thus increasing the number of parallel transmissions. 

\begin{figure}[ht!]
	\centering
	\includegraphics[width=.5\columnwidth]{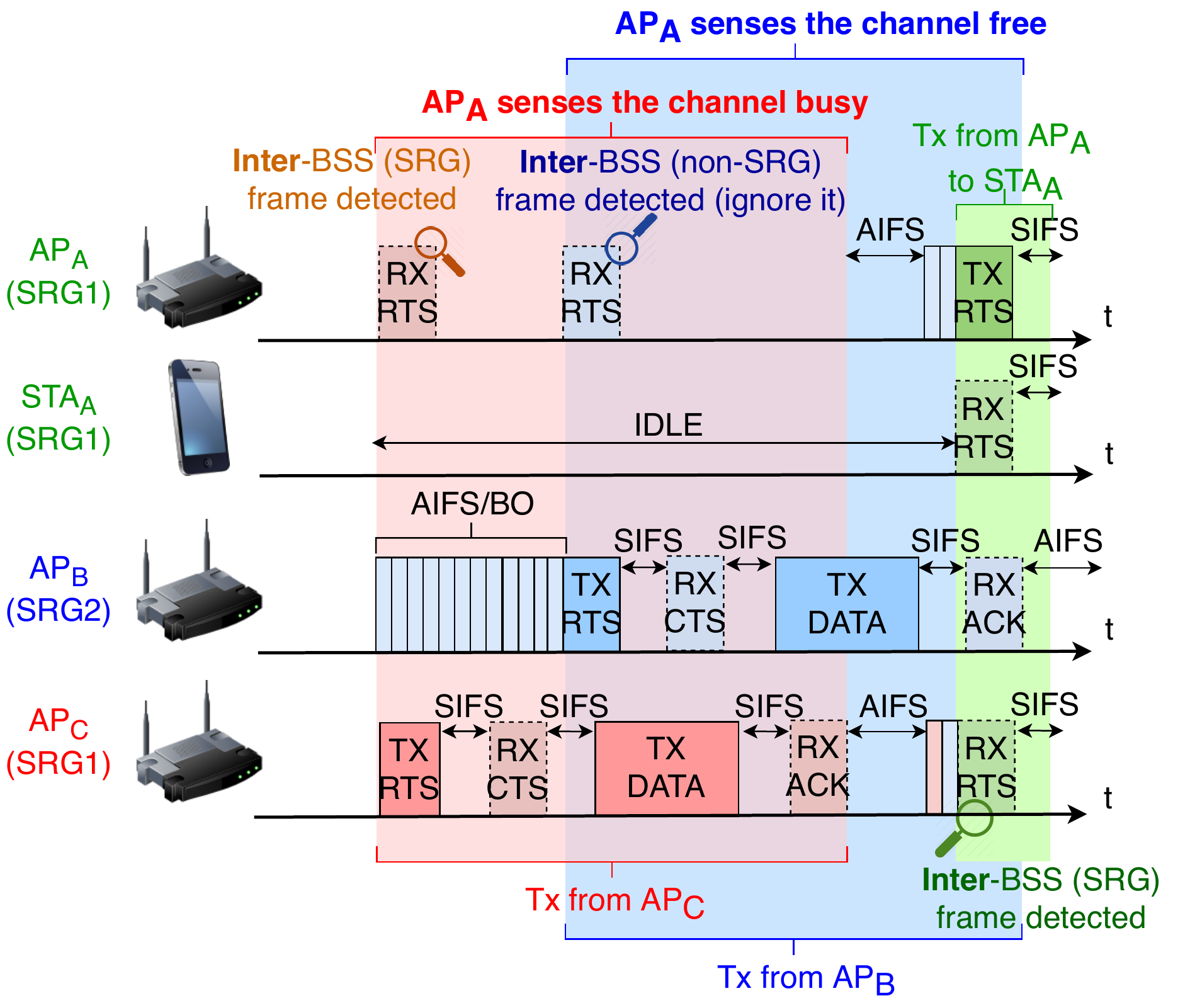}
	\caption{Packets exchange based on SRG channel access rules.}
	\label{fig:srg_channel_access}
\end{figure}

\textcolor{black}{To illustrate the SRG-based channel access rules, let us retake the scenario shown in Figure~\ref{fig:fig_6}. Figure~\ref{fig:srg_channel_access} shows the behavior of HE nodes when detecting transmissions from different SRGs. In particular, transmissions from $\text{BSS}_\text{C}$ (in blue) provoke that $\text{AP}_\text{A}$ senses the channel busy as a result of the SRG OBSS/PD-threshold. In contrast, frames detected from $\text{BSS}_\text{B}$ (in red) can be ignored by $\text{AP}_\text{A}$ because a less restrictive non-SRG OBSS/PD threshold is used. In this example, simultaneous transmissions between $\text{BSS}_\text{A}$ and $\text{BSS}_\text{B}$ are completely feasible, but the opposite occurs for $\text{BSS}_\text{A}$-$\text{BSS}_\text{C}$ interactions. Note that collisions may occur at STA$_\text{C}$ if $\text{AP}_\text{A}$ and $\text{AP}_\text{C}$ transmit simultaneously.}

%% TB communications
\subsection{Triggered-based communications}
\label{section:tb_communication}
\begin{figure}[ht!]
	\centering
	\includegraphics[width=.5\columnwidth]{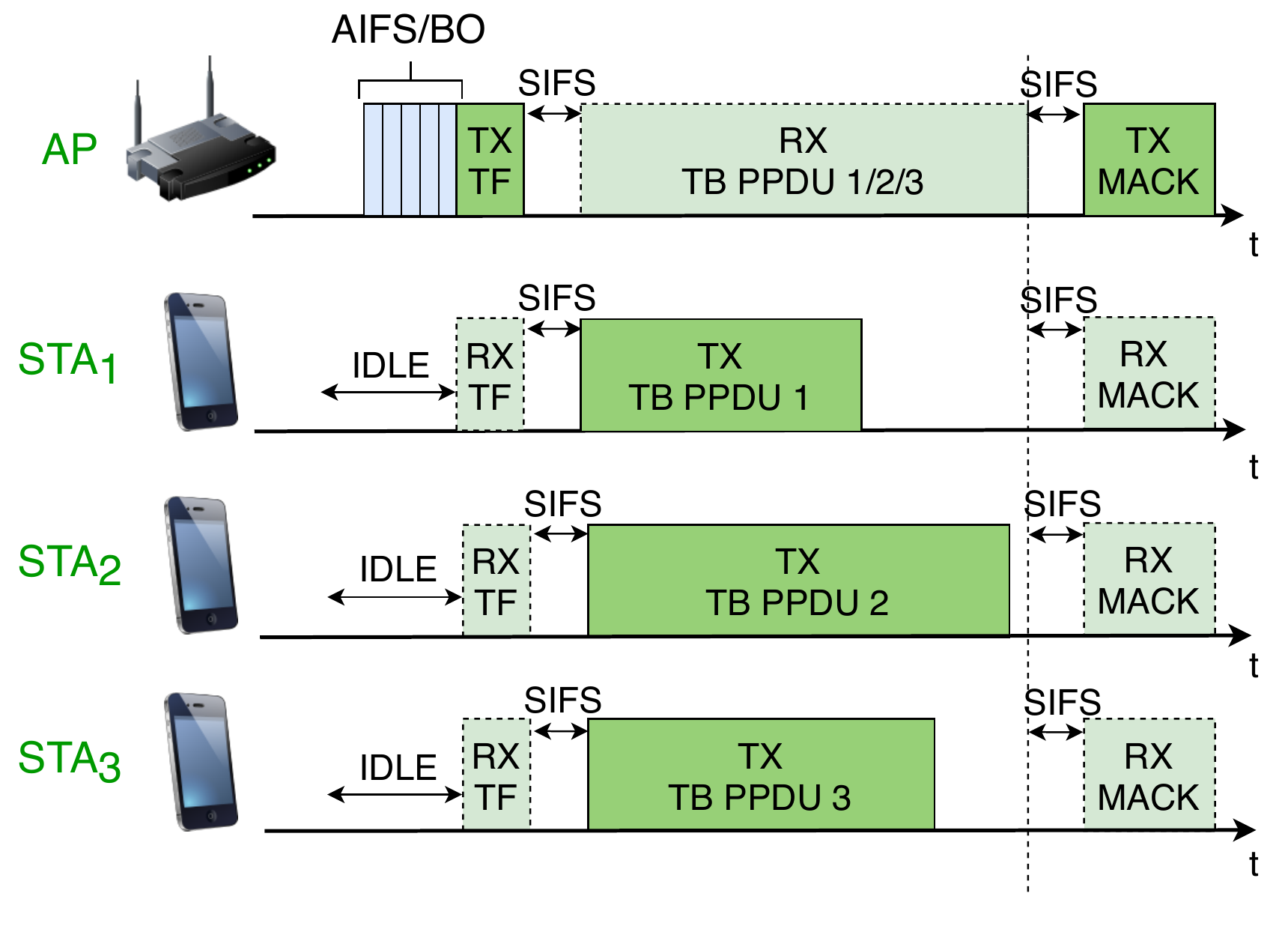}
	\caption{TB UL transmission held in a BSS.}
	\label{fig:TB_transmission_example}
\end{figure}

One of the 11ax SR mechanisms (i.e., PSR) relies on TB transmissions \cite{bellalta2019ap}. Roughly, in a TB communication, an AP schedules UL transmissions from one or more STAs. To that purpose, a Trigger Frame (TF) is sent by a given AP to indicate the group of users that are allowed to transmit during the current TXOP, along with other relevant information. Figure~\ref{fig:TB_transmission_example} illustrates an example of a TB transmission. Upon successful reception of the TF sent by the AP, STAs start simultaneous TB UL transmissions, which can be enabled by using multiple antenna technologies (i.e., MU-MIMO) or different OFDMA subcarriers. Once all the UL transmissions finish, the AP acknowledges all the packets with a multi-station block acknowledgment (MACK).

The PSR operation takes advantage of TB communications for detecting the so-called PSR opportunities. By inspecting an inter-BSS TF packet, an HE STA implementing PSR can determine the maximum allowed interference supported by the inter-BSS AP scheduling the transmission. As a result, it can transmit during the TXOP at a regulated transmission power. Further details on PSR are provided in Section \ref{section:srp_based}. Finally, it is worth mentioning that, before scheduling a UL transmission, APs can cancel the virtual carrier sensing of their STAs by sending a CF-End frame. This is done to reduce the idle periods provoked by inter-BSS transmissions, thus enhancing network efficiency. 

% ----------------------------------
% -
% 	-- IEEE 802.11ax --
% -
% ----------------------------------

\section{IEEE 802.11ax Spatial Reuse Operation}
\label{section:operation_sr_11ax}
The IEEE 802.11ax SR operation is divided into two different mechanisms: \emph{i)} \textbf{OBSS/PD-based SR} and \emph{ii)} \textbf{PSR}. So far, we have described the elements that enable both operations, thus providing insights on the potential of applying SR. In this Section, we show the technical details of IEEE 802.11ax SR, thus embodying the concepts that have been previously introduced in Section \ref{section:enablers_sr_11ax}.

%% OBSS_PD-BASED SR
\subsection{OBSS/PD-based Spatial Reuse}
\label{section:obss_pd_based}
\textcolor{black}{Upon PPDU reception, the MAC layer of a given device receives a notification from the PHY. At that moment, the node inspects the frame, and, among several operations, it determines whether the PPDU is an intra-BSS or an inter-BSS frame. The latter may be subdivided into SRG or non-SRG frames, provided that SRGs are enabled. By quickly identifying the source of an ongoing transmission, an HE STA can employ the appropriate OBSS/PD value to improve the probability of accessing the channel.}

\begin{figure}[ht!]
	\centering
	\includegraphics[width=0.4\columnwidth]{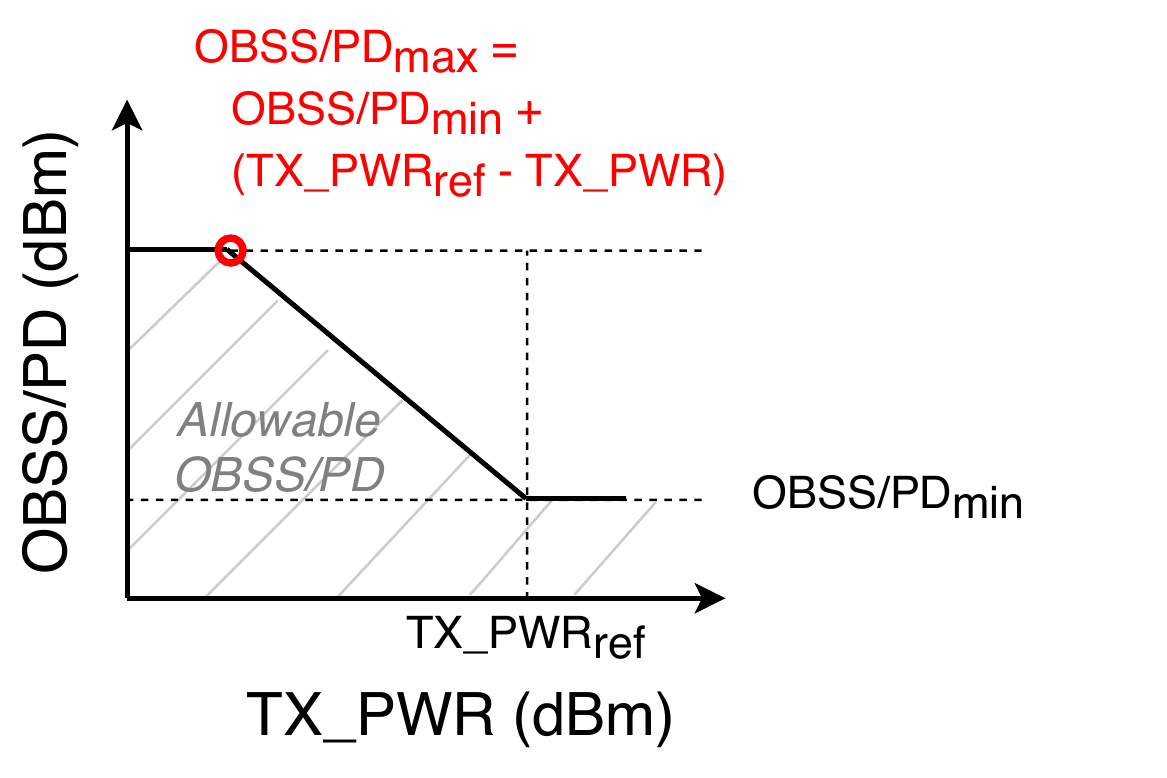}
	\caption{Graphical representation of the adjustment rules for OBSS/PD and transmission power \cite{tgax2019draft}.}
	\label{fig:fig_7}
\end{figure}

\textcolor{black}{Notice that the 11ax amendment does not provide any mechanism for setting the OBSS/PD threshold, thus remaining open (see, e.g., the proposal in \cite{tgax2016obss_pd_evaluation}). Nevertheless, the standard defines a set of rules for limiting the OBSS/PD threshold,\footnote{\textcolor{black}{The $\text{OBSS/PD}$ threshold is defined for 20 MHz PPDUs received on the primary channel, and it increases 3 dB each time the channel width is doubled.}} which is upper bounded as follows (also illustrated in Figure~\ref{fig:fig_7}):}
\begin{align}\nonumber \text{OBSS/PD} \leq  \max\Big(&\text{OBSS/PD}_{\min},\\ &  \min\big(\text{OBSS/PD}_{\max},\text{OBSS/PD}_{\min} + (\text{TX\_PWR}_{\text{ref}}-\text{TX\_PWR})\big)\Big), \nonumber \end{align}
where $\text{OBSS/PD}_{\min}$ and $\text{OBSS/PD}_{\max}$ are set to $-82$ dBm and $-62$ dBm, respectively, the reference power $\text{TX\_PWR}_{\text{ref}}$ is set to 21 or 25 dBm, based on the
capabilities of the device,\footnote{The $\text{TX\_PWR}_{\text{ref}}$ is set to 21 dBm at HE nodes which \texttt{Highest NSS Supported M1} field is equal or less than 1. Otherwise, the  $\text{TX\_PWR}_{\text{ref}}$ is set to 25 dBm. The \texttt{Highest NSS Supported M1} subfield is part of the \texttt{Tx Rx HE MCS Support} field of the \texttt{HE Capabilities element}.} and $\text{TX\_PWR}$ refers to the transmission power at the antenna connector in dBm of the HE node identifying the SR-based TXOP. \textcolor{black}{Besides the general rules for setting the OBSS/PD threshold, further constraints apply when considering SRGs (the details can be found in Appendix~\ref{section:srps}).}

% Tx Power restriction
\textcolor{black}{Together with sensitivity adjustment, the SR operation includes a transmit power limitation for any transmission occurring as a result of a detected SR TXOP (i.e., after ignoring a given inter-BSS frame through the OBSS/PD-based SR operation). The maximum allowed transmission power ($\text{TX\_PWR}_{\max}$) is defined as:}
\begin{equation}
\text{TX\_PWR}_{\max} = \text{TX\_PWR}_{\text{ref}} - (\text{OBSS/PD} -\text{OBSS/PD}_{\min})
\label{eq:power_restriction}
\end{equation}

The previous equation holds for $\text{OBSS/PD}_{\max} \geq \text{OBSS/PD} > \text{OBSS/PD}_{\min}$. Otherwise, the maximum transmission power is unconstrained. By applying a power restriction, the amendment aims to reduce the impact of concurrent transmissions taking place due to SR. Simply put, the higher the OBSS/PD threshold (more inter-BSS transmissions can be ignored), the lower the transmit power (less interference should be generated). The transmission power restriction lasts until the end of the SR TXOP identified by an HE node, which starts when its backoff reaches zero. Notice that this period depends on the duration of the active transmission(s) used for detecting the SR TXOPs. 

\begin{figure*}[ht!]
	\centering
	\raisebox{20mm}{\subfigure[Scenario]{\label{fig:fig_8_a}\includegraphics[width=0.35\columnwidth]{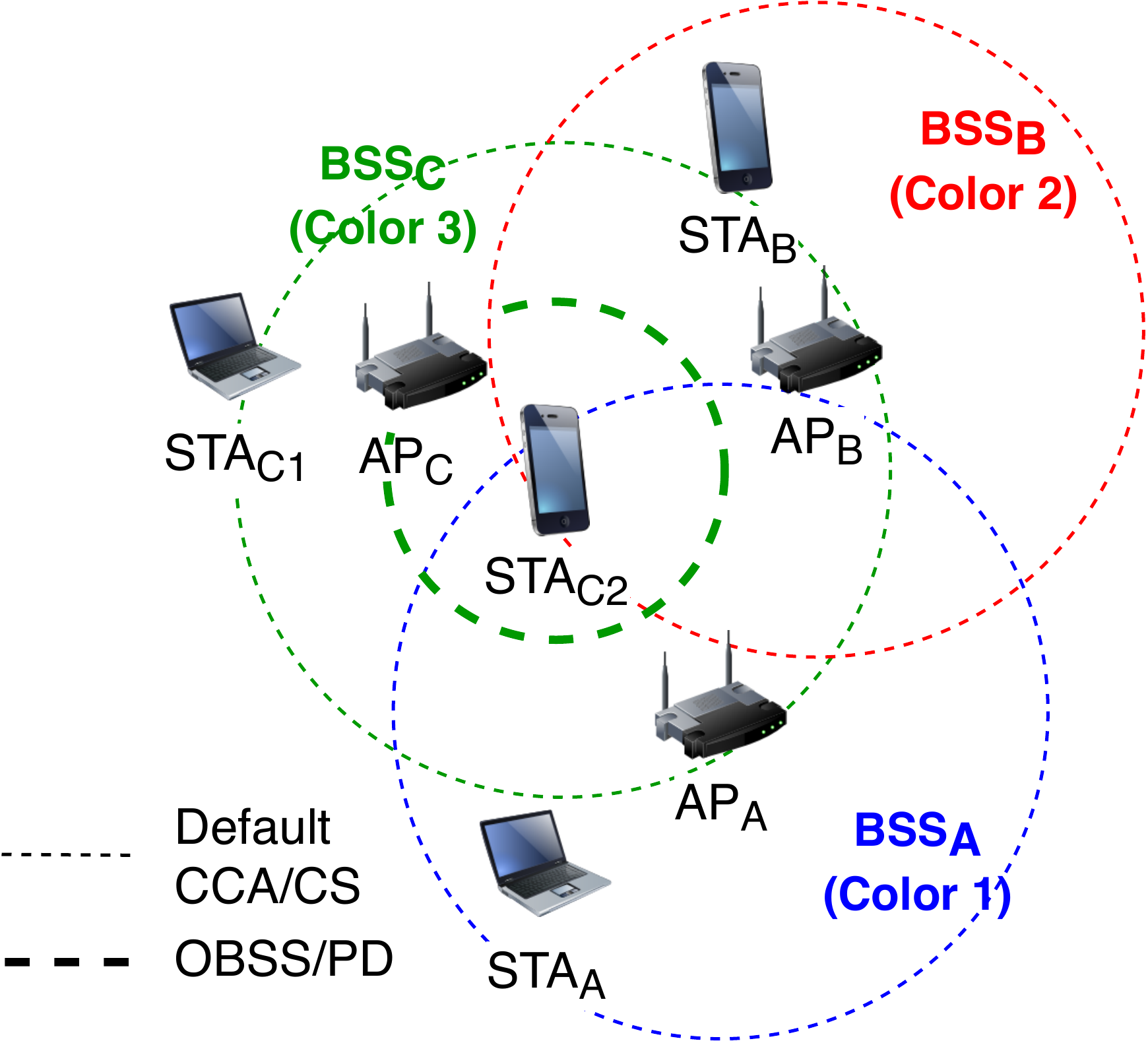}}}
	\hspace{1cm}
	\subfigure[Packets exchange]{\label{fig:fig_12}\includegraphics[width=.45\columnwidth]{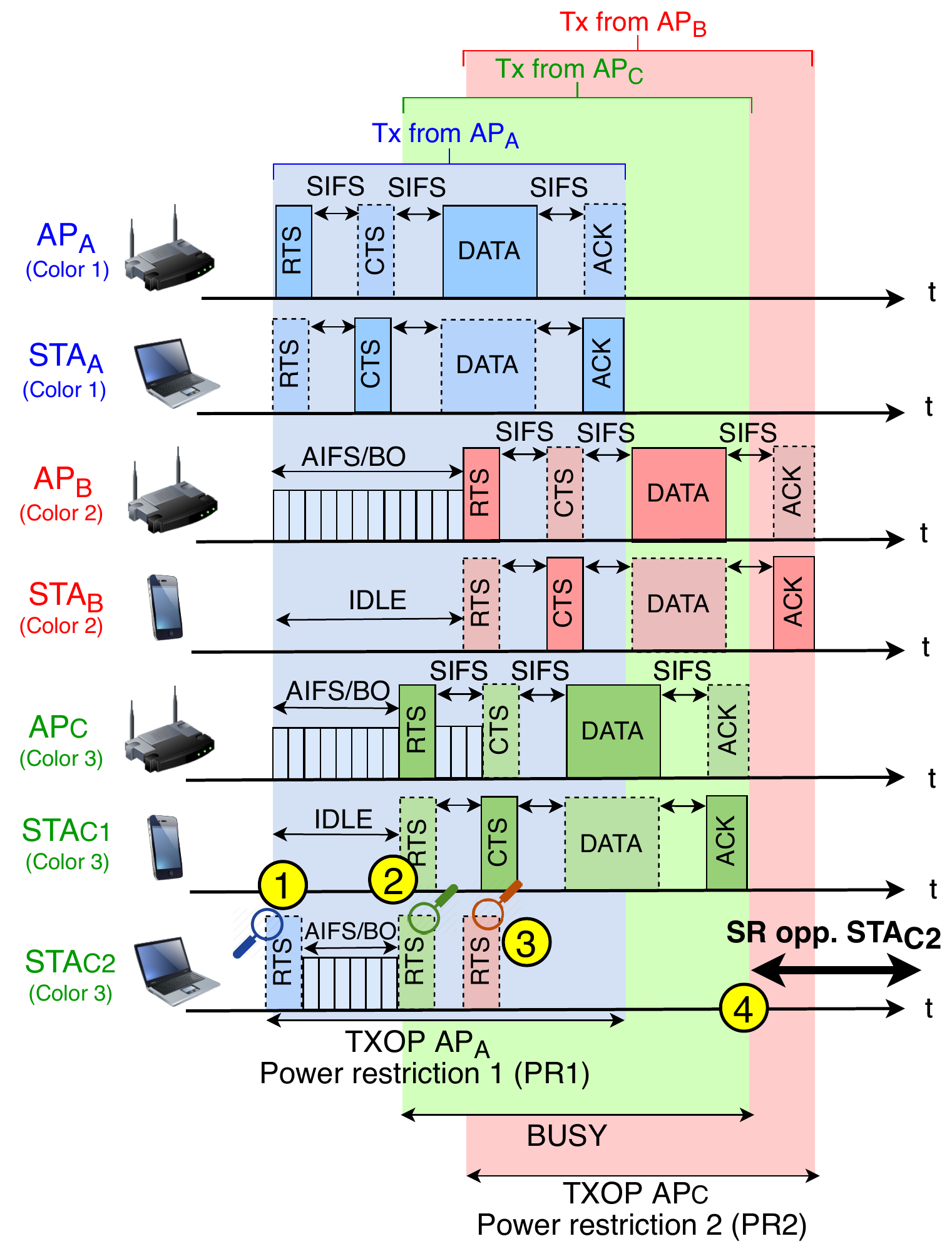}}
	\caption{Example of the OBSS/PD-based SR operation.}
	\label{fig:fig_10}
\end{figure*}

Finally, \textcolor{black}{we illustrate the OBSS/PD-based SR operation through the example in Figure~\ref{fig:fig_10}, where we put the focus on $\text{STA}_\text{C2}$}. \textcolor{black}{In the deployment depicted in Figure~\ref{fig:fig_8_a}}, several potential interfering devices (belonging to $\text{BSS}_\text{A}$ and $\text{BSS}_\text{B}$) surround $\text{STA}_\text{C2}$. \textcolor{black}{When} using the default CCA/CS threshold, all the APs are able to transmit simultaneously. However, $\text{STA}_\text{C2}$ may suffer flow starvation because of its unprivileged location. \textcolor{black}{To address this problem, $\text{STA}_\text{C2}$ applies an OBSS/PD threshold that allows ignoring inter-BSS transmissions.}

Figure~\ref{fig:fig_12} illustrates the behavior of $\text{STA}_\text{C2}$ when applying OBSS/PD-based SR. Notice that any detected SR TXOP is subject to a power restriction (see Eq.~\eqref{eq:power_restriction}), and that different power restrictions can be applied as a result of applying different OBSS/PD thresholds (of SRG and non-SRG type). That said, the following activity points (displayed in yellow) are observed from Figure~\ref{fig:fig_12}:
\begin{enumerate}
	\item $\text{STA}_\text{C2}$ analyzes the RTS frame sent by $\text{AP}_\text{A}$ and classifies it as an inter-BSS frame. As a result, it applies a given OBSS/PD value that allows sensing the channel idle ($\text{RSSI}_{\text{A} \rightarrow \text{C2}} < \text{OBSS/PD}$). \textcolor{black}{A first power restriction is considered by $\text{STA}_\text{C2}$.}
	\item The same procedure is followed at $\text{STA}_\text{C2}$ when detecting the RTS frame transmitted by $\text{AP}_\text{C}$. However, the transmission cannot be ignored this time because it is an intra-BSS transmission ($\text{RSSI}_{\text{C} \rightarrow \text{C2}} < \text{CCA/CS}$ ).
	\item As for points 1) and 2), $\text{AP}_\text{B}$'s transmission is ignored by $\text{STA}_\text{C2}$ because $\text{RSSI}_{\text{B} \rightarrow \text{C2}} < \text{OBSS/PD}$. Again, a new power restriction is considered.
	\item Finally, $\text{STA}_\text{C2}$ transmits by taking advantage of the detected SR TXOPs. \textcolor{black}{The transmission is nonetheless subject to the most restrictive limitation among all the collected power restrictions (PRs)}. In particular, $\text{TX PWR}_{max} = \min(\text{PR}_1, \text{PR}_2)$.\footnote{Notice that, once $\text{STA}_\text{C2}$ transmits under the power restriction, the ACK sent by $\text{STA}_\text{B}$ can be ignored, so that a new power restriction is not defined.}
\end{enumerate}

%\begin{figure}[ht!]
%	\centering
%	\includegraphics[width=.55\columnwidth]{fig_12}
%	\caption{Example of the OBSS/PD-based SR operation. $\text{STA}_\text{C2}$ applies different OBSS/PD values according to the source of detected transmissions.}
%	\label{fig:fig_12}
%\end{figure}

%%% PSR
\subsection{Parametrized Spatial Reuse}
\label{section:srp_based}	
\textcolor{black}{The PSR operation is defined as an alternative to OBSS/PD-based SR for TB transmissions. Although the analysis conducted in this paper focuses on OBSS/PD-based SR, we introduce PSR for the sake of tutorial completeness.}

In PSR, we find nodes taking advantage of PSR opportunities (i.e., the \emph{opportunists}). These nodes identify PSR opportunities from detected TB transmissions. On the other hand, we find the \emph{transmission holders}, which perform TB transmissions and indicate support for the PSR operation in the headers of the TF. To identify a PSR opportunity, an opportunist must check whether the TB PPDUs that follow a given TF packet can be ignored or not. To do so, the opportunist's intended transmission power must not exceed the requirements imposed by the transmission holder \textcolor{black}{(encapsulated in the \texttt{PSR\_INPUT} parameter).} 

Once an opportunist inspects the PSR value of the detected TF\footnote{The PSR can be extracted either from the \texttt{SPATIAL REUSE} field, which is included in the \texttt{Common Info} field of the Trigger frame, or the \texttt{SIG-A PSR} field of the HE TB PPDU.} and confirms that the intended transmission power is acceptable, it transmits during the duration of the TB PPDU(s) (indicated in the \texttt{Common Info} field). In particular, the intended transmission power must be below the value of PSR minus the Received Power Level (RPL), which is measured from the legacy portion of the TF (i.e., from PHY headers). The PSR value is computed as:
\begin{equation}
\text{PSR} = \text{TX PWR}_\text{AP} + \text{I}_\text{AP}^{\max},
\label{eq:srp_input}
\nonumber
\end{equation}
where $\text{TX PWR}_\text{AP}$ is the normalized transmit power in dBm at the output of the antenna connector, and $\text{I}_\text{AP}^{\max}$ is a normalized value in dB that captures the maximum allowed interference at the transmission holder. In particular, $\text{I}_\text{AP}^{\max}$ is computed as the target RSSI indicated in the TF minus the minimum SNR granting a 10\% PER (based on the highest MCS to be used for transmitting the UL HE TB PPDU). A safety margin (set by the AP) is also included not to exceed 5 dB.

%\begin{table}[ht!]
%	\centering			
%	\scriptsize{
%		\begin{tabular}{|c|c|c|c|}
%			\hline
%			\textbf{Value} & \textbf{Meaning} & \textbf{Value} & \textbf{Meaning} \\ \hline
%			0 & PSR\_DISALLOW & 8 & PSR = -44 dBm \\ \hline
%			1 & PSR = -80 dBm & 9 & PSR = -41 dBm \\ \hline
%			2 & PSR = -74 dBm & 10 & PSR = -38 dBm \\ \hline
%			3 & PSR = -68 dBm & 11 & PSR = -35 dBm \\ \hline
%			4 & PSR = -62 dBm & 12 & PSR = -32 dBm \\ \hline
%			5 & PSR = -56 dBm & 13 & PSR = -29 dBm \\ \hline
%			6 & PSR = -50 dBm & 14 & PSR $\geq$ -26 dBm \\ \hline		7 & PSR = -47 dBm & 15 & \begin{tabular}[c]{@{}c@{}}PSR\_AND\_NON-\\ SRG\_OBSS-PD\_\\ PROHIBITED\end{tabular} \\ \hline
%	\end{tabular}}
%	\caption{PSR subfield encoding for Trigger and HE TB PPDU frames \cite{tgax2019draft}.}
%	\label{tbl:sr_subfield_encoding_TB_ppdu}
%\end{table}

\begin{figure*}[ht!]
	\centering
	\raisebox{10mm}{\subfigure[Scenario]{\label{fig:fig_13a}\includegraphics[width=0.35\columnwidth]{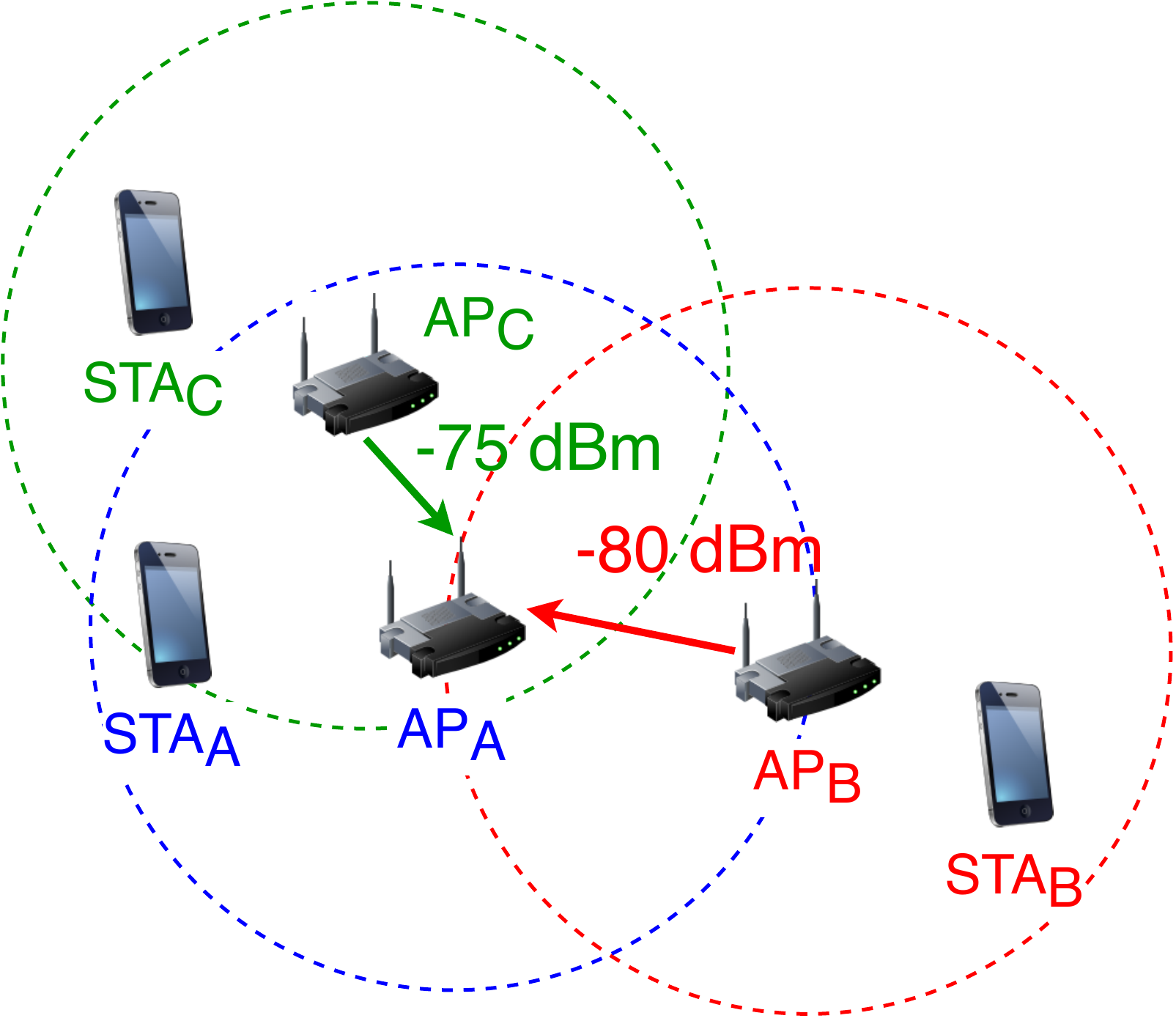}}}
	\hspace{1cm}
	\subfigure[Packets exchange]{\label{fig:fig_13b}\includegraphics[width=.45\columnwidth]{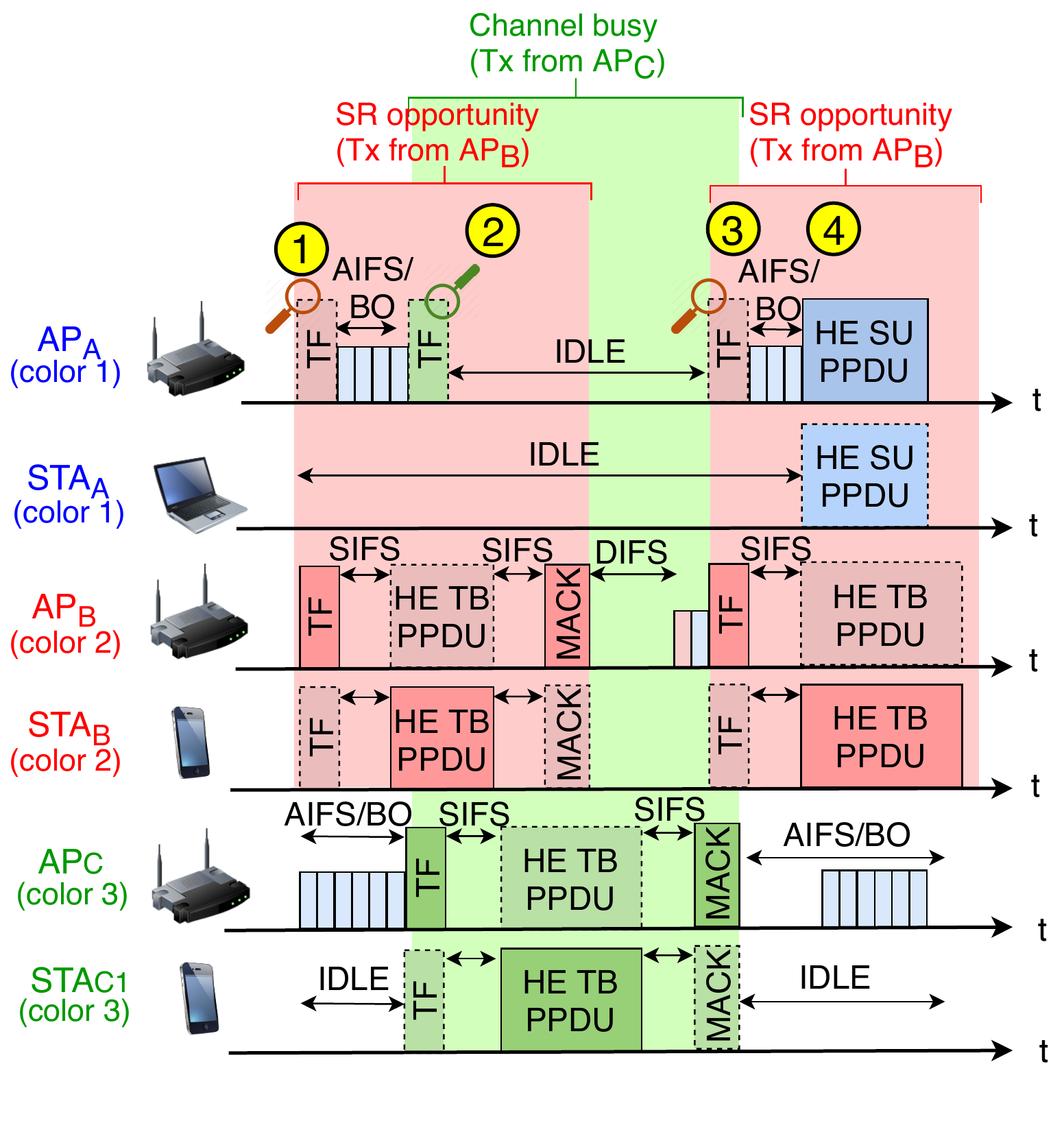}}
	\caption{Example of the PSR operation.}
	\label{fig:fig_13}
\end{figure*}

\textcolor{black}{To help understanding the PSR operation, we propose the example in Figure~\ref{fig:fig_13}. In particular:}
\begin{enumerate}
	\item $\text{AP}_\text{A}$ detects a PSR TXOP from $\text{AP}_\text{B}$'s TF packet. \textcolor{black}{Since $\text{AP}_\text{A}$'s intended transmit power for the next queued packet is below the PSR indicated by $\text{AP}_\text{B}$ minus the RPL, $\text{AP}_\text{A}$'s backoff keeps counting down.}
	\item As soon as $\text{AP}_\text{C}$ transmits a TF, the PSR TXOP previously detected by $\text{AP}_\text{A}$ is canceled because the transmit power condition no longer holds. As a result, the channel becomes busy and the backoff is frozen.
	\item \textcolor{black}{A new PSR TXOP is detected  from the new $\text{AP}_\text{B}$'s transmission, even if $\text{BSS}_\text{C}$ is still transmitting.}
	\item Once $\text{BSS}_\text{C}$'s transmission is over, $\text{AP}_\text{A}$ \textcolor{black}{consumes the backoff} and transmits according to the last detected PSR TXOP.
\end{enumerate}

% ----------------------------------
% -
% 	-- Analytical Framework --
% -
% ----------------------------------
\section{Analysis of OBSS/PD-based SR in Toy Scenarios}
\label{section:analytical_model}

\textcolor{black}{In this Section, we analyze the OBSS/PD-based SR operation described in Section \ref{section:obss_pd_based} through Continuous Time Markov Networks (CTMNs) \cite{bellalta2014throughput, bellalta2016throughput,barrachina2019dynamic, barrachina2019overlap}.\footnote{Notice that OBSS/PD-based SR has drawn much more attention than PSR. Therefore, from now onwards, we may refer to the OBSS/PD-based SR operation simply as SR. The implementation/modeling of PSR is left as future work.} The goal of the CTMN model is twofold: \emph{i)} it is useful to illustrate OBSS/PD-based SR in toy scenarios, thus helping to understand the new kind of inter-BSS interactions resulting from its operation, and \emph{ii)} it allows validating the implementation of SR done in the Komondor simulator, from which we present results later in Section \ref{section:performance_evaluation}.}

\textcolor{black}{A disadvantage of the CTMN model is its computational cost when characterizing crowded deployments. Modeling dense scenarios may become intractable since the number of feasible states increases in a combinatorial manner with the number of BSSs. For that reason, the usage of the CTMN model is limited to toy scenarios.}

\textcolor{black}{To the best of our knowledge, none of the previous literature has attempted to model the 11ax SR operation analytically. Nevertheless, we find some works that analyze the impact of sensitivity adjustment and transmit power control in wireless networks. The most prominent ones are SINR-based methods \cite{gupta2000capacity, guo2003spatial} for characterizing radio links, and Stochastic Geometry (SG) for modeling the random nature of dense wireless networks \cite{zhao2016stochastic, zhang2015stochastic, iwata2019stochastic, nguyen2007stochastic}. However, these methods typically consider symmetric deployments and worst-case interference (i.e., nodes are assumed to transmit permanently), which entails losing insights on the MAC operation. In turn, the CTMN model allows capturing both the PHY and MAC layers and estimate the throughput of CSMA/CA more accurately.} 

\subsection{Introduction to Continuous Time Markov Networks}
The CTMN model captures the CSMA/CA operation in IEEE 802.11 WLANs through states that represent the set of potentially overlapping BSSs that are active at a given moment. Transitions between states occur when BSSs become active (i.e., they gain access to the medium) or abandon the channel (i.e., their transmission finishes). While the arrival rate ($\lambda$) characterizes channel access, the time it takes for transmitting data is derived from the service rate ($\mu$). Based on the transition probabilities, it is straightforward to obtain every state's probability, \textcolor{black}{and thus compute} the long-term throughput experienced by the different BSSs.

\textcolor{black}{To characterize the throughput of an IEEE 802.11 WLAN, the CTMN model makes the following assumptions:}

\begin{assumption}[Continuous backoff]\label{as:1} \textcolor{black}{The backoff procedure for accessing the medium is continuous in time.}\end{assumption}

\begin{assumption}[Downlink transmissions]\label{as:2} \textcolor{black}{Data transmissions are downlink only, which are only held from APs to STAs.} \end{assumption}

\begin{assumption}[Uplink neglected]\label{as:3} \textcolor{black}{Uplink control packets (e.g., ACKs) are only considered for computing the total transmission time.}\end{assumption}

\begin{assumption}[Full-buffer traffic]\label{as:4} \textcolor{black}{Full-buffer traffic is considered, which means that APs always have data to transmit.}\end{assumption}

\textcolor{black}{Assumption \ref{as:1} allows capturing the backoff procedure in CSMA/CA but prevents modeling collisions due to backoff expiring at the same instant. Nevertheless, the CTMN model is accurate when the collision probability is small (either because the minimum contention window is large, or there is a small number of overlapping nodes) and is useful for depicting inter-AP interactions \cite{bellalta2016throughput}.} 

\textcolor{black}{Assumptions \ref{as:2} and \ref{as:3} are necessary to relax the complexity of the model, which would increase in a combinatorial manner if considering uplink transmissions as well.\footnote{\textcolor{black}{A state in the Markov chain represents a set of nodes transmitting. Thus, capturing uplink transmission would increase the number of states combinatorially.}} Considering only downlink traffic is a commonly accepted assumption since the asymmetry between downlink/uplink traffic is expected to increase even more in future networks \cite{union2015imt}. Since uplink control packets are small in size, their effects in terms of interference is neglected, but considered in the transmission duration \cite{bellalta2016throughput,michaloliakos2016performance,liew2010back}.} 

\textcolor{black}{Finally, we apply Assumption \ref{as:4} for analyzing the SR interactions more lucidly. Besides, this assumption is useful for characterizing the strict application requirements in next-generation wireless networks. In any case, the CTMN model can also be applied to represent unsaturated traffic conditions \cite{barrachina2019overlap}.}

%%% Simple OBSS/PD-based interactions
\subsection{SR-enabled inter-BSS Interactions}
\label{section:simple_interactions}

\textcolor{black}{Our CTMN model for SR\footnote{\textcolor{black}{The code is open and available in Github \cite{wilhelmi2019sfctm_spatial_reuse}.}} distinguishes between different types of states, which stem from the set of sensitivity levels that each BSS can apply (default CCA/CS, non-SRG OBSS/PD, and SRG OBSS/PD). Notice that sensitivity adjustment potentially leads to new interactions (e.g.,~concurrent transmissions). Moreover, the transmit power limitation, which depends on the OBSS/PD threshold, affects the MCS used for a given transmission.}

\textcolor{black}{To analyze the SR operation, we propose two toy scenarios: \emph{Toy scenario~1} and \emph{Toy scenario~2}. With these scenarios, we aim to illustrate the most important characteristics of SR in terms of inter-BSS interactions, and shed light on the strengthens and limitations of this technology. The details of both toy scenarios are described in Table \ref{table:toy_scenarios}, while simulation parameters are provided in Appendix \ref{section:simulation_parameters}.}

\begin{table}[]
	\centering
	\resizebox{.7\columnwidth}{!}{\begin{tabular}{|l|c|c|}
		\hline
		& \textbf{\begin{tabular}[c]{@{}c@{}}Toy scenario 1\\ BSS$_\text{A}$ / BSS$_\text{B}$\end{tabular}} & \textbf{\begin{tabular}[c]{@{}c@{}}Toy scenario 2\\ BSS$_\text{A}$ / BSS$_\text{B}$ / BSS$_\text{C}$\end{tabular}} \\ \hline
		\textbf{AP position {[}m{]}} & (4,0) / (6,0) & (4,0) / (8,0) / (6,5) \\ \hline
		\textbf{STA position {[}m{]}} & (0,0) / (8,0) & (0,0) / (12,0) / (6,9) \\ \hline
		\textbf{Default tx power {[}dBm{]}} & 20 / 20 & 20 / 20 / 20 \\ \hline
		\textbf{Default CCA/CS {[}dBm{]}} & -82 / -82 & -82 / -82 / -82 \\ \hline
		\textbf{Channel selected} & 1 / 1 & 1 / 1 / 1 \\ \hline
	\end{tabular}}
	\caption{\textcolor{black}{Toy scenarios' deployment details.}}
	\label{table:toy_scenarios}
\end{table}

\subsubsection{Toy scenario 1}
\begin{figure*}[h!]
	\centering
	% Situation 1
	\subfigure[Sensing area (CCA/CS = -82 dBm)]{\label{fig:cs_toy_scenario_1a}\includegraphics[width=0.4\textwidth]{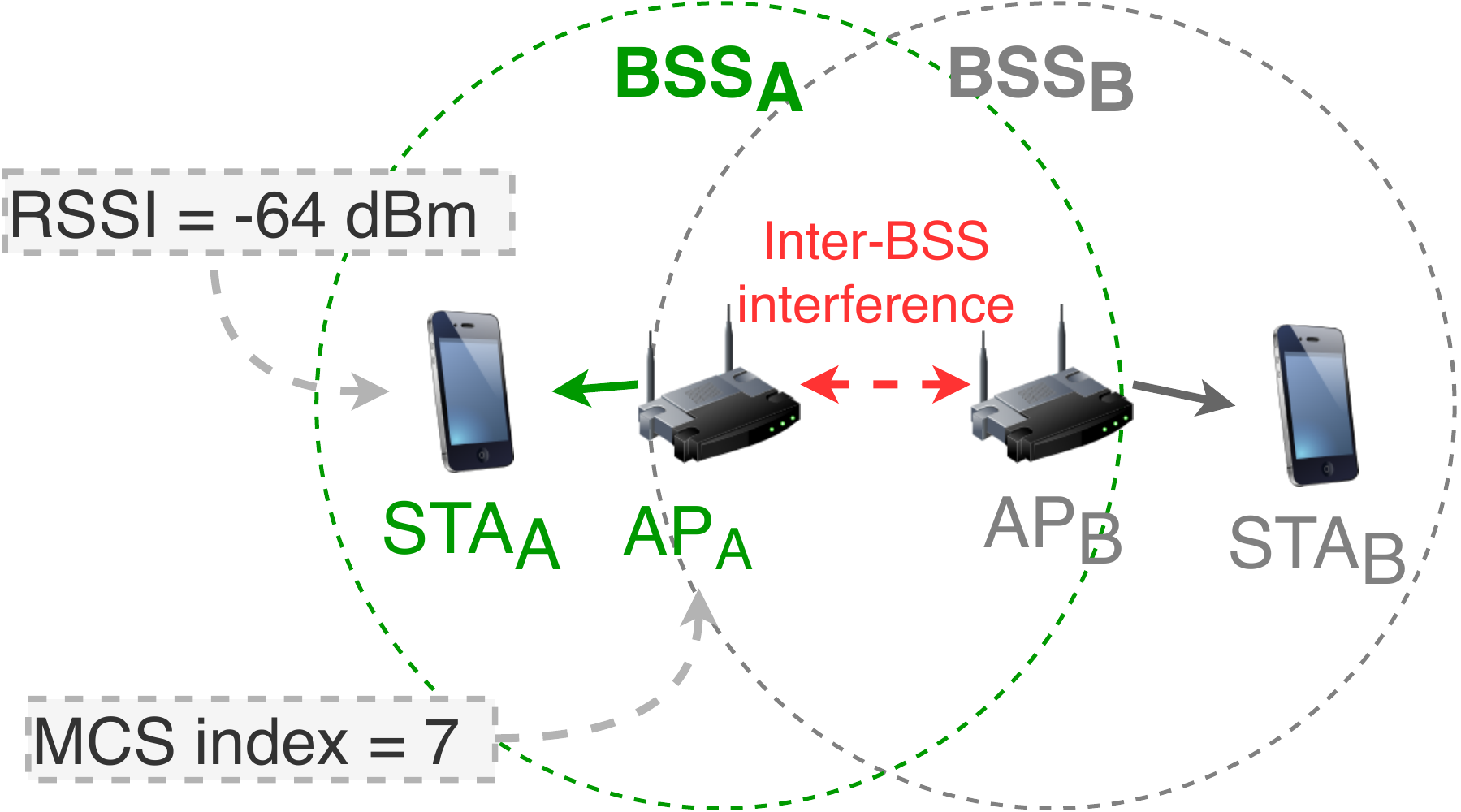}}
	\hspace{1cm}
	% CTMN 1
	\subfigure[CTMN (CCA/CS = -82 dBm)]{\label{fig:ctmn_toy_scenario_1a}\includegraphics[width=0.4\textwidth]{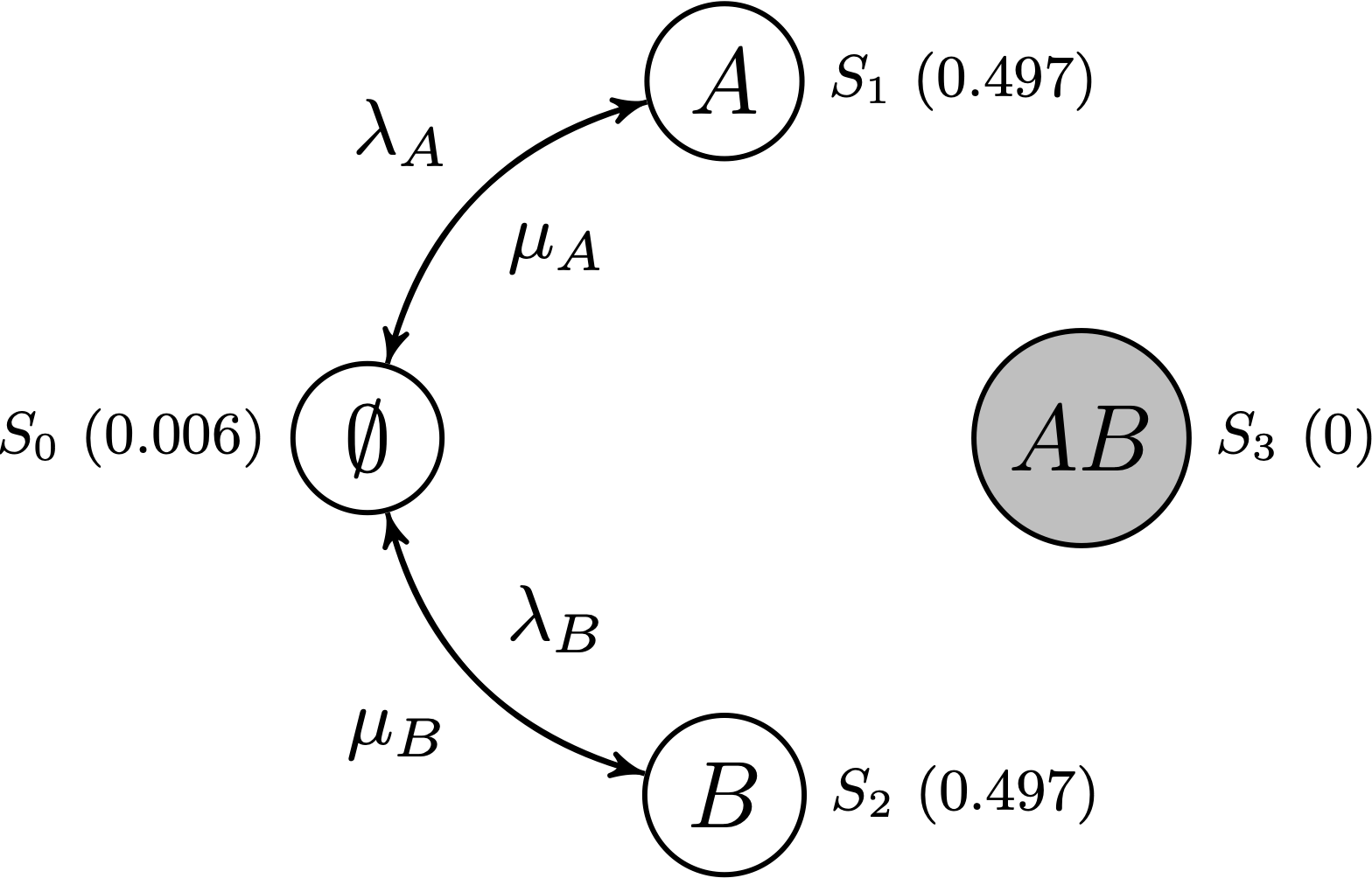}}\\
	% Situation 2
	\subfigure[Sensing area (OBSS/PD = -78 dBm)]{\label{fig:cs_toy_scenario_1b}\includegraphics[width=0.4\textwidth]{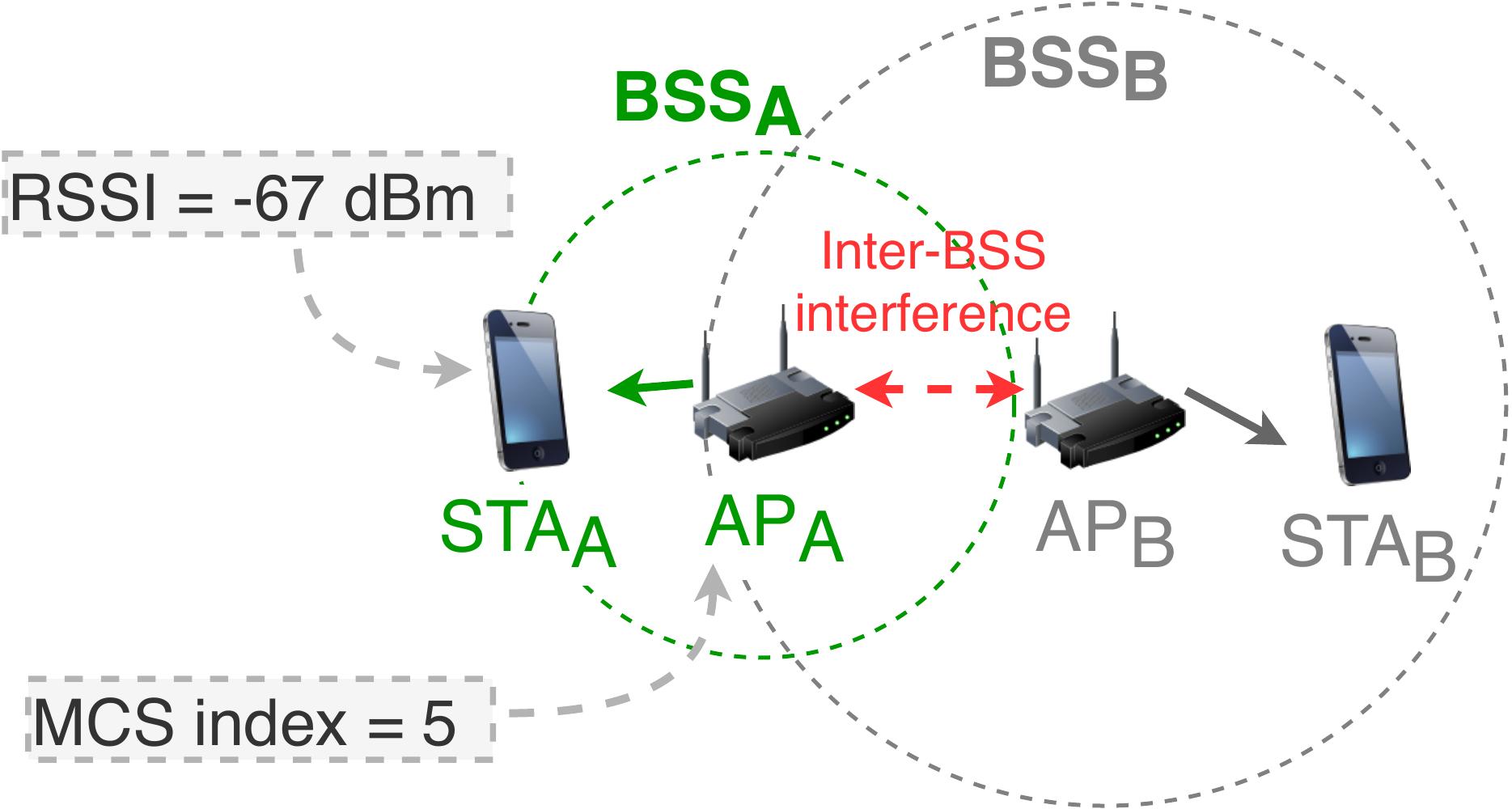}}
	\hspace{1cm}
	% CTMN 2
	\subfigure[CTMN (OBSS/PD = -78 dBm)]{\label{fig:ctmn_toy_scenario_1b}\includegraphics[width=0.4\textwidth]{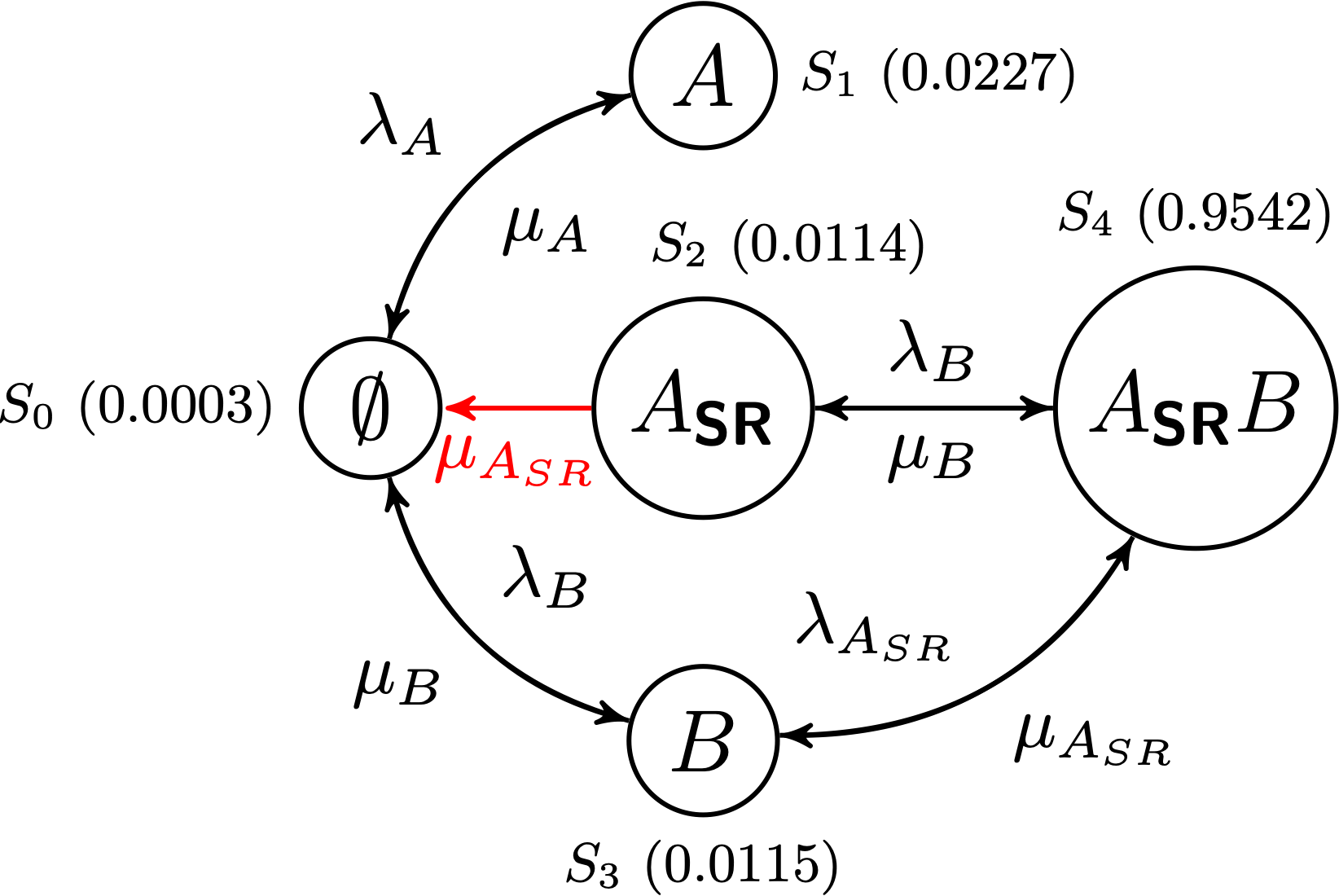}}%\\
	\caption{Representation of \emph{Toy scenario 1} for different settings. (a) and (c) illustrate the sensing area of each transmitter for CCA/CS = -82 dBm and OBSS/PD = -78 dBm, whereas (b) and (d) illustrate the corresponding inter-BSS interactions through CTMNs (unidirectional transitions are shown in red).}
	\label{fig:toy_scenario_1b}
\end{figure*}

\textcolor{black}{We start analyzing \emph{Toy scenario 1} by considering that only $\text{BSS}_\text{A}$ implements SR. Figure~\ref{fig:cs_toy_scenario_1a} and Figure~\ref{fig:cs_toy_scenario_1b} illustrate the sensitivity areas of the involved APs, which represent the default CCA/CS and the OBSS/PD thresholds, respectively. The resulting CTMNs are shown in Figures \ref{fig:ctmn_toy_scenario_1a} and \ref{fig:ctmn_toy_scenario_1b}, respectively (the long-term probability of each state is shown in parentheses).}

\textcolor{black}{When CCA/CS is applied at both BSSs, concurrent transmissions are not possible. This can be noticed in the corresponding CTMN representation (see Figure~\ref{fig:ctmn_toy_scenario_1a}), where state $s_3$ (AB) cannot be reached from any other state (we hence represented it in gray). Nevertheless, both APs transmit at a high rate (i.e., using a high MCS) because the maximum power is used in isolation.} 

\textcolor{black}{As for the SR setting, concurrent transmissions are possible if $\text{AP}_\text{A}$ uses an OBSS/PD value greater or equal than -78 dBm, \textcolor{black}{which allows ignoring transmissions} from $\text{AP}_\text{B}$. As shown in Figure~\ref{fig:ctmn_toy_scenario_1b}, using SR leads to new states (i.e., $s_2$ and $s_4$, denoted with subindex $\text{SR}$). It is worth pointing out that state $s_2$ ($A_\text{SR}$) can never be reached from the empty state ($s_0$) because $\text{AP}_\text{A}$ can only use the SR mode after identifying an SR-based TXOP. Although parallel transmissions can be held with SR, its imposed transmit power limitation results in poorer signal strength in STA$_\text{A}$ (RSSI = -67 dBm when the transmit power used by AP$_\text{A}$ is 17 dBm), thus requiring to use a more robust MCS.}

\begin{figure}[ht!]
	\centering
	\includegraphics[width=.5\columnwidth]{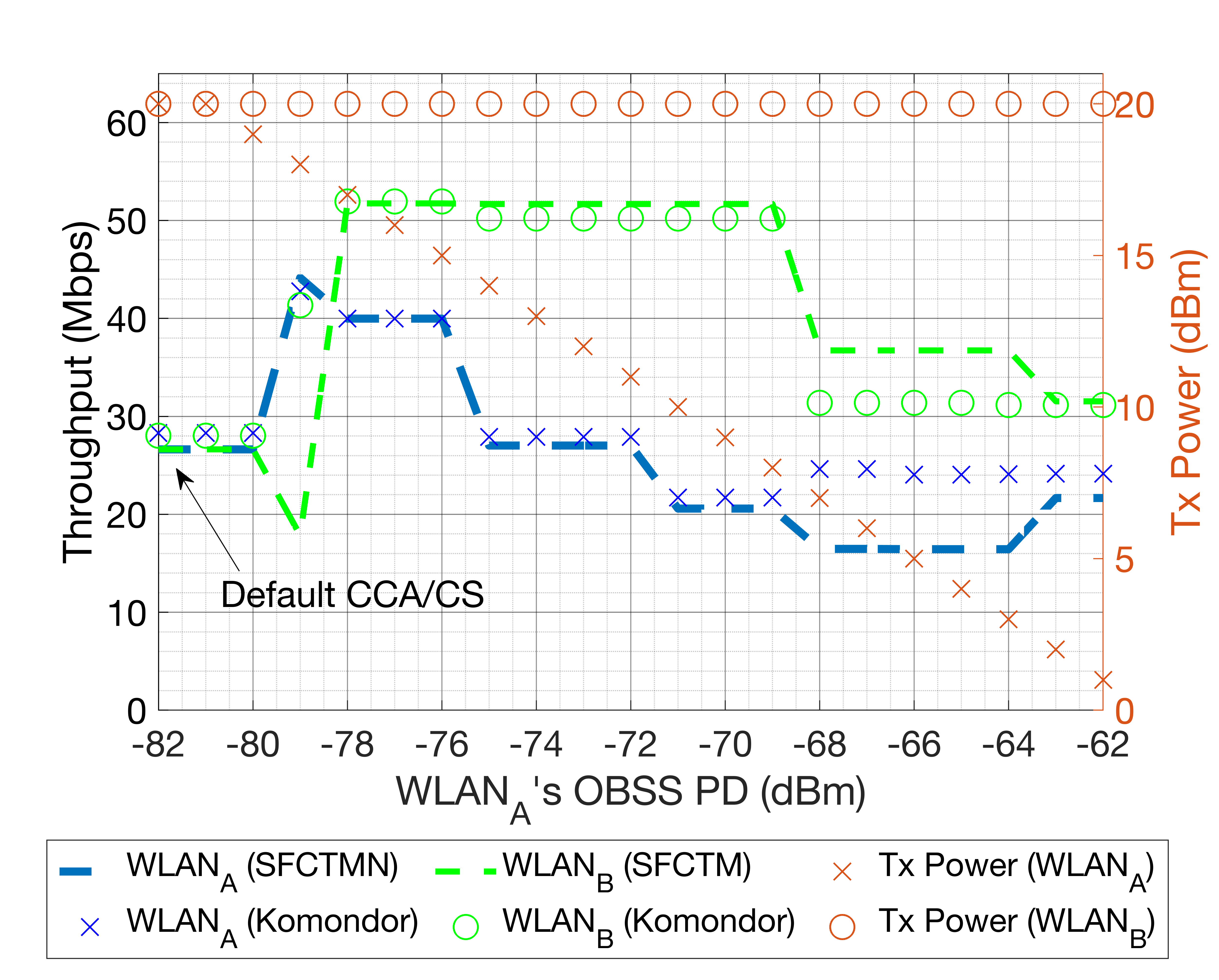}
	\caption{Effects of applying OBSS/PD-based SR in $\text{BSS}_\text{A}$ of \emph{Toy scenario 1}, for each possible OBSS/PD value. The transmission power is shown in red.}		
	\label{fig:toy_scenario_1_results}
\end{figure}

Figure~\ref{fig:toy_scenario_1_results} shows the throughput achieved by $\text{BSS}_\text{A}$ and $\text{BSS}_\text{B}$, for each possible OBSS/PD value. The transmit power limitation that is imposed to each BSS is also illustrated (right side). \textcolor{black}{For validation purposes, the results have been gathered from both the SFCTMN and the Komondor simulator. }

As shown, both BSSs achieve the same performance when OBSS/PD $<$ -79 dBm because they share the channel on equal terms. Above this value, $\text{BSS}_\text{A}$ is able to ignore $\text{BSS}_\text{B}$'s transmissions and transmit concurrently. However, what might seem a worthy strategy for $\text{BSS}_\text{A}$ turns out to be more beneficial to $\text{BSS}_\text{B}$. \textcolor{black}{The fact is that $\text{AP}_\text{A}$ uses a lower transmission power as a result of applying SR, which makes $\text{AP}_\text{B}$ sensing the channel idle. A particular phenomenon occurs for OBSS/PD = -79 dBm, whereby $\text{AP}_\text{A}$'s transmission power limitation is insufficient to make $\text{AP}_\text{B}$ detecting the channel idle. Therefore, both APs need to transmit in isolation.} 

\textcolor{black}{Regarding the accuracy of the model with respect to the simulator, we find a region (from OBSS/PD = -68 dBm to -64 dBm) in which CTMNs are less accurate.\footnote{\textcolor{black}{Although differing in performance, the CTMN model still represents the behavior of BSSs applying SR.}} In the model, states are independent and the overall throughput is computed under this assumption (the throughput of a BSS in a given state depends on the configuration and sensed interference in that state). However, the performance achieved by BSS$_\text{A}$ in state $A_\text{SR}$ depends on the interactions that occur in state $A_\text{SR}B$. The fact is that, in state $A_\text{SR}B$, $\text{AP}_\text{A}$ is expected to abandon its transmission as soon as a timeout is noticed due to an insufficient signal strength at the STA (as a result of the interference sensed from BSS$_\text{B}$). In consequence, the throughput of BSS$_\text{A}$ in state $A_\text{SR}$ is inherited by the spoiled packet transmission initiated in state $A_\text{SR}B$. In contrast, the CTMN treats the performance of BSS$_\text{A}$ in states $A_\text{SR}$ and $A_\text{SR}B$ independently, thus spending more time in state $A_\text{SR}$ and leading to higher performance.}

%, thus spending a shorter time in SR mode (transition $A_\text{SR}B$ to $A_\text{SR}$ is unlikely). In contrast, the model considers that much more time is spent at state $A_\text{SR}$ since transmissions at that point are successful (but slow due to the low MCS used).
%For these OBSS/PD values, $\text{STA}_\text{A}$ cannot decode any transmission from $\text{AP}_\text{A}$ in state $A_\text{SR}B$. In turn, it can do that for the $A_\text{SR}$ state. In particular, the transmit power limitation used by $\text{AP}_\text{A}$ in the SR mode makes that $\text{STA}_\text{A}$ perceives an insufficient signal-to-noise-plus-interference ratio (SINR) when $\text{BSS}_\text{B}$ is also occupying the channel. 

\begin{figure}[ht!]
	\centering    
	\includegraphics[width=.45\columnwidth]{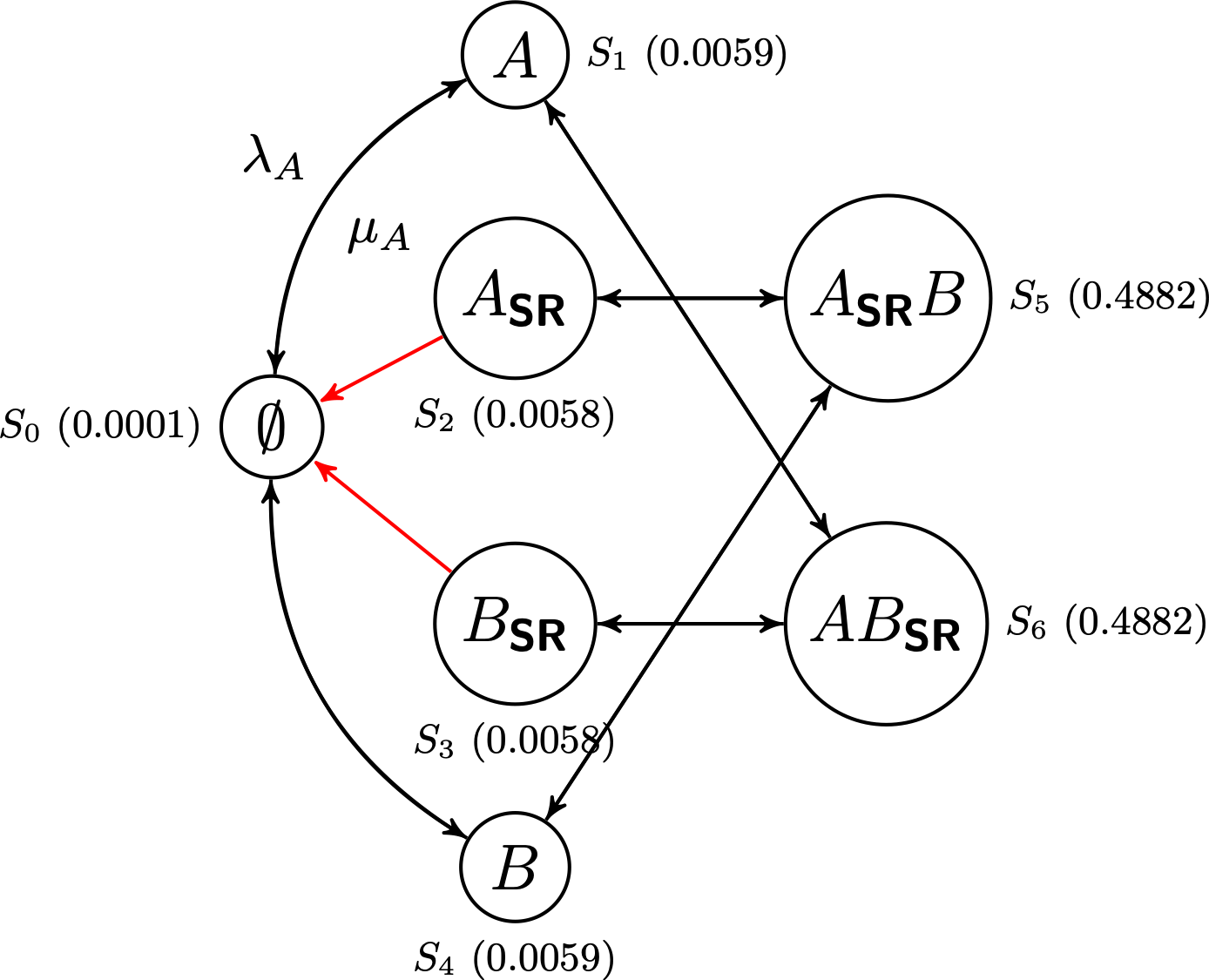}
	\caption{CTMN of \emph{Toy scenario 1} when both BSSs apply OBSS/PD-based SR with OBSS/PD~$\geq$~-79~dBm (unidirectional transitions are marked in red).}
	\label{fig:ctmn_toy_scenario_1c}
\end{figure}

\begin{figure}[ht!]
	\centering
	\includegraphics[width=.5\columnwidth]{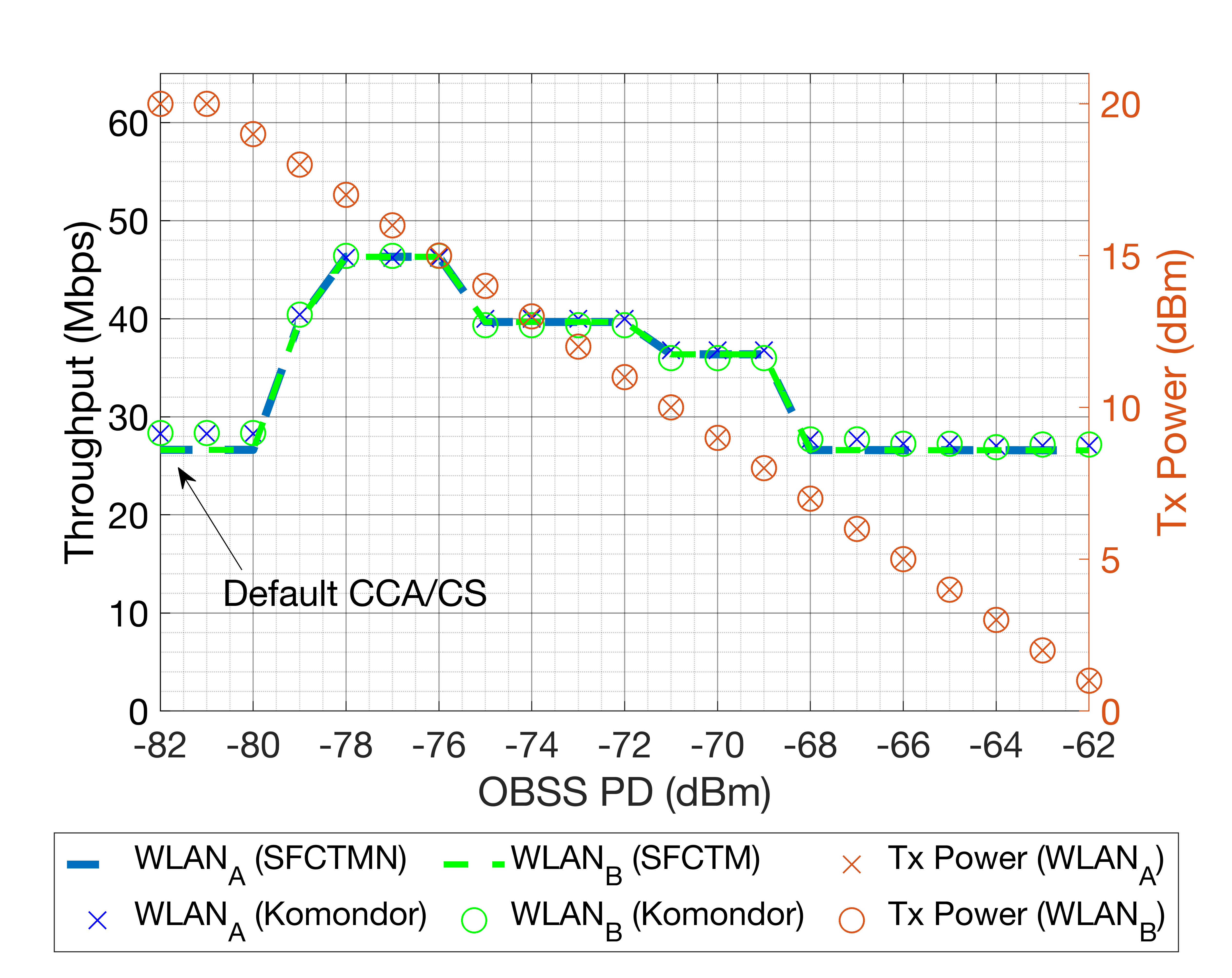}
	\caption{Effects of applying OBSS/PD-based SR in both BSSs of \emph{Toy scenario 1}, for each possible OBSS/PD value. The transmission power is shown in red.}		
	\label{fig:toy_scenario_1c_results}
\end{figure}

\textcolor{black}{Now, we consider the case where both BSSs apply SR and concurrent transmissions are possible for OBSS/PD $\geq$ -79 dBm. As shown in Figure~\ref{fig:ctmn_toy_scenario_1c}, the CTMN is symmetric and dominated by states $s_5$  and $s_6$. The high probability of being in those states ($\approx$0.9) indicates that both BSSs alternate the default and the SR modes.} The results obtained under this setting are shown in Figure~\ref{fig:toy_scenario_1c_results}, for each OBSS/PD threshold. Again, the results have been extracted from both SFCTMN and Komondor, which in this case  match almost perfectly.

%In order to show the long-term performance of each BSS in the Komondor simulator, we have displayed the average values obtained from 100 simulations.  
%
%However, in reality, one of the BSSs may monopolize the channel \textcolor{black}{through} the default mode, while the other operates under the transmit power-constrained SR mode. This phenomenon is properly captured in the Komondor simulator, where SR-based TXOPs are identified on a per-packet basis. In this case, the BSS that accesses the channel for the first time (e.g., $\text{BSS}_\text{A}$) is most likely to monopolize the channel. In contrast, the other BSS (e.g., $\text{BSS}_\text{B}$) is prone to transmit under the SR mode (as a result of $\text{BSS}_\text{A}$'s activity), until they alternate roles. Notice that either state $s_5$  or $s_6$ is more likely to be monopolized as the transmission time becomes longer than the idle periods. In our case, we have very long transmission times compared to the idle time since we assume full-buffer traffic, packet aggregation, and short Contention Window (CW) values.

\textcolor{black}{For each considered OBSS/PD threshold, both BSSs obtain the same throughput due to the deployment's symmetry. The BSSs alternate the default and the SR modes, which allows improving the aggregate throughput with respect to the case in which only one BSS applies SR. This gives a first indication of the potential for SR operation when applied jointly.}

%%% Complex OBSS/PD-based interactions
%\subsection{Interactions among Spatial Reuse groups}
\subsubsection{Toy scenario 2}
\label{section:advanced_interactions}
Differentiating between SRGs may potentially enhance spectral efficiency. In practice, devices belonging to the same SRG use a dedicated OBSS/PD threshold, namely SRG OBSS/PD. For the rest of the inter-BSS transmissions, the non-SRG OBSS/PD threshold is used instead. One possible use case may lie in residential building apartments, where BSSs belonging to the same building form an SRG. For the rest of \textcolor{black}{the} networks (e.g., public Wi-Fi in the street), other SRGs can be considered. 

\begin{figure}[ht!]
	\centering
	\subfigure[Deployment]{\label{fig:toy_scenario_2}\includegraphics[width=0.35\textwidth]{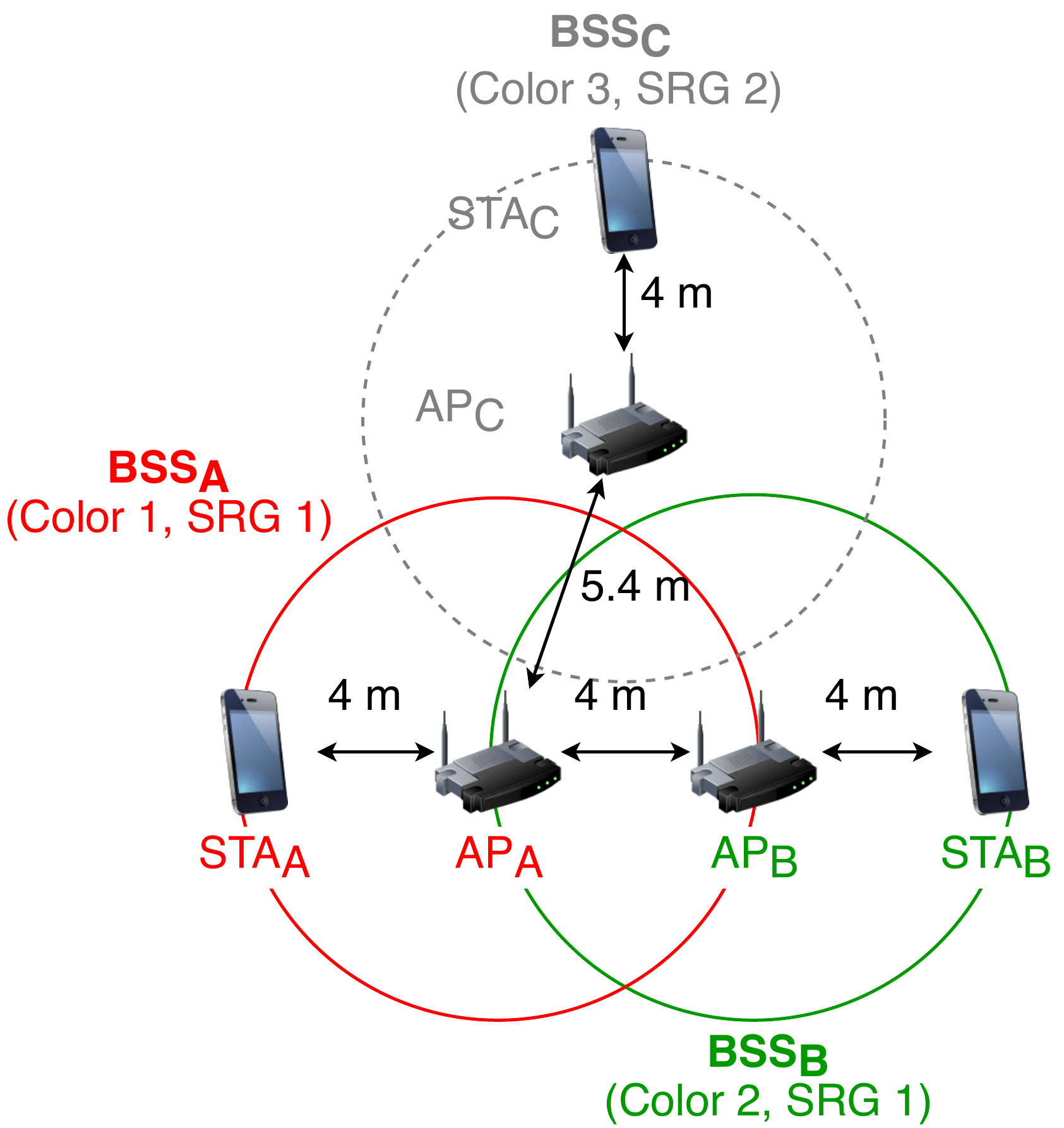}}
	\subfigure[Results]{\label{fig:SIM_1_3_individual}\includegraphics[width=0.5\textwidth]{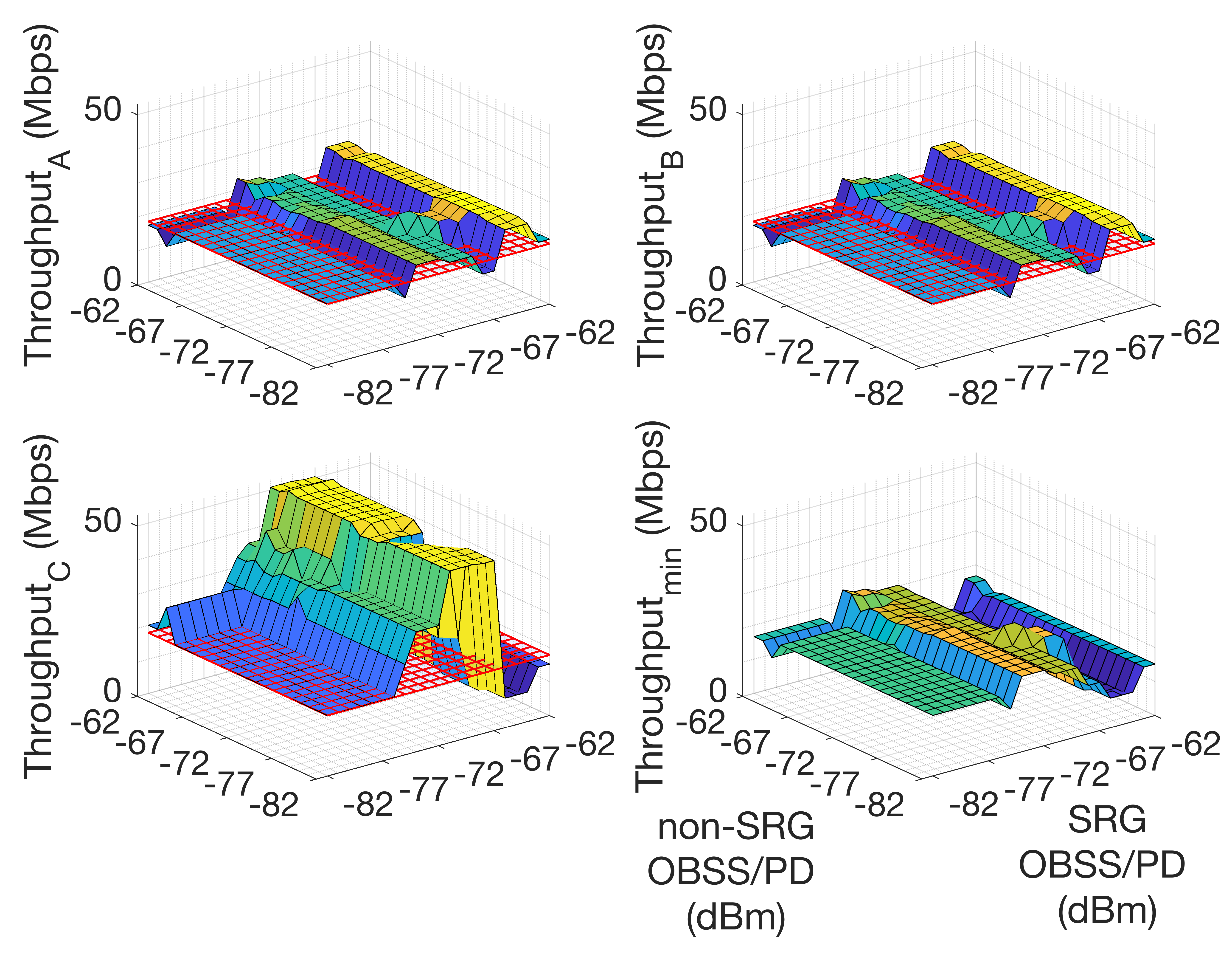}}
	\caption{Results of applying the OBSS/PD-based SR operation in \emph{Toy scenario 2}. In the right figure, the individual and minimum throughput are shown for each SRG and non-SRG OBSS/PD threshold. The red mesh indicates the performance achieved by using the default CCA/CS.}
	\label{fig:SIM_1_3}
\end{figure}

To illustrate the implications of using SR with SRGs, we propose \emph{Toy scenario 2}, which is depicted in Figure~\ref{fig:toy_scenario_2}. In this deployment, two different SRGs are considered for applying SR in all the BSSs. In particular, BSSs belonging to SRG1 (i.e., $\text{BSS}_\text{A}$ and $\text{BSS}_\text{B}$) are close to each other, \textcolor{black}{like} in a residential building. \textcolor{black}{Conversely}, $\text{BSS}_\text{C}$ belongs to SRG2. Figure~\ref{fig:SIM_1_3_individual} shows the individual and minimum throughput obtained by jointly applying OBSS/PD-based SR in \emph{Toy scenario 2}. \textcolor{black}{To address the scenario's complexity, we have considered that all the BSSs of an SRG use the same OBSS/PD threshold.}

\begin{figure}[ht]
	\centering
	\includegraphics[width=.45\columnwidth]{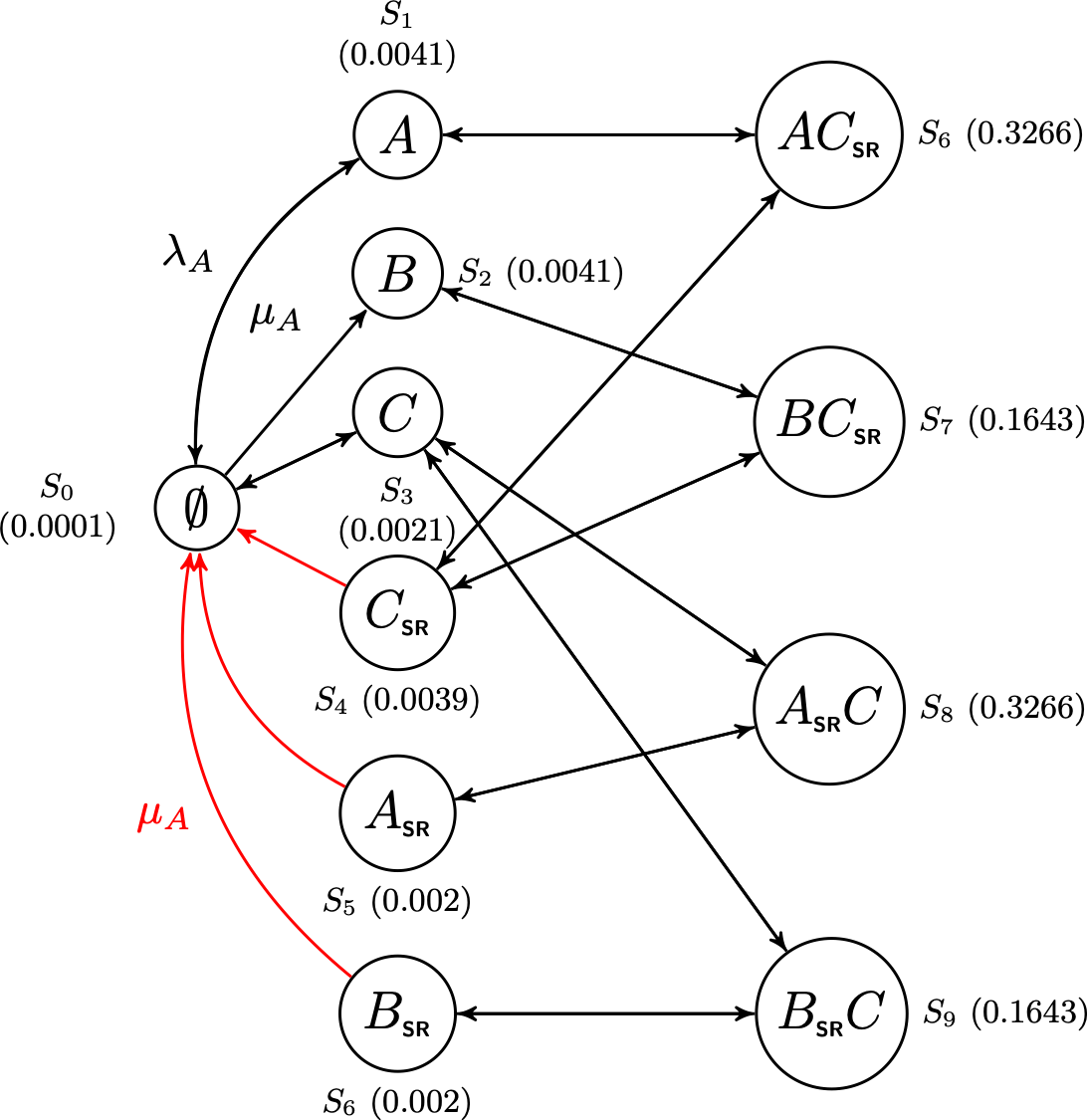}
	\caption{CTMN of \emph{Toy scenario 2}, for non-SRG OBSS/PD~=~-73~dBm and SRG OBSS/PD~=~-82~dBm. The unidirectional transitions are marked in red, and subindex \emph{SR} indicates the use of the non-SRG OBSS/PD threshold.}
	\label{fig:ctmn_scenario_2}
\end{figure}

\textcolor{black}{As a result of the complex inter-BSS interactions taking place in this deployment, the throughput functions are non-convex and contain different local minimum/maximum. This reveals the competitive nature of this deployment, where BSSs' individual performance is sometimes maximized at the expense of reducing the others' throughput. For instance, $\text{BSS}_\text{C}$ obtains the maximum performance only when $\text{BSS}_\text{A}$ suffers flow-in-the-middle throughput starvation (the same occurs for $\text{BSS}_\text{B}$). The CTMN that represents this situation is shown in Figure~\ref{fig:ctmn_scenario_2}, where all the BSSs use non-SRG OBSS/PD = -82 dBm and SRG OBSS/PD = -73 dBm.\footnote{\textcolor{black}{The CTMN model captures the utilization of different sensitivity thresholds: \emph{i)} default CCA/CS, \emph{ii)} SRG OBSS/PD, and \emph{iii)} non-SRG OBSS/PD.}}}

\textcolor{black}{Concerning the optimal max-min performance,\footnote{The max-min throughput corresponds to the solution that maximizes the minimum throughput achieved by a set of BSSs.} \textcolor{black}{it is given} when all the BSSs can ignore every single detected inter-BSS transmission, i.e., all the inter-BSS transmissions are equally treated regardless of its source. This situation is fair and, simultaneously increases the overall performance. As shown, OBSS/PD-based SR with SRGs can improve the performance of specific nodes (from a single group), but it also potentially leads to unfair situations. }

Finally, Table \ref{tbl:cross_validation} compares the results obtained in \emph{Toy scenario 2} from both the CTMN model and Komondor. In particular, we provide the \textcolor{black}{Mean Absolute Error (MAE) and the Mean Absolute Deviation (MAD)} for all the considered sensitivity thresholds. As shown, the error for  $\text{BSS}_\text{A}$ and $\text{BSS}_\text{B}$ is small, but significantly higher for $\text{BSS}_\text{C}$. This is strongly related to the fact that $\text{BSS}_\text{C}$ belongs to a separate SRG, which leads to different inter-BSS interactions. Moreover, dominant states may lead to situations that cannot be captured by the CTMNs, as previously shown for \emph{Toy scenario 1}. In particular, $\text{BSS}_\text{C}$ in \emph{Toy scenario 2} is prone to participate in these states because of its asymmetric location with respect to $\text{BSS}_\text{A}$ and $\text{BSS}_\text{B}$.

\begin{table}[ht!]
	\centering
	\resizebox{.32\columnwidth}{!}{\begin{tabular}{|c|c|c|c|}
			\hline
			& $\text{BSS}_\text{A}$ & $\text{BSS}_\text{B}$ & $\text{BSS}_\text{C}$ \\ \hline
			\begin{tabular}[c]{@{}c@{}}MAE \\ (Mbps)\end{tabular} & 4.72 & 4.74 & 14.81\\ \hline
			\begin{tabular}[c]{@{}c@{}}MAD \\ (Mbps)\end{tabular} & 2.63 & 2.63 & 8.17\\ \hline
	\end{tabular}}
	\caption{\textcolor{black}{Verification of the results obtained in \emph{Toy scenario 2} from the CTMN model and Komondor.}}
	\label{tbl:cross_validation}
\end{table}

% ----------------------------------
% -
% 	-- Performance Evaluation --
% -
% ----------------------------------

\section{Performance Evaluation}
\label{section:performance_evaluation}

%The Komondor simulator was conceived to allow the low-complexity integration of novel mechanisms included in new IEEE 802.11 standards, and to simulate massively crowded next-generation deployments in a reasonable amount of time. In particular, the implementation of SR in Komondor is expected to showcase the effects of using SR in dense scenarios. Besides, simulating SR allows us to validate the inter-BSS interactions studied through the analytical model, and to address the assumptions done by CTMNs for representing the behavior of WiFi devices more realistically. Notice that the 11ax SR operation has not been yet fully implemented in any other well-known simulator. At the time of publishing this article, SR is still being developed for ns-3.\footnote{It is planned to be included in the following repository: \url{https://gitlab.com/nsnam/ns-3-dev}.}

%\textit{Apart from the analytical model, we characterize SR in the Komondor simulator,\footnote{The implementation of SR can be found in Komondor v3.0, available in \url{https://github.com/wn-upf/Komondor/releases/tag/v3.0}.} The DCF's implementation in Komondor has been validated against ns-3 and the CTMN and Bianchi models \cite{barrachina2019komondor}. }

\begin{figure}[ht!]
	\centering
	\includegraphics[width=0.4\columnwidth]{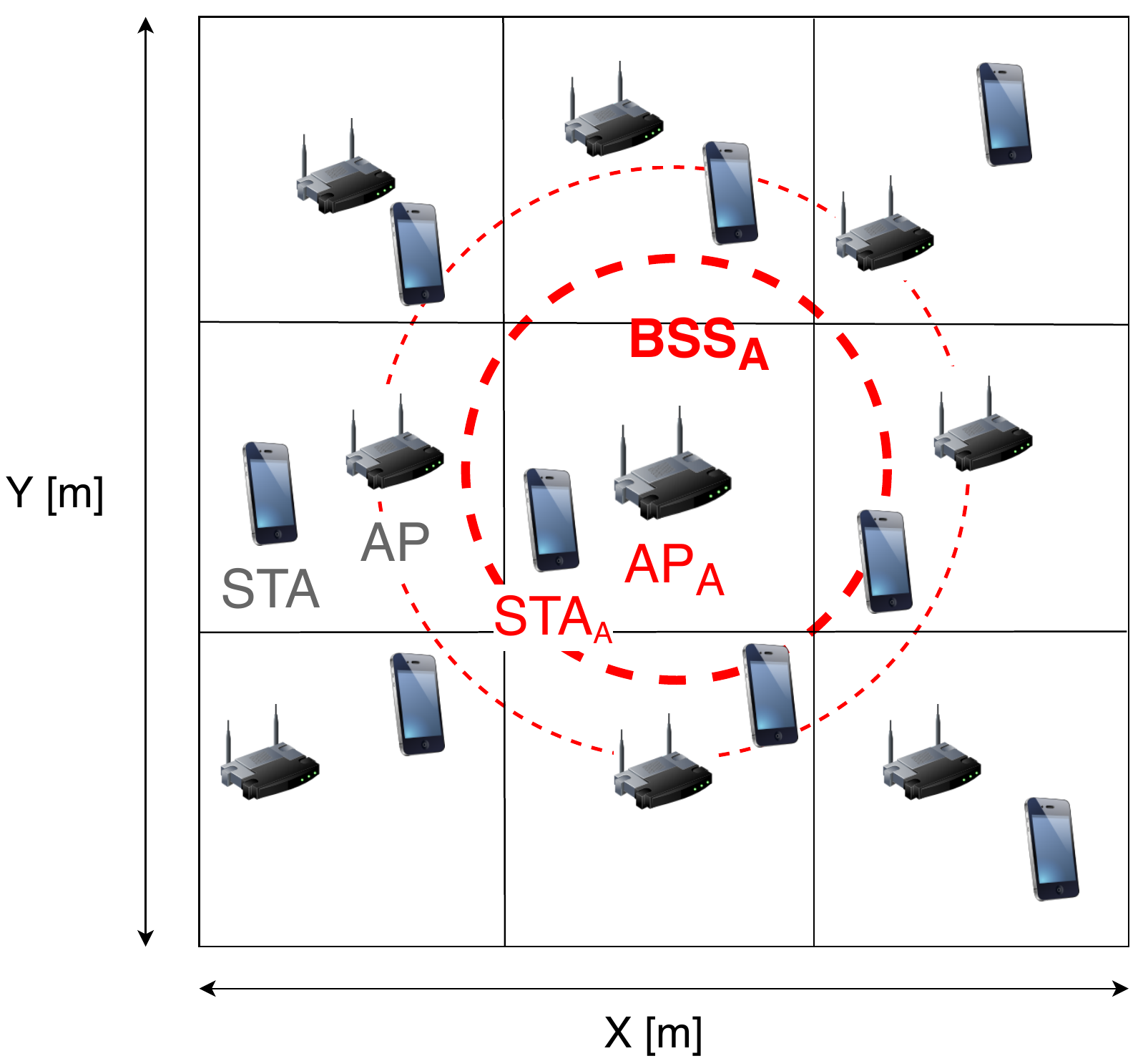}
	\caption{Random grid scenario containing 9 BSSs. The location of $\text{BSS}_\text{A}$ is fixed at the center for the sake of analysis.}
	\label{fig:random_scenario}
\end{figure}

In this Section, we study the performance gains of SR in large-scale WLAN scenarios. To that purpose, we leave the CTMNs-based analysis out and concentrate on simulation results. For the rest of this Section, each BSS is considered to be composed by an AP and a single STA, which are placed uniformly at random, as shown in Figure~\ref{fig:random_scenario}. 

\textcolor{black}{The simulation parameters are provided in Table \ref{table:parameters} of Appendix \ref{section:simulation_parameters}.}\footnote{\textcolor{black}{For more details on the Komondor's packet reception and interference models, we refer the interested reader to \cite{barrachina2019komondor}.}} The scenario is divided into nine cells, but the location of $\text{BSS}_\text{A}$ is always fixed at the center of the scenario for the sake of analysis. For the rest of the APs and STAs, their position is randomly selected within their corresponding cell. The configuration of each BSS is set homogeneously, i.e., they all use the same channel, the default sensitivity is set to -82 dBm, and the default transmission power is set to 20 dBm. Notice that, for dense deployments, $\text{BSS}_\text{A}$ is expected to suffer a higher level of interference than the others, which allows us to assess the effectiveness of the SR operation in crowded environments.

%%% DENSITY
\subsection{Network Density}
\label{section:random_scenarios_density}
To analyze SR based on the network density, we consider four different map sizes: sparse ($25\times25$ m), semi-dense ($20\times20$ m), dense ($15\times15$ m) and ultra-dense ($10\times10$ m). For each type of scenario, we provide 50 different deployments, in which APs and STAs are placed uniformly at random within their corresponding cell. $\text{BSS}_\text{A}$ is the only one applying the SR operation. Since we compute all the possible OBSS/PD values to be used by $\text{BSS}_\text{A}$, each random deployment leads to $21\times4\times50$ = 4,200 different scenarios.

\begin{figure}[ht!]
	\centering		
	\includegraphics[width=.48\columnwidth]{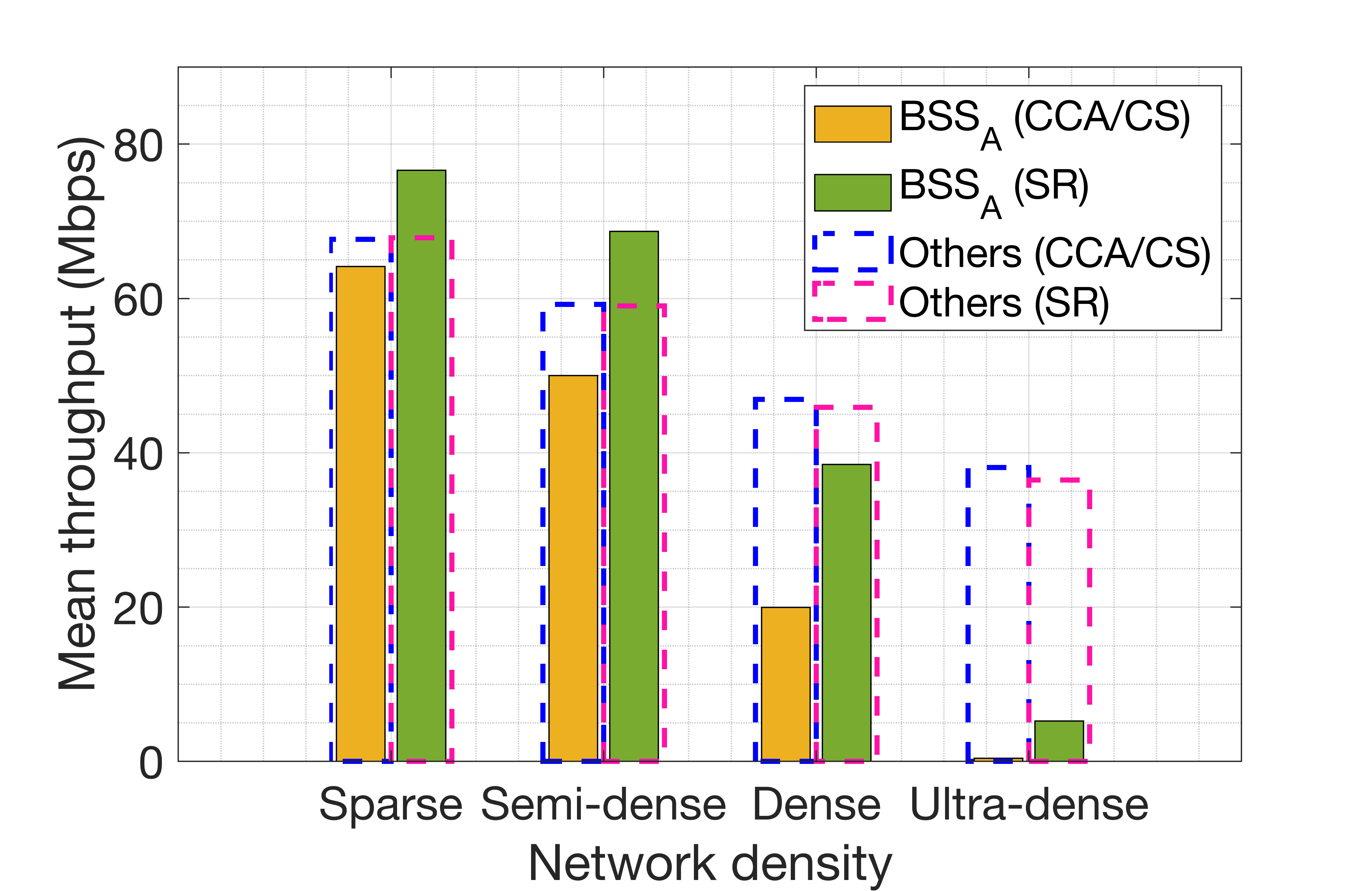}
	\caption{Mean throughput achieved with and without applying the SR operation in $\text{BSS}_\text{A}$, for each network density. Results show the mean throughput achieved by $\text{BSS}_\text{A}$ and the rest of BSSs.}
	\label{fig:SIM_2_1}
\end{figure}

Figure~\ref{fig:SIM_2_1} shows the average throughput achieved under the default and the SR settings\textcolor{black}{, respectively}. In particular, we differentiate between the individual throughput of $\text{BSS}_\text{A}$ and the average throughput of the other BSSs. For each network density, we have tried all the possible OBSS/PD values and compared the best one to the default CCA/CS.

First of all, we focus on the throughput that $\text{BSS}_\text{A}$ experiences by default (solid amber bars). We notice a dramatic decrease as network density increases. Nevertheless, the SR operation allows $\text{BSS}_\text{A}$ to significantly improve the throughput (displayed by the solid green bars). Note, as well, that the maximum improvement is experienced at the dense scenario ($15\times15$ m). While the default performance is quite high for sparser scenarios, channel re-utilization cannot be significantly improved at the ultra-dense scenario due to the high level of inter-BSS interference.

Apart from that, we observe that the average performance of the other BSSs (dashed bars) does not suffer radical changes for any of the network densities when $\text{BSS}_\text{A}$ applies SR. \textcolor{black}{This is a remarkably positive result, which indicates that SR allows improving the individual performance in a non-intrusive manner. This aspect is further analyzed in Section~\ref{section:random_scenarios_collaborative}, where we study the effects of simultaneously applying SR at multiple BSSs.}

%%% TRAFFIC LOAD
\subsection{Traffic load}
\label{section:random_scenarios_traffic_load}
\textcolor{black}{We now analyze the impact of the traffic load on the SR operation and the effect of frame aggregation. For that, we consider three different traffic loads ($l$), which are the same for all the BSSs: \emph{i)} low (1,000 packets/s, i.e., 12 Mbps), \emph{ii)} medium (2,000 packets/s, i.e., 24 Mbps), and \emph{iii)} high (10,000 packets/s, i.e., 120 Mbps). Besides, regarding packet aggregation, we consider a maximum of 1 and 64 aggregated frames, respectively.\footnote{\textcolor{black}{Notice that the maximum number of aggregated packets in a given transmission is bounded by the maximum PPDU duration (5,484 $\mu$s). So, depending on the employed data rate, the number of aggregated frames may be lower than 64.}} The traffic generated by APs in the downlink follows a Poisson process with average rate $\lambda$.}

\begin{figure}[ht!]
	\centering		
	\includegraphics[width=.9\columnwidth]{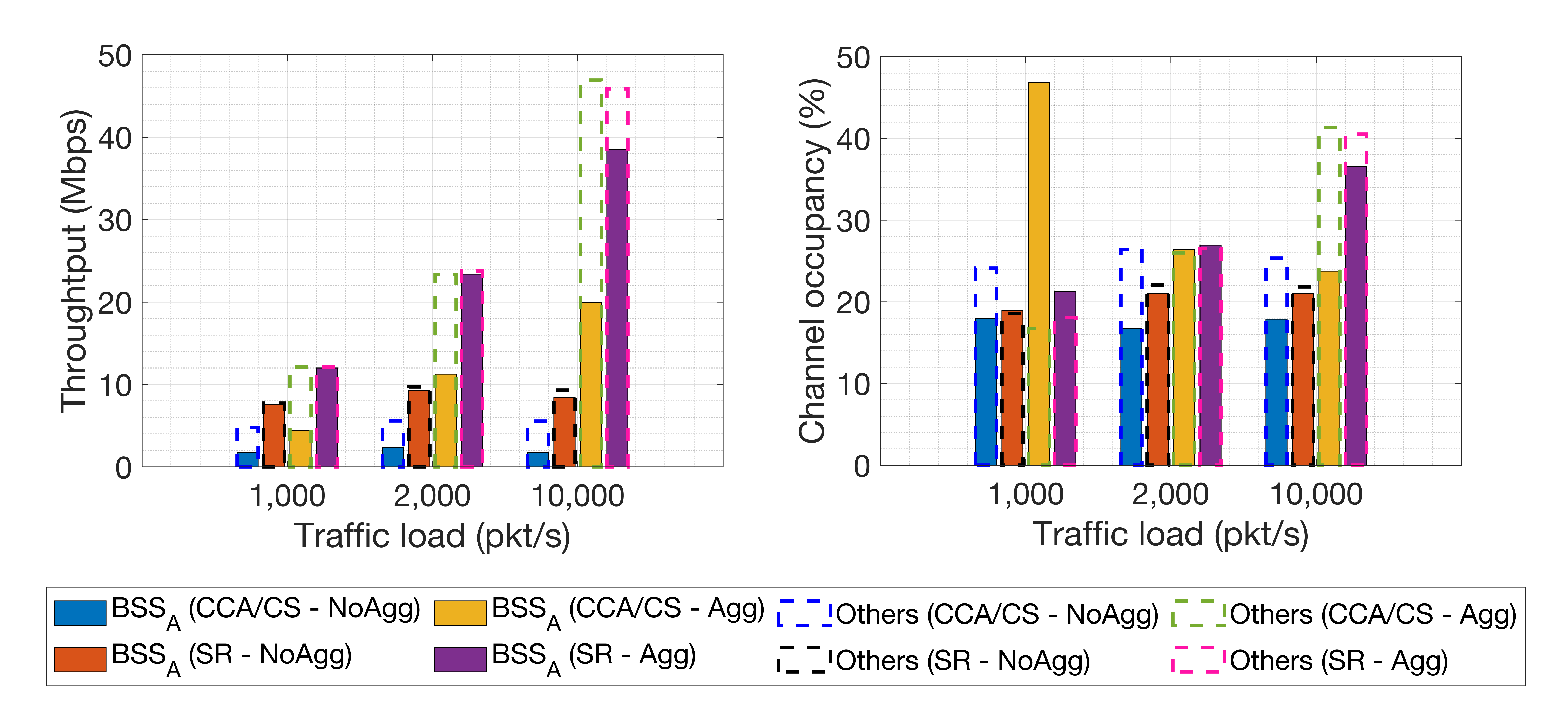}	
	\caption{\textcolor{black}{Mean channel occupancy and throughput achieved with and without applying the SR operation in $\text{BSS}_\text{A}$, for each traffic load. Results are shown for $\text{BSS}_\text{A}$ and for the rest of BSSs (others) when frame aggregation is enabled (\textit{Agg.}) and disabled (\textit{NoAgg.}).}}
	\label{fig:SIM_2_2}
\end{figure}

\textcolor{black}{Under the aforementioned conditions, Figure~\ref{fig:SIM_2_2} compares the performance achieved by default and SR configurations in the second densest scenario, which has been previously shown to achieve the maximum gains of the SR operation. As done before, the results show the individual performance of $\text{BSS}_\text{A}$ and the average performance of the rest of BSSs (again, only $\text{BSS}_\text{A}$ applies SR). In particular, Figure~\ref{fig:SIM_2_2} shows the maximum improvements achieved by BSS$_\text{A}$ in terms of throughput (in the left) and the corresponding channel occupancy (in the right). Notice that the SR configuration considers the OBSS/PD values that maximize BSS$_\text{A}$'s throughput.}

\textcolor{black}{We observe that $\text{BSS}_\text{A}$ obtains higher throughput gains as the traffic load increases, which are exacerbated when frame aggregation is enabled. In particular, the highest gain is noticed for the largest traffic load (120 Mbps), which constitutes a saturation regime. This is quite a remarkable result since the interference noticed by $\text{BSS}_\text{A}$ is much higher when all the surrounding devices are constantly transmitting due to their high traffic load. In turn, with the default CCA/CS threshold, BSS$_\text{A}$ is not able to deliver all the generated traffic, not even for the lightest traffic load (12 Mbps). The fact is that Figure~\ref{fig:SIM_2_2} shows the results of a dense scenario, which entails that the probability for BSS$_\text{A}$ to sense the channel busy is still significantly high.}

Regarding channel occupancy, an interesting phenomenon is observed for the lowest traffic load \textcolor{black}{and packet aggregation}. The fact is that the legacy CCA/CS configuration provides a higher channel occupancy than the SR one. However, this is not translated into higher throughput due to the high number of experienced collisions. Notice that collisions entail a high number of re-transmissions, which cause the observed increase in the occupancy. Finally, it is worth pointing out that the performance of the other BSSs is not affected when $\text{BSS}_\text{A}$ applies SR.

\textcolor{black}{Finally, Table \ref{tbl:summary_results} provides a summary of the results in all the scenarios for completeness. This table shows the gains in throughput (in Mbps) achieved by both BSS$_A$ and the rest of BSSs by applying SR  when compared to the legacy configuration. The results are provided for all the considered random deployments, including different network densities, traffic loads, and frame aggregation values.}

\begin{table}[ht!]
	\resizebox{\textwidth}{!}{
		\begin{tabular}{c|c|c|c|c|c|c|c|c|c|}
			\cline{2-10}
			\multicolumn{1}{c|}{} & \multirow{2}{*}{\textbf{Load (Mbps)}} & \multicolumn{2}{c|}{\textbf{Sparse}} & \multicolumn{2}{c|}{\textbf{Semi-dense}} & \multicolumn{2}{c|}{\textbf{Dense}} & \multicolumn{2}{c|}{\textbf{Ultra-dense}} \\ \cline{3-10} 
			\multicolumn{1}{l|}{} &  & \textit{A} & \textit{Others} & \textit{A} & \textit{Others} & \textit{A} & \textit{Others} & \textit{A} & \textit{Others} \\ \hline
			\multicolumn{1}{|c|}{\multirow{3}{*}{\textit{\textbf{Agg.}}}} & 
			\textbf{12} & 2.50 & 0.00 & 3.57 & 0.05 & 7.58 & 0.00 & 10.33 & 0.13 \\ \cline{2-10} 
			\multicolumn{1}{|c|}{} & 
			\textbf{24} & 4.26 & 0.01 & 10.21 & 0.13 & 12.13 & 0.41 & 10.50 & -0.35 \\ \cline{2-10} 
			\multicolumn{1}{|c|}{} & 
			\textbf{120} & 12.48 & 0.18 & 18.64 & -0.20 & 18.52 & -1.03 & 4.80 & -1.63 \\ \hline
			\multicolumn{1}{|c|}{\multirow{3}{*}{\cellcolor{mygray}}} & \cellcolor{mygray}\textbf{12} & \multicolumn{1}{c|}{\cellcolor{mygray}2.91} & \multicolumn{1}{c|}{\cellcolor{mygray}1.79} & \multicolumn{1}{c|}{\cellcolor{mygray}4.57} & \multicolumn{1}{c|}{\cellcolor{mygray}2.04} & \multicolumn{1}{c|}{\cellcolor{mygray}5.86} & \multicolumn{1}{c|}{\cellcolor{mygray}2.96} & \multicolumn{1}{c|}{\cellcolor{mygray}1.47} & \multicolumn{1}{c|}{\cellcolor{mygray}3.00} \\ \cline{2-10} 
			%\rowcolor{mygray}
			\multicolumn{1}{|c|}{\cellcolor{mygray}\textit{\textbf{NoAgg.}}} & \textbf{\cellcolor{mygray}24} & \multicolumn{1}{c|}{\cellcolor{mygray}5.84} & \multicolumn{1}{c|}{\cellcolor{mygray}2.36} & \multicolumn{1}{c|}{\cellcolor{mygray}7.41} & \multicolumn{1}{c|}{\cellcolor{mygray}3.74} & \multicolumn{1}{c|}{\cellcolor{mygray}6.96} & \multicolumn{1}{c|}{\cellcolor{mygray}4.11} & \multicolumn{1}{c|}{\cellcolor{mygray}1.25} & \multicolumn{1}{c|}{\cellcolor{mygray}3.94} \\ \cline{2-10} 
			%\rowcolor{mygray}
			\multicolumn{1}{|c|}{\cellcolor{mygray}} & \cellcolor{mygray}\textbf{120} & \multicolumn{1}{c|}{\cellcolor{mygray}5.16} & \multicolumn{1}{c|}{\cellcolor{mygray}2.99} & \multicolumn{1}{c|}{\cellcolor{mygray}7.02} & \multicolumn{1}{c|}{\cellcolor{mygray}3.23} & \multicolumn{1}{c|}{\cellcolor{mygray}6.65} & \multicolumn{1}{c|}{\cellcolor{mygray}3.76} & \multicolumn{1}{c|}{\cellcolor{mygray}1.13} & \multicolumn{1}{c|}{\cellcolor{mygray}3.89} \\ \hline
		\end{tabular}
	}
	\caption{\textcolor{black}{Mean throughput difference (in Mbps) achieved by applying SR in the considered random deployments, for each network density and traffic load. Results are displayed for BSS$_A$ (\textit{A}) and the rest of BSSs (\textit{Others}) when packet aggregation is enabled (\textit{Agg.}) and disabled (\textit{NoAgg.}).}}
\label{tbl:summary_results}
\end{table}

%%% COLLABORATIVE SR
\subsection{Joint Spatial Reuse Operation}
\label{section:random_scenarios_collaborative}

So far, we have studied the effects of applying SR at a single BSS (i.e., $\text{BSS}_\text{A}$). \textcolor{black}{Now, we focus on the concurrent behavior of the SR operation. Provided that $\text{BSS}_\text{A}$ always applies SR, we propose three study cases according to the set of \textcolor{black}{neighboring BSSs} that apply it: }
\begin{itemize}
	\item \textbf{Legacy:} all the other BSSs employ the default CCA/CS.
	\item \textbf{Mixed SR:} at the beginning of the simulation, each BSS randomly decides (with the same probability) whether to apply the SR operation or to remain using the default configuration.
	\item \textbf{All SR:} all the BSSs apply the SR operation. 
\end{itemize}

In particular, we \textcolor{black}{now consider} the \textcolor{black}{second densest scenario ($25\times25$m$^2$)} and the highest traffic load \textcolor{black}{(120 Mbps)}, thus representing the worst-case situation, where all APs become saturated \textcolor{black}{(the buffer limit is set to 100 packets)}. As done before, we have generated 50 random scenarios for averaging purposes, and, for each of them, we have tried all the possible OBSS/PD values to be used homogeneously by the BSSs applying the SR operation. Accordingly, we have used the best value to extract the maximum average improvement of SR with respect to the legacy configuration. In every situation (\emph{legacy}, \emph{mixed} and \emph{all SR}), we select the best OBSS/PD threshold from $\text{BSS}_\text{A}$'s point of view, which is also used to assess its impact on the others. Again, the SR configuration used for the channel occupancy is the one whereby $\text{BSS}_\text{A}$'s throughput is maximized.

\begin{figure*}[ht!]
	\centering		
	\subfigure[BSS$_\text{A}$]{\includegraphics[width=.52\columnwidth]{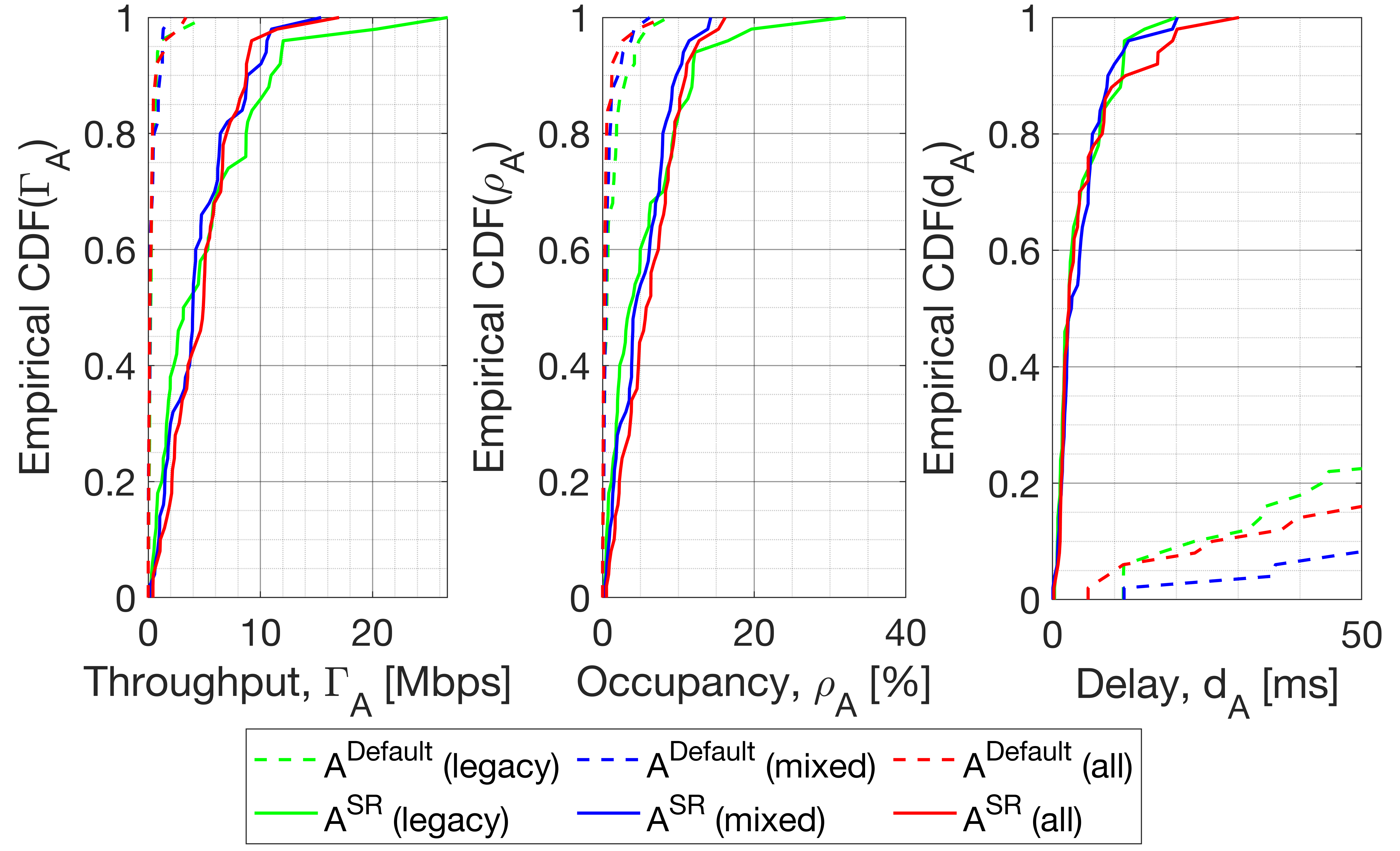}\label{fig:SIM_2_3_1}}%
	\subfigure[Others]{\includegraphics[width=.52\columnwidth]{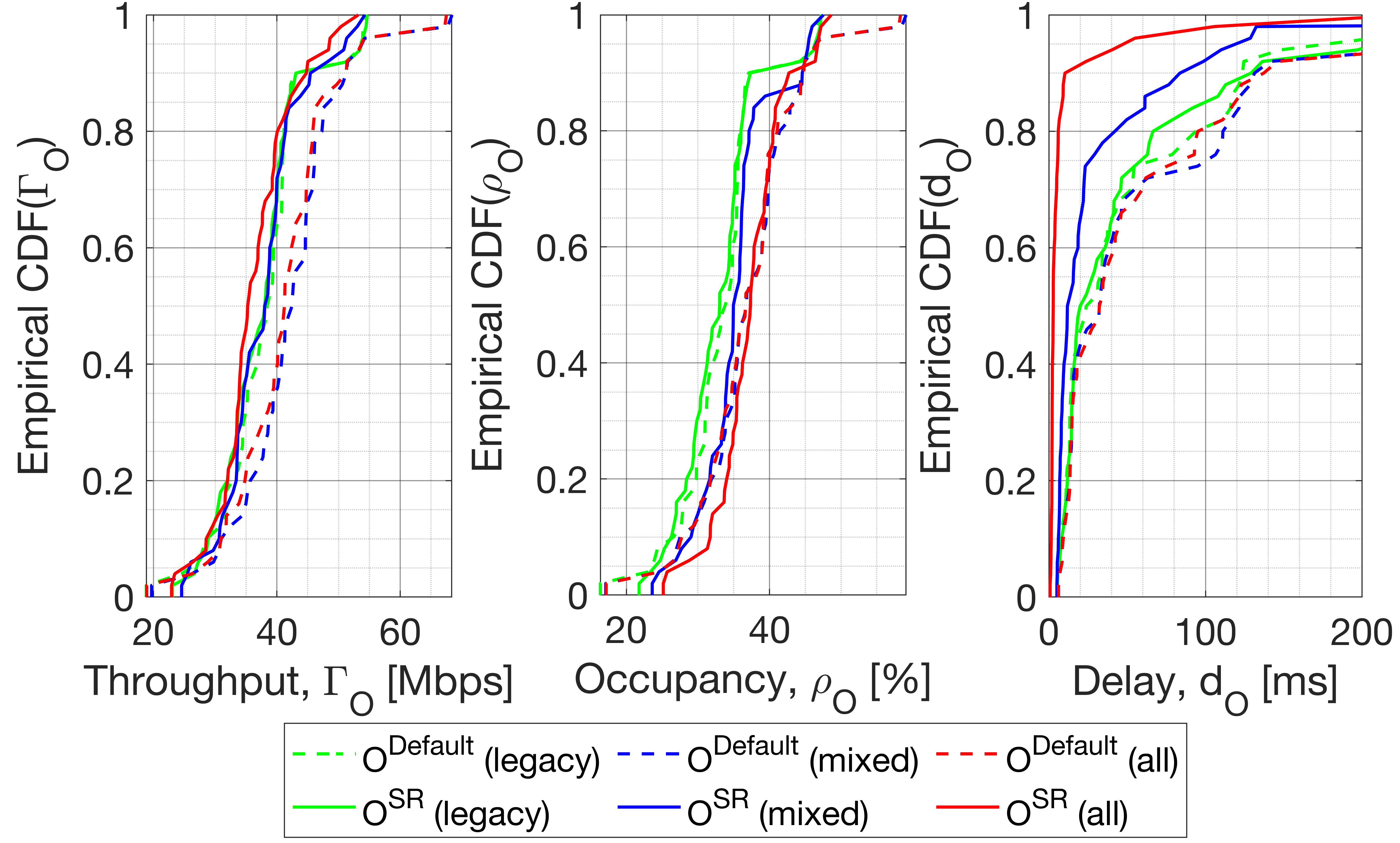}\label{fig:SIM_2_3_2}}
	\caption{Mean performance improvements achieved for each SR setting by BSS$_\text{A}$ ($A$) and the others ($O$). The results are shown for the OBSS/PD values that maximize the performance of BSS$_\text{A}$.}\label{fig:SIM_2_3}
\end{figure*}

Figure~\ref{fig:SIM_2_3} shows the empirical cumulative distribution function (CDF) of the following performance metrics: \emph{i)} throughput ($\Gamma$), \emph{ii)} percentage of time occupying the channel ($\rho$), and \emph{iii)} average delay for transmitting a packet once it arrives at the queue ($d$). Notice that, for each metric, we consider the performance improvements achieved by $\text{BSS}_\text{A}$ (indicated with subindex \emph{A}), and the average across the rest of BSSs (indicated with subindex \emph{O}). While Figure~\ref{fig:SIM_2_3_1} shows the performance of BSS$_\text{A}$, Figure~\ref{fig:SIM_2_3_2} focuses on the performance of the others.

As shown in Figure~\ref{fig:SIM_2_3_1}, $\text{BSS}_\text{A}$ achieves similar performance improvements, regardless of whether the environment applies SR or not. In particular, a high gain is noticed on the average delay. Moreover, regarding the others' performance (Figure~\ref{fig:SIM_2_3_2}), a null improvement is observed on the throughput, even for the \emph{all SR} context. \textcolor{black}{Notice that we are considering the SR configuration that maximizes $\text{BSS}_\text{A}$'s throughput, which is also used by the other BSSs to reduce the degrees of freedom and to better illustrate the effects of applying the SR operation. As a result, the best configuration from $\text{BSS}_\text{A}$'s perspective does not necessarily improve the others' performance (especially due to the imposed transmit power limitation).} Nevertheless, the delay is notably reduced as the number of BSSs using SR increases.

% ----------------------------------
% -
% 	-- Gaps --
% -
% ----------------------------------
\section{Ways Forward and Research Opportunities}
\label{section:ways_forwad}
The IEEE 802.11ax SR operation can potentially increase spectral efficiency in dense deployments. However, it is \textcolor{black}{still} in an early stage, and further developments are expected to sustain progress towards next-generation wireless deployments. 

\subsection{Unexplored Areas within the Spatial Reuse Operation}
In the context of the SR operation, the following areas have not been fully exploited yet:
\begin{itemize}
	\item \textbf{Assignment of BSS colors:} as discussed in Sections \ref{section:bss_coloring} and \ref{section:obss_pd_based}, BSS coloring is key for the OBSS/PD-based SR operation since it allows differentiating between intra and inter-BSS frames. However, the way BSS colors are assigned to BSSs is not specified, thus leading to potential collisions and miss-behaviors regarding the SR operation.
	\item \textbf{Election of SRGs:} similarly to the BSS color, the SRG is used to sub-classify inter-BSS frames, so that different PD policies can be applied to increase spectral efficiency. However, forming SRGs is not trivial since inter-BSS interactions must be carefully captured to \textcolor{black}{take advantage} of the SR operation. The set of policies regarding SRGs may be decided by the APs, as a result of monitoring phases.% (e.g., after experiencing several packet losses).
	\item \textbf{Establishment of OBSS/PD thresholds:} the election of OBSS/PD thresholds for each type of frame (SRG, and non-SRG) entails a set of trade-offs. On the one hand, too low values may lead to null improvement, thus framing the legacy operation whereby the channel is shared. On the other hand, too high values may generate performance anomalies, such as the hidden terminal problem or flow starvation. A potential solution to properly establish each OBSS/PD threshold is to capture all the inter-BSS interactions on a per-STA basis.
	\item \textbf{Optimal transmit power:} the current transmit power restriction is useful to prevent the accentuation of unfair situations. However, the SR operation's performance may be further increased in case of properly leveraging the transmit power according to the noticed interactions among nodes.
	\item \textbf{Disabling the SR operation:} there are situations in which the SR operation may be harmful to certain devices (e.g., in terms of fairness). Therefore, a given BSS must be able to identify whether the SR operation must be disabled or not. This can be achieved by setting the OBSS/PD threshold to the default CCA/CS value. Alternatively, the SR operation can only be disabled at STAs, leading to an AP-only SR setting. In this regard, AP-AP interactions would be mostly targeted.
\end{itemize}

Solving most of the problems mentioned above is not straightforward and requires an in-depth analysis to offer optimal or close-to-optimal solutions. While BSS color assignment may appear to be straightforward (e.g., through graph coloring techniques), defining OBSS/PD thresholds is a complex task that embraces many variables. In particular, \textcolor{black}{inter-BSS interactions have been shown in this paper to vary depending on the selected OBSS/PD values significantly}. Since the performance of IEEE 802.11 BSSs is not linear with the sensitivity and the transmission power (due to the nature of CSMA/CS), the optimal OBSS/PD threshold cannot be computed explicitly. Notice that the number of total combinations in an N-BSS scenario is $C = 21^\text{N}$, for 21 different non-SRG OBSS/PD thresholds. Therefore, the problem is intractable. If considering SRGs, the problem becomes even more \textcolor{black}{complicated} since the number of combinations is $C = (21\times21)^\text{N}$ (provided that we have 21 values to be used for SRG and non-SRG OBSS/PD thresholds).

\subsection{Integration of the Spatial Reuse Operation with other Techniques}

% Other ways forward
In addition to problems specific to the SR operation, the integration with many other novel mechanisms remains unexplored. Among them, we highlight OFDMA \cite{bankov2018ofdma, dovelos2018optimal}, multiple antenna systems \cite{liao2016mu}, and scheduled transmissions \cite{nurchis2019target}. The potential of SR goes further when combined with other techniques. 

\begin{figure}[ht!]
	\centering        
	\includegraphics[width=0.4\columnwidth]{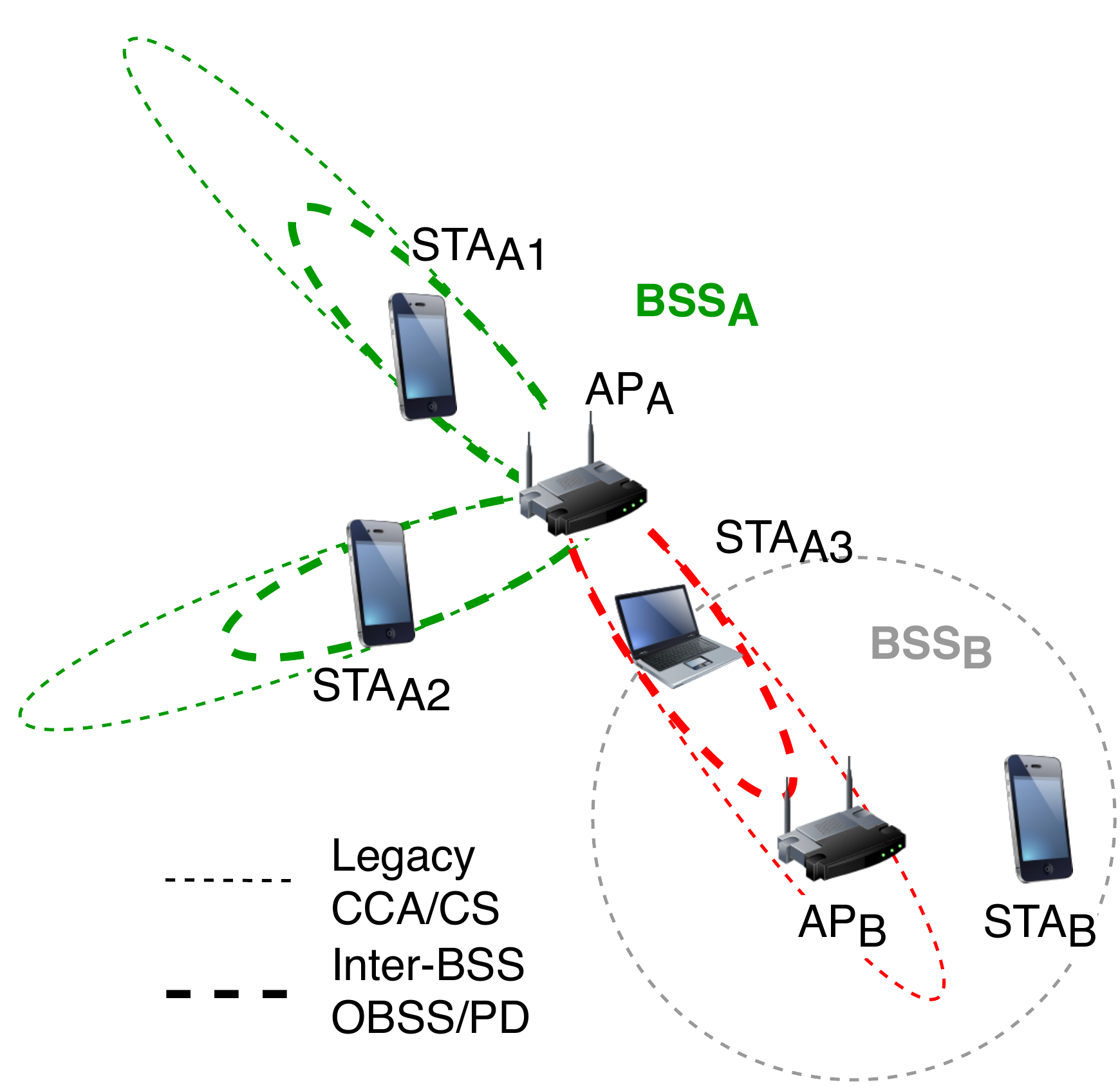}
	\caption{Potential application of SR combined with directional transmissions.}
	\label{fig:sr_and_beamforming}
\end{figure}

For instance, the combination of SR with directional transmissions may lead to efficient and \textcolor{black}{optimized} communications, where SR is applied on a per-beam basis. Figure~\ref{fig:sr_and_beamforming} devises the potential of combining SR with directional transmissions. As illustrated, $\text{BSS}_\text{A}$ applies the SR operation on a per-beam basis, while $\text{BSS}_\text{B}$ \textcolor{black}{continues} using the default CCA/CS. In particular, collisions by hidden nodes may be experienced for $\text{STA}_\text{A3}$, in case of using the inter-BSS OBSS/PD. However, channel reuse can be enhanced for transmissions to $\text{STA}_\text{A1}$ and $\text{STA}_\text{A2}$, which are out of range of $\text{AP}_\text{B}$. Therefore, the inter-BSS OBSS/PD can be used only for transmissions involving those two STAs, while a more conservative threshold can be employed for $\text{STA}_\text{A3}$.

\begin{figure*}[ht!]
	\centering        
	\subfigure[Scenario]{\includegraphics[width=0.4\columnwidth]{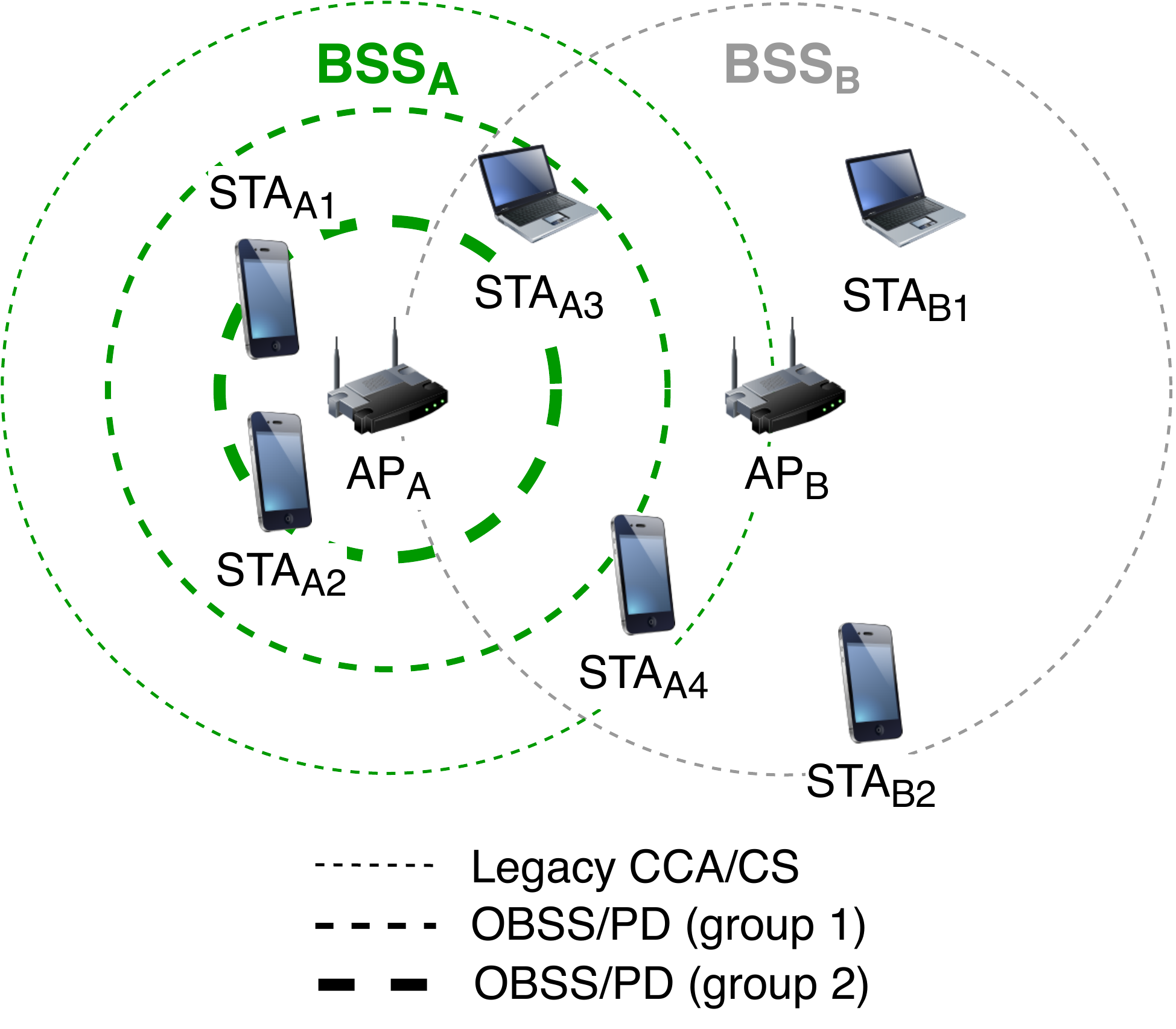}\label{fig:sr_and_tb_a}}
	\subfigure[Packets exchange]{\includegraphics[width=.45\columnwidth]{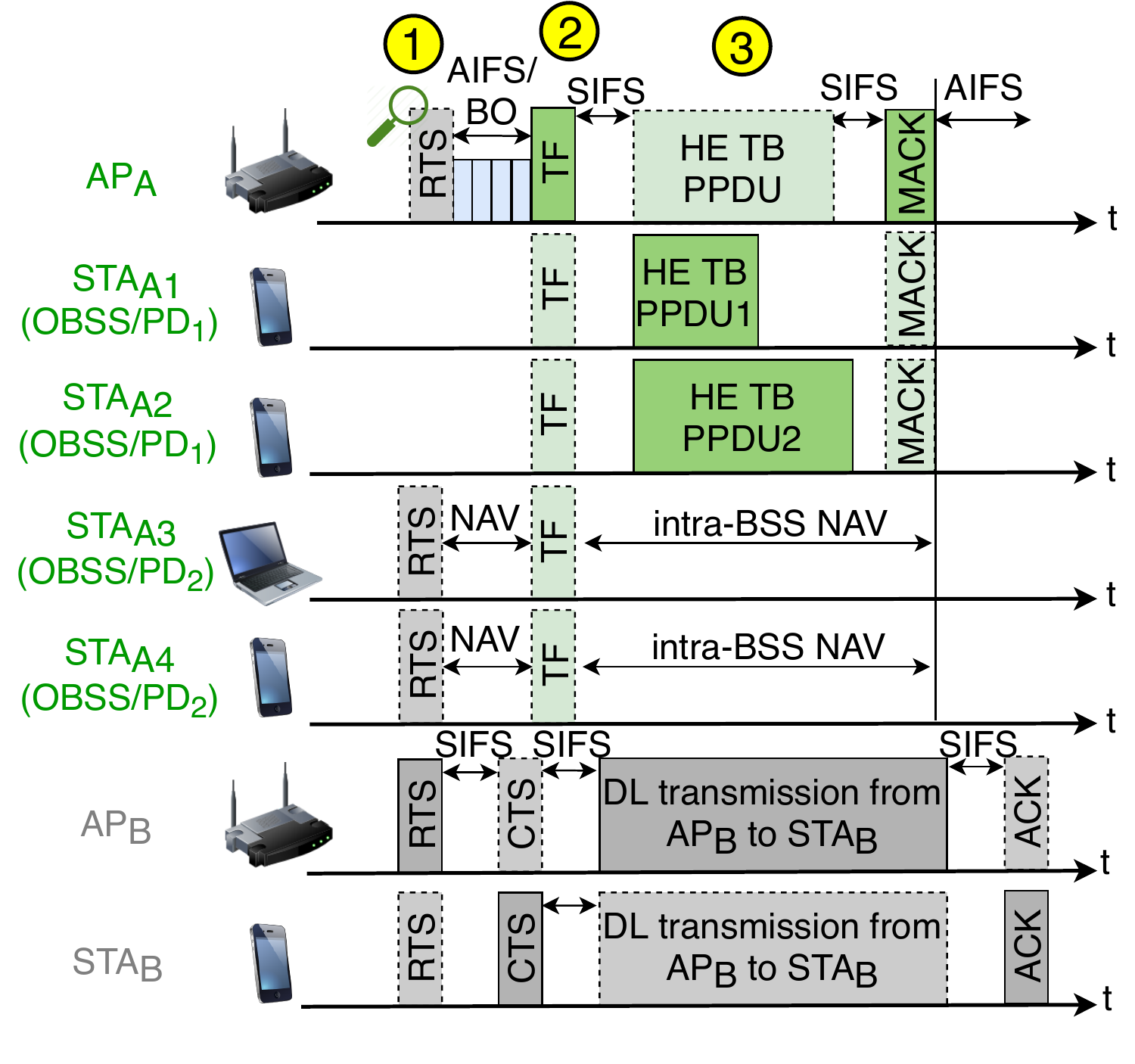}\label{fig:sr_and_tb_b}}
	\caption{Potential application of SR combined with TB communications.}        
	\label{fig:sr_and_tb}
\end{figure*}

Similarly to the integration with directional antennas, the potential of SR can be further exploited through TB communications. In this case, users of a given BSS can be categorized into different types, so that different inter-BSS OBSS/PD values are assigned to them. Figure \ref{fig:sr_and_tb_a} shows how users can be grouped based on different OBSS/PD thresholds. As a result, transmissions within the same BSS can be scheduled \textcolor{black}{differently}, thus improving spectral efficiency. In the proposed example, $\text{STA}_\text{A1}$ and $\text{STA}_\text{A2}$ belong to the first group because of their privileged position with respect to $\text{AP}_\text{A}$. Therefore, a more aggressive OBSS/PD threshold is employed when scheduling transmissions to these stations. The same reasoning can be applied to $\text{STA}_\text{A3}$, which, in this case, requires the usage of a more conservative OBSS/PD threshold for being scheduled in combination with SR. Finally, the legacy CCA/CS is used for $\text{STA}_\text{A4}$, in order to prevent negative interactions \textcolor{black}{concerning} $\text{BSS}_\text{B}$. It is worth pointing out that users belonging to different groups can be scheduled together, provided that the most restrictive OBSS/PD threshold is used. 

In Figure \ref{fig:sr_and_tb_b}, we show a data transmission resulting from the combination of TB communications and SR. In the yellow point \#1, $\text{AP}_\text{A}$ detects an inter-BSS transmission from $\text{AP}_\text{B}$, which can be ignored by using the most aggressive OBSS/PD, i.e., the one devoted for STAs in group 1. Accordingly, it schedules an uplink transmission from $\text{STA}_\text{A1}$ and $\text{STA}_\text{A2}$ (yellow point \#2). Finally, $\text{AP}_\text{A}$ receives the scheduled transmissions from group 1 (yellow point \#3).

\subsection{Artificial Intelligence for Spatial Reuse}
In light of the challenges posed by the 11ax SR operation, Artificial Intelligence (AI) emerges as a potential solution. In particular, WLANs are characterized by being highly varying in terms of users and channel dynamics. Moreover, we typically find decentralized deployments, at which none or little coordination is allowed. Hence, online learning stands as a suitable technique to address the optimization of SR in WLANs. In fact, many works on OBSS/PD adjustment, such as DSC \cite{smith2017dynamic} and COST \cite{selinis2018control}, are based on iterative methods. 

\textcolor{black}{Machine Learning (ML), and more precisely, Reinforcement Learning (RL), can improve the existing methods' performance. RL has been shown to fit with the decentralized nature of IEEE 802.11 WLANs \cite{long2007non, naddafzadeh2010distributed, zhou2011reinforcement, ghadimi2017reinforcement, wilhelmi2019collaborative, wilhelmi2019potential}. In particular, the usage of RL allows capturing subtle information that cannot be predicted before-hand (for instance, regarding inter-BSS interactions). Such information enables conducting a learning-based procedure, which is aimed at increasing performance while reducing the number of undesired situations (e.g., poor fairness).}

% ----------------------------------
% -
% 	-- Conclusions --
% -
% ----------------------------------
\section{Conclusions}
\label{section:conclusions}
In this paper, we have provided an extensive tutorial of the IEEE 802.11ax SR operation, which aims to maximize the performance of next-generation WLANs by increasing the number of parallel transmissions. Our purpose has been to do so in a clear and easy-to-understand manner. Thus, significant efforts have been made in providing meaningful examples of the different specifications related to SR. %First of all, we have presented the concepts that enable such an operation, which mostly refer to BSS coloring, SRGs, and scheduled transmissions. From there, we described the 11ax SR specification, which has been supported with illustrative examples. 

Apart from the tutorial, we have modeled the SR operation analytically using CTMNs. Through this model, we have analyzed the new kind of inter-BSS interactions that may result from applying SR in an OBSS. In particular, we have considered BSSs with a single STA, but more complex interactions are expected to happen when applying the SR operation in BSSs with multiple STAs. Apart from the analytical analysis, we have implemented the 11ax SR operation in the Komondor simulator. The potential of SR in large-scale scenarios has been evaluated through extensive simulations. \textcolor{black}{These results open the door to new contributions concerning the evaluation of the 11ax SR operation in more sophisticated environments, including nodes mobility, complex uplink/downlink traffic models, or heterogeneous deployments.}	

\textcolor{black}{Besides the} significant improvements achieved by the SR operation, other important aspects have been identified. First of all, it is important to highlight the non-intrusive characteristic of the SR operation. In particular, devices using SR can increase their performance without affecting other overlapping networks or preventing them \textcolor{black}{from transmitting}. This is a key feature to sustain performance growth. Moreover, the SR operation has been shown to perform better in scenarios with a high level of interference, i.e., high-density scenarios with a high traffic load. This confirms the utility of the SR for dense next-generation wireless networks.

However, finding the best SR configuration is far from trivial (it is a combinatorial problem), and remains an open problem. Indeed, the 11ax amendment does not provide \textcolor{black}{any specifications and guidelines} on this matter. We left as future work the design of mechanisms able to find the optimal parameters within the IEEE 802.11ax SR operation. For that purpose, the usage of RL can be particularly targeted. \textcolor{black}{Besides}, SR can evolve and be combined with other novel techniques such as directional transmissions or distributed OFDMA \textcolor{black}{to achieve further performance gains.}

\appendices
\section{IEEE 802.11ax Frames}
\label{section:frames}
In this Section, we introduce the type of frames that are considered in the 11ax amendment. Such information is key \textcolor{black}{to understand the SR operation better.}

% HE PPDUs
\subsection{HE PPDU formats}
Below, we briefly describe the Physical Protocol Data Unit (PPDU) formats available in the 11ax:
\begin{itemize}
	\item SU (Single User) HE PPDU: are meant for single user communications.
	\item  HE Extended Range HE PPDU: are meant for single user long-range transmissions, hence only contemplate 20 MHz bandwidths in a single spatial stream.
	\item  MU (Multi-User) HE PPDU: due to the OFDMA operation, such kind of PPDUs are meant for multiple transmissions to one or more users.
	\item Trigger-Based (TB) HE PPDU: in this case, MU UL transmissions are scheduled by the AP, which decides which STAs are expected to transmit during a specific elapse of time. The TB HE PPDUs can make use of OFDMA and/or MU-MIMO.
\end{itemize}

\begin{figure}[ht!]
	\centering
	\epsfig{file=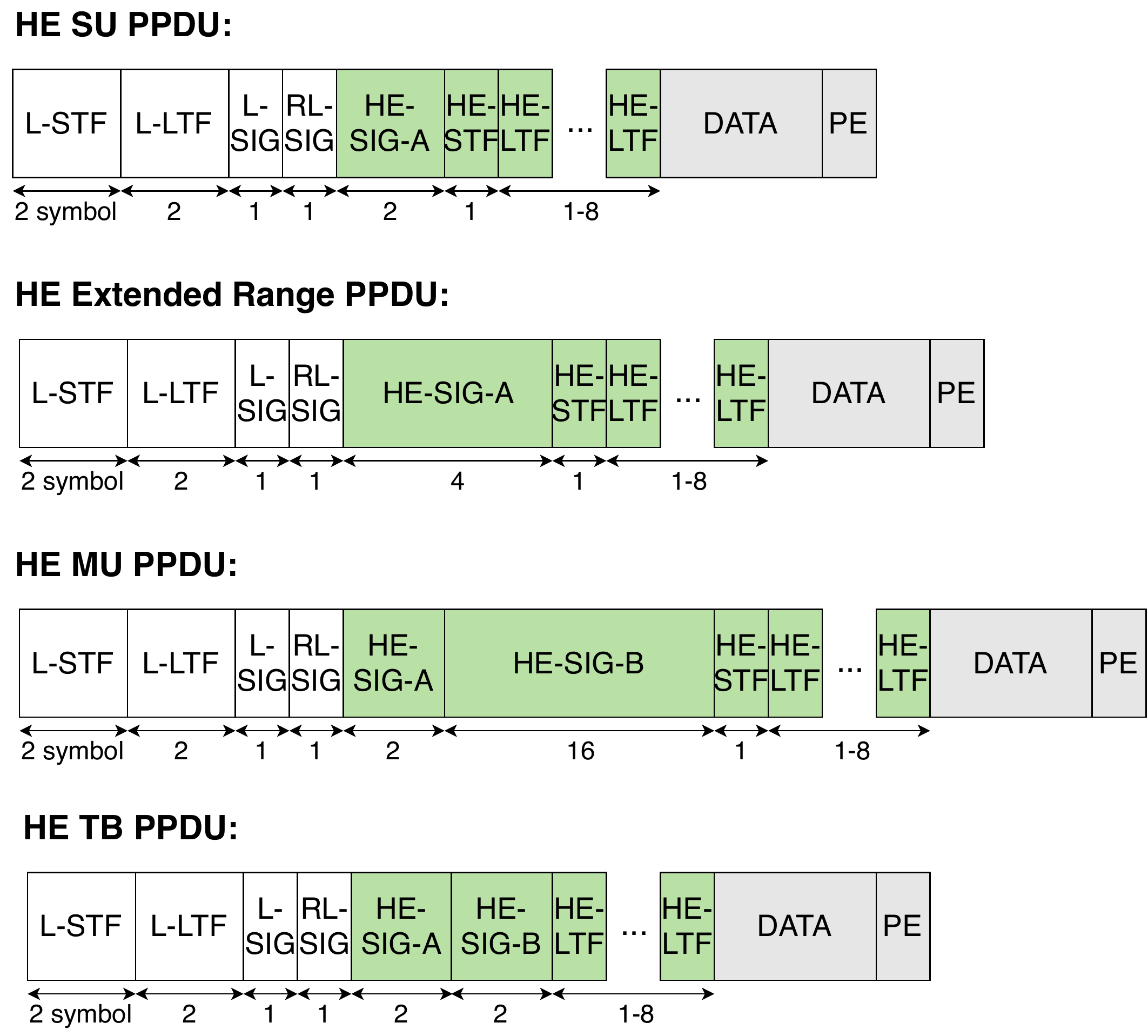, width=.55\columnwidth}
	\caption{HE PPDU formats. New IEEE 802.11ax fields are highlighted in green.}
	\label{fig:appendix_1}
\end{figure}

The new fields included in the abovementioned HE PPDU formats are HE Signal A Field (HE-SIG-A), HE Signal B Field (HE-SIG-B), HE Short Training Field (HE-STF), and HE Long Training Field (HE-LTF), which are shown in Figure \ref{fig:appendix_1}.

Among the new fields, we highlight HE-SIG-A, which includes the following elements related to the SR operation:
\begin{itemize}
	\item BSS color: it is used as an identifier of the BSS (refer to Section \ref{section:bss_coloring}).
	\item Spatial Reuse: this field indicates whether the HE node supports the SR operation. If this is the case, the field also indicates the limit on the transmission power to be used during the SR opportunities that can potentially be detected. Notice that a single Spatial Reuse field (of length of 4 bits) is carried in HE SU/MU/ER PPDUs,  while HE TB PPDUs may include up to four Spatial Reuse fields. In particular, each field is meant for the SR operation in each allowed channel width (i.e., 20 MHz, 40 MHz, 80 MHz, and 160 MHz).
\end{itemize}

Besides supporting HE PPDU formats, HE STAs are required to be compatible with legacy formats. More information regarding HE PPDU formats can be found in \cite{rhode2017whitepaper}. 

% Other
\subsection{Management Fields for Spatial Reuse}
Some operations auxiliary to SR are enabled by control frames, which are Beacon, Probe Response, and (Re)Association Response frames. Beacons are used by APs to announce the presence of a BSS and to provide details of it. \textcolor{black}{In particular, by means of Beacons, an AP may request an STA to gather information regarding the environment:} information of BSSs matching a particular BSSID and/or SSID, channel-specific report, or HE Operation element of neighboring HE APs. With this information, the AP can make decisions related to the SR operation. Regarding Probe Responses, they are meant to carry the information requested by devices scanning the area through Probe Requests. Finally, (Re)Association Response frames are sent by APs to which an STA attempts to associate.

The abovementioned kind of frames are important to the SR operation because they carry, among other fields, the following information:
\begin{itemize}
	\item \textbf{HE Capabilities:} it is used by HE STAs to announce support for certain HE capabilities.
	\item \textbf{HE Operation:} it defines the operation of HE STAs. For instance, it indicates whether BSS coloring is enabled or not.
	\item \textbf{BSS Color Change Announcement:} it is used by HE APs to indicate the utilization of a new BSS color so that the associated STAs and the surrounding devices can be aware of the change.
	\item \textbf{Spatial Reuse Parameter Set (SRPS) element:} this element provides the necessary information to carry out the OBSS/PD-based SR operation, which is defined in Section \ref{section:obss_pd_based}. The SRPS element is further defined in Appendix \ref{section:srps}.
\end{itemize}

\subsubsection{Spatial Reuse Parameter Set element}
\label{section:srps}
\begin{figure}[ht!]
	\centering
	\epsfig{file=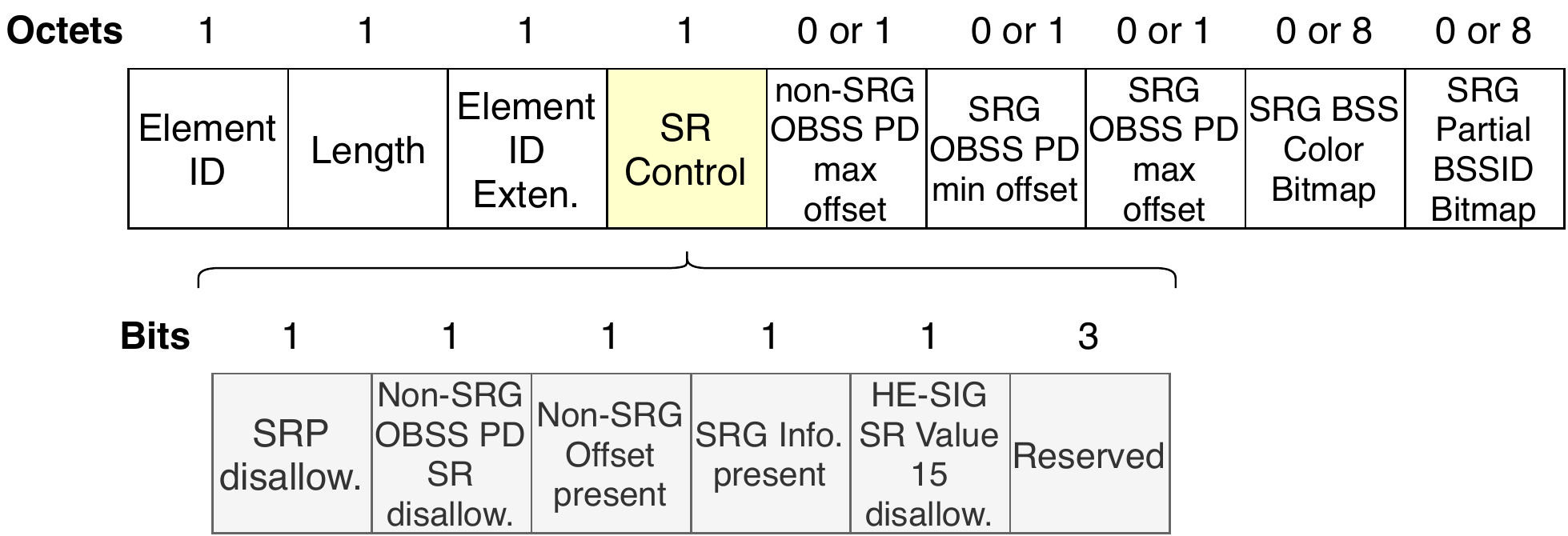, width=.7\columnwidth}
	\caption{Spatial Reuse Parameter Set element.}
	\label{fig:appendix_2}
\end{figure}

The format of the SRPS element is optionally present in Beacons, Probe Responses and (Re)Association responses. Figure \ref{fig:appendix_2} shows the SRPS element in detail.
% SR PARAMETER SET ELEMENT

Each item in the SRPS element is next described:
\begin{itemize}
	\item Element ID: set to 255.
	\item Length: not defined.
	\item Element ID extension: set to 39.
	\item SR Control field: contains the following parameters:
	\begin{itemize}
		\item PSR Disallowed: indicates whether PSR transmissions are allowed or not at non-AP STAs associated with the AP that transmitted this element.
		\item Non-SRG OBSS/PD SR Disallowed: indicates whether non-SRG OBSS/PD SR transmissions are allowed or not at non-AP STAs associated with the AP that transmitted this element.
		\item Non-SRG Offset Present: indicates whether the Non-SRG OBSS/PD Max Offset subfield is present in the element.
		\item SRG Information Present: indicates whether the SRG OBSS/PD Min Offset, SRG OBSS/PD Max Offset, SRG BSS Color Bitmap, and SRG Partial BSSID Bitmap subfields are present in the element.
		\item HE-SIG-A Spatial Reuse Value 15 disallow: indicates whether non-AP STAs associated with the AP that transmitted this element may set the TXVECTOR parameter SPATIAL\_REUSE to PSR\_AND\_NON-SRG-OBSS-PD\_PROHIBITED to avoid PSR transmissions.
	\end{itemize}
	\item Non-SRG OBSS/PD Max Offset: integer to generate the maximum Non-SRG OBSS/PD threshold.
	\item Non-SRG OBSS/PD Min Offset: integer to generate the minimum Non-SRG OBSS/PD threshold.
	\item SRG OBSS/PD Max Offset: integer to generate the maximum SRG OBSS/PD threshold.
	\item SRG BSS Color Bitmap: indicates which BSS Color values are used by the members of the SRG.
	\item SRG Partial BSSID Bitmap: indicates which partial BSSID values are used by members of the SRG.
\end{itemize}

\textcolor{black}{The offsets for setting both SRG and non-SRG OBSS/PD thresholds are limited by the following conditions:
\begin{enumerate}
	\item -82 dBm $\leq$ -82 dBm + SRG OBSS/PD Min Offset $\leq$ -62 dBm 
	\item SRG OBSS/PD Min Offset $\leq$ SRG OBSS/PD Max Offset
	\item SRG OBSS/PD Max Offset + -82 dBm $\leq$ -62 dBm 
	\item Non-SRG OBSS/PD Max Offset + -82 dBm $\leq$  -62 dBm
\end{enumerate}
Based on the aforementioned offsets, both non-SRG and SRG OBSS/PD thresholds, respectively, are bounded as shown in Tables \ref{tbl:non-srg} and \ref{tbl:srg}.}

\begin{table}[ht!]
	\centering
	\scriptsize{\resizebox{.8\columnwidth}{!}{\begin{tabular}{|c|c|c|c|}
				\hline
				\begin{tabular}[c]{@{}c@{}}\textbf{OBSS/PD SR} \\\textbf{disallowed}\end{tabular} & \textbf{Non-SRG Offset} & \begin{tabular}[c]{@{}c@{}}\textbf{Non-SRG} \\\textbf{OBSS/PD Min}\end{tabular} & \begin{tabular}[c]{@{}c@{}}\textbf{Non-SRG} \\\textbf{OBSS/PD Max}\end{tabular} \\ \hline
				Unspecified & Unspecified & -82 & -62 \\ \hline
				0 & 0 & -82 & -62 \\ \hline
				0 & 1 & -82 & \begin{tabular}[c]{@{}c@{}}-82 + Non-SRG\\ OBSS/PD Max off.\end{tabular}\\ \hline
				1 & Don't care & -82 & -82 \\ \hline
	\end{tabular}}}
	\caption{\textcolor{black}{Minimum and maximum allowed non-SRG OBSS/PD thresholds (in dBm), based on the value of \texttt{OBSS/PD SR Disallowed} and \texttt{Non-SRG Offset Present} fields.}}
	\label{tbl:non-srg}
\end{table}

\begin{table}[ht! ]
	\centering
	%\resizebox{\columnwidth}{!}{}
	\scriptsize{\begin{tabular}{|c|c|c|}
			\hline
			\textbf{SRG field} & \textbf{SRG OBSS/PD Min} & \textbf{SRG OBSS/PD Max} \\ \hline	\begin{tabular}[c]{@{}c@{}}Unspecified\end{tabular} & N/A & N/A \\ \hline
			0 & N/A & N/A \\ \hline
			1 & \begin{tabular}[c]{@{}c@{}}-82 + SRG OBSS/PD\\ Min Offset\end{tabular} & \begin{tabular}[c]{@{}c@{}}-82 + SRG OBSS/PD\\ Max Offset\end{tabular} \\ \hline
	\end{tabular}}
	\caption{\textcolor{black}{Minimum and maximum allowed SRG OBSS/PD values (in dBm), based on the value of the \texttt{SRG} field. If SRG is not activated (or its value is unspecified), PPDU frames cannot be classified as SRG frames.}}
	\label{tbl:srg}
\end{table}

\section{Model and Simulation Parameters}
\label{section:simulation_parameters}
The simulation parameters used to obtain results from both CTMN model and Komondor are collected in Table \ref{table:parameters}.

\begin{table}[h!]
	\centering
	\resizebox{.6\columnwidth}{!}{
		\begin{tabular}{c|l|l|}
			\cline{2-3}
			\multicolumn{1}{l|}{} & \textbf{Parameter} & \textbf{Value}
			\\ \hline
			% PHY
			\multicolumn{1}{|c|}{\multirow{9}{*}{\rotatebox[origin=c]{90}{PHY}}} & Central frequency, $f_c$ & 5 GHz \\ \cline{2-3} 
			\multicolumn{1}{|c|}{} & Transmission/Reception gain, $G_{tx}/G_{rx}$ & 0/0 dB \\ \cline{2-3} 
			\multicolumn{1}{|c|}{} & Path-loss (residential scenario), $\text{PL}(d)$ & See (\cite{pathloss11ax})  \\ \cline{2-3}
			\multicolumn{1}{|c|}{} & Background noise level, $N$ & -95 dBm \\ \cline{2-3}
			\multicolumn{1}{|c|}{} & Legacy OFDM symbol duration, $\sigma_\text{leg}$ & 4 \textmu s \\
			\cline{2-3}
			\multicolumn{1}{|c|}{} & OFDM symbol duration (GI-32), $\sigma$ & 16 \textmu s \\ 				\cline{2-3}
			\multicolumn{1}{|c|}{} & Number of subcarriers (20 MHz), $N_{sc}$ & 234   \\
			\cline{2-3}
			\multicolumn{1}{|c|}{} & Number of spatial streams, $N_{ss}$ & 1  \\
			%\cline{2-3}
			%\multicolumn{1}{|c|}{} & Transmit power levels, $\mathcal{T}$ & 1 to 20 dBm (1 dBm steps) \\
			\hline
			% MAC
			\multicolumn{1}{|c|}{\multirow{16}{*}{\rotatebox[origin=c]{90}{MAC}}} & Empty slot duration, $\text{T}_e$ & 9 $\mu$s\\ 
			\cline{2-3} 
			\multicolumn{1}{|c|}{} & SIFS duration, $T_\text{SIFS}$ & 16 \textmu s  \\
			\cline{2-3} 
			\multicolumn{1}{|c|}{} & DIFS/AIFS duration, $T_\text{DIFS/AIFS}$ & 34 \textmu s \\
			\cline{2-3} 
			\multicolumn{1}{|c|}{} & PIFS duration, $T_\text{PIFS}$  & 25 \textmu s \\
			\cline{2-3} 
			\multicolumn{1}{|c|}{} & Legacy preamble duration, $T_\text{PHY-leg}$ & 20 \textmu s  \\
			\cline{2-3}
			\multicolumn{1}{|c|}{} & HE single-user field duration, $T_\text{HE-SU}$ & 100 \textmu s \\
			\cline{2-3} 
			\multicolumn{1}{|c|}{} & ACK duration, $T_\text{ACK}$ & 28 \textmu s\\
			\cline{2-3} 
			\multicolumn{1}{|c|}{} & Block ACK duration, $T_\text{BACK}$ & 32 \textmu s \\
			\cline{2-3} 
			\multicolumn{1}{|c|}{} &  Size OFDM symbol (legacy), $L_{s,l}$ & 24 bits \\
			\cline{2-3} 
			\multicolumn{1}{|c|}{} & Length of data packets, $\text{L}_{d}$ & 12,000 bits \\
			\cline{2-3} 
			\multicolumn{1}{|c|}{} & No. of frames in an A-MPDU, $N_{\text{agg}}$ & 64 \\
			\cline{2-3} 
			\multicolumn{1}{|c|}{} & Length of an RTS packet, $L_\text{RTS}$ & 160 bits \\
			\cline{2-3} 
			\multicolumn{1}{|c|}{} & Length of a CTS packet, $L_\text{CTS}$ & 112 bits \\
			\cline{2-3} 
			\multicolumn{1}{|c|}{} & Length of service field, $L_\text{SF}$ & 16 bits  \\
			\cline{2-3} 
			\multicolumn{1}{|c|}{} & Length of MAC header, $L_\text{MH}$ & 320 bits \\
			\cline{2-3} 
			\multicolumn{1}{|c|}{} & Contention window (fixed), $\text{CW}$ & 15 \\
			\cline{2-3} 
			\multicolumn{1}{|c|}{} & Allowed sensitivity levels, $\mathcal{S}$ & -82 to -62 (1 dBm steps) \\
			\hline
%			% Other
%			\multicolumn{1}{|c|}{\multirow{2}{*}{\centering\rotatebox[origin=c]{90}{Misc.  }}} & Traffic model, $\Lambda$ & Downlink (UDP)\\
%			\cline{2-3} 
%			\multicolumn{1}{|c|}{} & Traffic generation ratio, $l$ & 1,000, 2,000, 10,000 pkts/s\\ 
%			\cline{2-3} 
%			\multicolumn{1}{|c|}{} & Map area (random scenario), $A$ & 625, 400, 225, 100 m$^2$\\
%			\hline
	\end{tabular}}
	\caption{Simulation parameters.}
	\label{table:parameters}
\end{table}

\section*{Acknowledgment}
This  work  has  been  partially  supported  by  the  Spanish Ministry of Economy and Competitiveness under the Maria de Maeztu  Units  of  Excellence  Programme  (MDM-2015-0502), by PGC2018-099959-B-100 (MCIU/AEI/FEDER,UE), by the Catalan Government under SGR grant for research support (2017-SGR-11888), by SPOTS project (RTI2018-095438-A-I00) funded by the Spanish Ministry of Science, Innovation and Universities, and  by a Gift from the Cisco University Research Program (CG\#890107, Towards Deterministic Channel Access in High-Density WLANs) Fund, a corporate advised fund of Silicon Valley Community Foundation.
The authors would like to thank Malcolm Smith and Adrian Garcia for their thorough reviews and insightful comments.
	
%%%%%%%%%%%%%%%%%%%%%%%%%
%%%  BIBLIOGRAPHY    
%%%%%%%%%%%%%%%%%%%%%%%%%

\bibliographystyle{unsrt}
\bibliography{references}

\begin{thebibliography}{10}

\bibitem{tgax2019draft}
TGax.
\newblock {IEEE P802.11ax/D4.0}, 2019.

\bibitem{bellalta2016ieee}
Boris Bellalta.
\newblock {IEEE 802.11 ax: High-efficiency WLANs}.
\newblock {\em IEEE Wireless Communications}, 23(1):38--46, 2016.

\bibitem{afaqui2016ieee}
M~Shahwaiz Afaqui, EG~Villegas, and EL~Aguilera.
\newblock {IEEE 802.11 ax: Challenges and requirements for future high
  efficiency WiFi}.
\newblock {\em IEEE Wireless Communications}, 99:2--9, 2016.

\bibitem{qu2018survey}
Qiao Qu, Bo~Li, Mao Yang, Zhongjiang Yan, Annan Yang, Der-Jiunn Deng, and
  Kwang-Cheng Chen.
\newblock Survey and performance evaluation of the upcoming next generation
  {WLANs} standard-{IEEE} 802.11 ax.
\newblock {\em Mobile Networks and Applications}, 24(5):1461--1474, 2019.

\bibitem{khorov2018tutorial}
Evgeny Khorov, Anton Kiryanov, Andrey Lyakhov, and Giuseppe Bianchi.
\newblock {A Tutorial on IEEE 802.11 ax High Efficiency WLANs}.
\newblock {\em IEEE Communications Surveys \& Tutorials}, 2018.

\bibitem{merlin2009methods}
Simone Merlin and Santosh Abraham.
\newblock {Methods for improving medium reuse in IEEE 802.11 networks}.
\newblock In {\em Consumer Communications and Networking Conference, 2009. CCNC
  2009. 6th IEEE}, pages 1--5. IEEE, 2009.

\bibitem{guo2003spatial}
Xingang Guo, Sumit Roy, and W~Steven Conner.
\newblock Spatial reuse in wireless ad-hoc networks.
\newblock In {\em 2003 IEEE 58th Vehicular Technology Conference. VTC 2003-Fall
  (IEEE Cat. No. 03CH37484)}, volume~3, pages 1437--1442. IEEE, 2003.

\bibitem{zhu2004adapting}
Jing Zhu, Xingang Guo, L~Lily~Yang, W~Steven~Conner, Sumit Roy, and Mousumi~M
  Hazra.
\newblock Adapting physical carrier sensing to maximize spatial reuse in 802.11
  mesh networks.
\newblock {\em wireless communications and mobile computing}, 4(8):933--946,
  2004.

\bibitem{mhatre2007interference}
Vivek~P Mhatre, Konstantina Papagiannaki, and Francois Baccelli.
\newblock Interference mitigation through power control in high density 802.11
  wlans.
\newblock In {\em IEEE INFOCOM 2007-26th IEEE International Conference on
  Computer Communications}, pages 535--543. IEEE, 2007.

\bibitem{zhou2005balancing}
Yihong Zhou and Scott~M Nettles.
\newblock {Balancing the hidden and exposed node problems with power control in
  CSMA/CA-based wireless networks}.
\newblock In {\em Wireless Communications and Networking Conference, 2005
  IEEE}, volume~2, pages 683--688. IEEE, 2005.

\bibitem{wilhelmi2019potential}
Francesc Wilhelmi, Sergio Barrachina-Mu{\~n}oz, Boris Bellalta, Cristina Cano,
  Anders Jonsson, and Gergely Neu.
\newblock {Potential and pitfalls of multi-armed bandits for decentralized
  spatial reuse in WLANs}.
\newblock {\em Journal of Network and Computer Applications}, 127:26--42, 2019.

\bibitem{mori2014performance}
Masahito Mori et~al.
\newblock {Performance Analysis of BSS Color and DSC}.
\newblock {\em Nov}, 3:11--14, 2014.

\bibitem{shen2018research}
Zhao Shen, Bo~Li, Mao Yang, Zhongjiang Yan, Xiaobo Li, and Yi~Jin.
\newblock {Research and Performance Evaluation of Spatial Reuse Technology for
  Next Generation WLAN}.
\newblock In {\em International Wireless Internet Conference}, pages 41--51.
  Springer, 2018.

\bibitem{barrachina2019komondor}
Sergio Barrachina-Mu{\~n}oz, Francesc Wilhelmi, Ioannis Selinis, and Boris
  Bellalta.
\newblock {Komondor: a Wireless Network Simulator for Next-Generation
  High-Density WLANs}.
\newblock In {\em 2019 Wireless Days (WD)}, pages 1--8. IEEE, 2019.

\bibitem{li2011achieving}
Wei Li, Yong Cui, Xiuzhen Cheng, Mznah~A Al-Rodhaan, and Abdullah Al-Dhelaan.
\newblock {Achieving proportional fairness via AP power control in multi-rate
  WLANs}.
\newblock {\em IEEE Transactions on Wireless Communications},
  10(11):3784--3792, 2011.

\bibitem{jamil2016novel}
Imad Jamil, Laurent Cariou, and Jean-Fran H{\'e}lard.
\newblock {Novel learning-based spatial reuse optimization in dense WLAN
  deployments}.

\bibitem{nakahira2014centralized}
Toshiro Nakahira, Koichi Ishihara, Yusuke Asai, Yasushi Takatori, Riichi Kudo,
  and Masato Mizoguchi.
\newblock {Centralized control of carrier sense threshold and channel bandwidth
  in high-density WLANs}.
\newblock In {\em Microwave Conference (APMC), 2014 Asia-Pacific}, pages
  570--572. IEEE, 2014.

\bibitem{chevillat2005dynamic}
Pierre Chevillat, Jens Jelitto, and Hong~Linh Truong.
\newblock {Dynamic data rate and transmit power adjustment in IEEE 802.11
  wireless LANs}.
\newblock {\em International Journal of Wireless Information Networks},
  12(3):123--145, 2005.

\bibitem{tang2011improving}
Suhua Tang, Akio Hasegawa, Riichiro Nagareda, Akito Kitaura, Tatsuo Shibata,
  and Sadao Obana.
\newblock {Improving throughput of wireless LANs with transmit power control
  and slotted channel access}.
\newblock In {\em Personal Indoor and Mobile Radio Communications (PIMRC), 2011
  IEEE 22nd International Symposium on}, pages 834--838. IEEE, 2011.

\bibitem{chau2017effective}
Chi-Kin Chau, Ivan~WH Ho, Zhenhui Situ, Soung~Chang Liew, and Jialiang Zhang.
\newblock {Effective static and adaptive carrier sensing for dense wireless
  CSMA networks}.
\newblock {\em IEEE Transactions on Mobile Computing}, 16(2):355--366, 2017.

\bibitem{wilhelmi2019collaborative}
Francesc Wilhelmi, Cristina Cano, Gergely Neu, Boris Bellalta, Anders Jonsson,
  and Sergio Barrachina-Mu{\~n}oz.
\newblock {Collaborative spatial reuse in wireless networks via selfish
  multi-armed bandits}.
\newblock {\em Ad Hoc Networks}, 88:129--141, 2019.

\bibitem{afaqui2015evaluation}
M~Shahwaiz Afaqui, Eduard Garcia-Villegas, Elena Lopez-Aguilera, Graham Smith,
  and Daniel Camps.
\newblock {Evaluation of dynamic sensitivity control algorithm for IEEE 802.11
  ax}.
\newblock In {\em Wireless Communications and Networking Conference (WCNC),
  2015 IEEE}, pages 1060--1065. IEEE, 2015.

\bibitem{afaqui2016dynamic}
M~Shahwaiz Afaqui, Eduard Garcia-Villegas, Elena Lopez-Aguilera, and Daniel
  Camps-Mur.
\newblock {Dynamic sensitivity control of access points for IEEE 802.11 ax}.
\newblock In {\em Communications (ICC), 2016 IEEE International Conference on},
  pages 1--7. IEEE, 2016.

\bibitem{kulkarni2015taming}
Parag Kulkarni and Fengming Cao.
\newblock {Taming the densification challenge in next generation wireless LANs:
  An investigation into the use of dynamic sensitivity control}.
\newblock In {\em Wireless and Mobile Computing, Networking and Communications
  (WiMob), 2015 IEEE 11th International Conference on}, pages 860--867. IEEE,
  2015.

\bibitem{selinis2016evaluation}
Ioannis Selinis, Marcin Filo, Seiamak Vahid, Jonathan Rodriguez, and Rahim
  Tafazolli.
\newblock {Evaluation of the DSC algorithm and the BSS color scheme in dense
  cellular-like IEEE 802.11 ax deployments}.
\newblock In {\em Personal, Indoor, and Mobile Radio Communications (PIMRC),
  2016 IEEE 27th Annual International Symposium on}, pages 1--7. IEEE, 2016.

\bibitem{selinis2017exploiting}
Ioannis Selinis, Konstantinos Katsaros, Seiamak Vahid, and Rahim Tafazolli.
\newblock {Exploiting the Capture Effect on DSC and BSS Color in Dense IEEE
  802.11 ax Deployments}.
\newblock In {\em Proceedings of the Workshop on ns-3}, pages 47--54. ACM,
  2017.

\bibitem{selinis2018control}
Ioannis Selinis, Konstantinos Katsaros, Seiamak Vahid, and Rahim Tafazolli.
\newblock {Control OBSS/PD Sensitivity Threshold for IEEE 802.11 ax BSS Color}.
\newblock In {\em 2018 IEEE 29th Annual International Symposium on Personal,
  Indoor and Mobile Radio Communications (PIMRC)}, pages 1--7. IEEE, 2018.

\bibitem{ropitault2018evaluation}
Tanguy Ropitault.
\newblock Evaluation of {RTOT} algorithm: A first implementation of
  {OBSS\_PD}-based {SR} method for {IEEE 802.11} ax.
\newblock In {\em 2018 15th IEEE Annual Consumer Communications \& Networking
  Conference (CCNC)}, pages 1--7. IEEE, 2018.

\bibitem{lee2007experimental}
Jeongkeun Lee, Wonho Kim, Sung-Ju Lee, Daehyung Jo, Jiho Ryu, Taekyoung Kwon,
  and Yanghee Choi.
\newblock An experimental study on the capture effect in 802.11 a networks.
\newblock In {\em Proceedings of the second ACM international workshop on
  Wireless network testbeds, experimental evaluation and characterization},
  pages 19--26, 2007.

\bibitem{durvy2007modeling}
Mathilde Durvy, Olivier Dousse, and Patrick Thiran.
\newblock Modeling the 802.11 protocol under different capture and sensing
  capabilities.
\newblock In {\em IEEE INFOCOM 2007-26th IEEE International Conference on
  Computer Communications}, pages 2356--2360. IEEE, 2007.

\bibitem{bellalta2019ap}
Boris Bellalta and Katarzyna Kosek-Szott.
\newblock {AP-initiated multi-user transmissions in IEEE 802.11 ax WLANs}.
\newblock {\em Ad Hoc Networks}, 85:145--159, 2019.

\bibitem{tgax2016obss_pd_evaluation}
Matthew Fischer.
\newblock {Simulation-based evaluation of OBSS PD based SR default parameters,
  doc.: IEEE 802.11-16/1161r1}, September 2016.

\bibitem{bellalta2014throughput}
Boris Bellalta, Alessandro Zocca, Cristina Cano, Alessandro Checco, Jaume
  Barcelo, and Alexey Vinel.
\newblock {Throughput analysis in CSMA/CA networks using continuous time Markov
  networks: a tutorial}.
\newblock In {\em Wireless Networking for Moving Objects}, pages 115--133.
  Springer, 2014.

\bibitem{bellalta2016throughput}
Boris Bellalta.
\newblock Throughput analysis in high density {WLANs}.
\newblock {\em IEEE Communications Letters}, 21(3):592--595, 2016.

\bibitem{barrachina2019dynamic}
Sergio Barrachina-Mu{\~n}oz, Francesc Wilhelmi, and Boris Bellalta.
\newblock {Dynamic Channel Bonding in Spatially Distributed High-Density
  WLANs}.
\newblock {\em IEEE Transactions on Mobile Computing}, 2019.

\bibitem{barrachina2019overlap}
Sergio Barrachina-Mu{\~n}oz, Francesc Wilhelmi, and Boris Bellalta.
\newblock To overlap or not to overlap: Enabling channel bonding in
  high-density {WLANs}.
\newblock {\em Computer Networks}, 152:40--53, 2019.

\bibitem{gupta2000capacity}
Piyush Gupta and Panganmala~R Kumar.
\newblock The capacity of wireless networks.
\newblock {\em IEEE Transactions on information theory}, 46(2):388--404, 2000.

\bibitem{zhao2016stochastic}
Xiaoguang Zhao, Xiangming Wen, Tao Lei, Zhaoming Lu, and Biao Zhang.
\newblock On stochastic geometry analysis of dense wlan with dynamic carrier
  sense threshold and rate control.
\newblock In {\em 2016 19th International Symposium on Wireless Personal
  Multimedia Communications (WPMC)}, pages 211--216. IEEE, 2016.

\bibitem{zhang2015stochastic}
Zhiwei Zhang, Yunzhou Li, Kaizhi Huang, and Chen Liang.
\newblock On stochastic geometry modeling of wlan capacity with dynamic
  sensitive control.
\newblock In {\em 2015 13th International Symposium on Modeling and
  Optimization in Mobile, Ad Hoc, and Wireless Networks (WiOpt)}, pages 78--83.
  IEEE, 2015.

\bibitem{iwata2019stochastic}
Motoki Iwata, Koji Yamamoto, Bo~Yin, Takayuki Nishio, Masahiro Morikura, and
  Hirantha Abeysekera.
\newblock Stochastic geometry analysis of individual carrier sense threshold
  adaptation in ieee 802.11 ax wlans.
\newblock {\em IEEE Access}, 7:161916--161927, 2019.

\bibitem{nguyen2007stochastic}
Huu~Quynh Nguyen, Fran{\c{c}}ois Baccelli, and Daniel Kofman.
\newblock A stochastic geometry analysis of dense ieee 802.11 networks.
\newblock In {\em IEEE INFOCOM 2007-26th IEEE International Conference on
  Computer Communications}, pages 1199--1207. IEEE, 2007.

\bibitem{union2015imt}
IT~Union.
\newblock Imt traffic estimates for the years 2020 to 2030.
\newblock {\em Report ITU}, pages 2370--0, 2015.

\bibitem{michaloliakos2016performance}
Antonios Michaloliakos, Ryan Rogalin, Yonglong Zhang, Konstantinos Psounis, and
  Giuseppe Caire.
\newblock Performance modeling of next-generation wifi networks.
\newblock {\em Computer Networks}, 105:150--165, 2016.

\bibitem{liew2010back}
Soung~Chang Liew, Cai~Hong Kai, Hang~Ching Leung, and Piu Wong.
\newblock Back-of-the-envelope computation of throughput distributions in csma
  wireless networks.
\newblock {\em IEEE Transactions on Mobile Computing}, 9(9):1319--1331, 2010.

\bibitem{wilhelmi2019sfctm_spatial_reuse}
Francesc Wilhelmi and Sergio Barrachina-Mu\~noz.
\newblock {IEEE 802.11ax Single-Channel Spatial Reuse in the Spatial Flexible
  Continuous Time Markov Network (11axSR-SFCTMN)}.
\newblock \url{https://github.com/sergiobarra/SFCTMN}, 2019.

\bibitem{bankov2018ofdma}
Dmitry Bankov, Andre Didenko, Evgeny Khorov, and Andrey Lyakhov.
\newblock {OFDMA Uplink Scheduling in IEEE 802.11 ax Networks}.
\newblock In {\em 2018 IEEE International Conference on Communications (ICC)},
  pages 1--6. IEEE, 2018.

\bibitem{dovelos2018optimal}
Konstantinos Dovelos and Boris Bellalta.
\newblock {Optimal Resource Allocation in IEEE 802.11ax Uplink OFDMA with
  Scheduled Access}.
\newblock {\em arXiv preprint arXiv:1811.00957}, 2019.

\bibitem{liao2016mu}
Ruizhi Liao, Boris Bellalta, Miquel Oliver, and Zhisheng Niu.
\newblock {MU-MIMO MAC protocols for wireless local area networks: A survey}.
\newblock {\em IEEE Communications Surveys \& Tutorials}, 18(1):162--183, 2016.

\bibitem{nurchis2019target}
Maddalena Nurchis and Boris Bellalta.
\newblock {Target wake time: scheduled access in IEEE 802.11 ax WLANs}.
\newblock {\em IEEE Wireless Communications}, 26(2):142--150, 2019.

\bibitem{smith2017dynamic}
Graham~Kenneth Smith.
\newblock Dynamic sensitivity control for wireless devices, March~14 2017.
\newblock US Patent 9,596,702.

\bibitem{long2007non}
Chengnian Long, Qian Zhang, Bo~Li, Huilong Yang, and Xinping Guan.
\newblock Non-cooperative power control for wireless ad hoc networks with
  repeated games.
\newblock {\em IEEE Journal on Selected Areas in Communications},
  25(6):1101--1112, 2007.

\bibitem{naddafzadeh2010distributed}
Ghasem Naddafzadeh-Shirazi, Peng-Yong Kong, and Chen-Khong Tham.
\newblock Distributed reinforcement learning frameworks for cooperative
  retransmission in wireless networks.
\newblock {\em IEEE Transactions on Vehicular Technology}, 59(8):4157--4162,
  2010.

\bibitem{zhou2011reinforcement}
Pan Zhou, Yusun Chang, and John~A Copeland.
\newblock Reinforcement learning for repeated power control game in cognitive
  radio networks.
\newblock {\em IEEE Journal on Selected Areas in Communications}, 30(1):54--69,
  2011.

\bibitem{ghadimi2017reinforcement}
Euhanna Ghadimi, Francesco~Davide Calabrese, Gunnar Peters, and Pablo Soldati.
\newblock A reinforcement learning approach to power control and rate
  adaptation in cellular networks.
\newblock In {\em 2017 IEEE International Conference on Communications (ICC)},
  pages 1--7. IEEE, 2017.

\bibitem{rhode2017whitepaper}
Lisa Ward.
\newblock {1MA222: IEEE 802.11ax Technology Introduction}.
\newblock {\em Rhode \& Schwarz. White paper 10.2016 - 1MA222\_0e}, 2017.

\bibitem{pathloss11ax}
Simone Merlin, G~Barriac, H~Sampath, et~al.
\newblock Tgax simulation scenarios.
\newblock {\em IEEE802}, pages 11--14, 2015.

\end{thebibliography}

\end{document}